\documentclass[print, aps, prd, nofootinbib, superscriptaddress]{revtex4-2}

\usepackage{amssymb, amsmath, graphicx, subcaption, epsfig, bm, appendix}
\usepackage{slashed}
\usepackage{cancel}

\DeclareMathOperator{\tr}{Tr}
\renewcommand{\d}{\mathrm{d}}

\newcommand{\xB}{x_{\scriptscriptstyle B}}
\newcommand{\sT}{{\scriptscriptstyle T}}
\newcommand{\sL}{{\scriptscriptstyle L}}

\newcommand{\qs}{q \!\!\! /}
\newcommand{\ks}{k \!\!\! /}

\newcommand{\Kos}{{K_1} \!\!\!\!\! \! \! / ~\,}
\newcommand{\Kts}{{K_2} \!\!\!\!\! \! \! / ~\,}

\usepackage[usenames,dvipsnames]{color}
\usepackage[normalem]{ulem} 
\renewcommand\sout{\bgroup \color[rgb]{0.55,0.00,0.99} \ULdepth=-.5ex \ULset}

\newcommand{\hide}[1]{}

\begin{document}

\title{Gluon GTMDs in the exclusive electroproduction of heavy-quark pairs}

\author{Mattia Bellotti}
\email{m.bellotti@studenti.unica.it}
\affiliation{Dipartimento di Fisica, Università di Cagliari, Cittadella Universitaria, I-09042 Monserrato (CA), Italy}
\affiliation{INFN, Sezione di Cagliari, Cittadella Universitaria, I-09042 Monserrato (CA), Italy}

\author{Daniël Boer}
\email{d.boer@rug.nl}
\affiliation{Van Swinderen Institute for Particle Physics and Gravity, University of Groningen, Nijenborgh 4, 9747 AG Groningen, The Netherlands}

\author{Cristian Pisano}
\email{cristian.pisano@unica.it}
\affiliation{Dipartimento di Fisica, Università di Cagliari, Cittadella Universitaria, I-09042 Monserrato (CA), Italy}
\affiliation{INFN, Sezione di Cagliari, Cittadella Universitaria, I-09042 Monserrato (CA), Italy}

\begin{abstract}
We study exclusive electroproduction of heavy quark-antiquark pairs off nucleons in the framework of generalized transverse momentum dependent parton distributions (GTMDs) for gluons. The short-distance part of the process is treated at leading order in perturbative Quantum Chromodynamics and in first order in a collinear expansion, which allows identification with the description in terms of Generalized Parton Distributions (GPDs). For the results for the structure functions in terms of GTMDs and GPDs we consider only unpolarized (spin-averaged) nucleons, but include all possible azimuthal modulations that can arise. The presented results extend known expressions in the literature and are relevant for experimental studies of this exclusive process at the future Electron Ion Collider. Furthermore, we introduce a convenient decomposition of the gluon-gluon correlation matrix in terms of GTMDs, expanded in a Lorentz basis of symmetric traceless tensors obtained from the partonic momentum $k_\sT$ and the momentum transfer $\Delta_\sT$. The adopted notation for the GTMDs relates to the nucleon helicity states at the amplitude level, rather than to polarization states of the incoming nucleon or of the gluons, which makes it more transparent which contributions from helicity difference and helicity flip matrix elements can be accessed with unpolarized nucleon beams. 
\end{abstract}

\date{\today}

\maketitle

\section{Introduction}

Hard exclusive processes, like Deeply Virtual Compton Scattering, have been considered extensively for the study of Generalized Parton Distributions (GPDs), 
mostly quark GPDs. 
Exclusive dijet production in coherent diffractive scattering has been suggested for its sensitivity to gluon GPDs
specifically \cite{Braun:2005rg}. Apart from being exclusive, one selects those events that have a rapidity gap in order to select the gluon distributions,
and in which the proton stays intact (coherent diffraction), otherwise one is 
not probing a GPD but some more general transition amplitude. Besides dijets also heavy quarkonium production has been studied which, together with restriction to the small-$x$ region, enhances the contribution from gluons, allowing studies of gluon saturation like in Ref.~\cite{Kowalski:2006hc}. Analysis of the small-$x$ region shows however that the 
distributions involved are generally not collinear distributions such as GPDs, but rather are transverse momentum dependent \cite{Kowalski:2006hc,Altinoluk:2015dpi,Hatta:2016dxp}. This requires the process to be expressed in terms of more general 
quantities, i.e.\ Wigner distributions or Generalized Transverse Momentum Dependent parton distributions (GTMDs), which are of particular interest to 
the study of orbital angular momentum and spin-orbit correlations inside the nucleon \cite{Bhattacharya:2022vvo,Bhattacharya:2024sck}. 
Although the processes do not directly probe the transverse momentum, in 
Refs.\ \cite{Hatta:2016dxp,Boer:2021upt,Boer:2023mip} it was shown that one is sensitive to transverse momentum integrals of GTMDs with weights that are dependent on the kinematical variables of the process and can therefore be varied, allowing a study of the GTMDs in this way. Different processes moreover come with different weights and therefore, the more processes that can be considered the better. The present paper adds a further process to this list: open heavy quark pair production or in practice, $D$-meson or $B$-meson pair production. We will perform two calculations of the differential cross section: one in terms of GTMDs and one in terms of GPDs to study whether the two match when the former is considered to the first order in a collinear expansion of the hard scattering. In that case one is effectively restricting to expressions involving GPDs, but expressed in terms of particular integrals of GTMDs\footnote{GPDs can be viewed as specific integrals of GTMDs, but they also determine the large transverse momentum behavior of the GTMDs, see Refs.~\cite{Bertone:2022awq,Bertone:2025vgy} for the one-loop expressions.}. We confirm that the two ways of calculating indeed agree. This also makes it clear which terms in the cross section provide constraints on which GTMDs. The expressions extend available GPD results \cite{Braun:2005rg,Chall:2026oes,Pang:2026lsr}. 
Although we restrict to unpolarized (spin averaged) incoming nucleons, we pay explicit 
attention to the helicity states of the outgoing nucleons to show the helicity difference and helicity flip contributions to the scattering off unpolarized nucleons. This extends 
the observation of Ref.~\cite{Boussarie:2019vmk} that one can become sensitive to the Sivers function in exclusive scattering with an unpolarized nucleon, or rather to the GTMD that in the forward limit becomes the Sivers Transverse Momentum Dependent distribution function (TMD). We also provide all possible angular distributions of the final state heavy quark pair, 
thereby providing more options to access specific GTMDs.  

Although there have been other suggested processes to access GTMDs, such as exclusive double Drell-Yan that probes quark GTMDs \cite{Bhattacharya:2017bvs} and exclusive double $\eta_c$ production that probes gluon GTMDs \cite{Bhattacharya:2018lgm}, 
the latter process is described by a product of four 
gluon GTMDs that have a gauge link structure that is referred to as $[+,+]$ or in a small-$x$ context as Weizs\"acker-Williams. 
In the process considered in the present 
paper the expressions involve products of just two gluon GTMDs that have a dipole link structure $[+,-]$ (for another process see Ref.~\cite{Boer:2018vdi}). The dipole gluon GTMD correlator 
is of particular interest because in the small-$x$ limit it becomes a Wilson loop correlator \cite{Dominguez:2010xd,Dominguez:2011wm,Boer:2015pni,Boer:2016xqr,Boer:2018vdi} which is most directly linked 
to pomeron and odderon operator matrix elements appearing in small-$x$ treatments based on the dipole scattering amplitude and the Color Glass Condensate (CGC)
framework (see also Ref.~\cite{Benic:2026idy}). In the present paper we do not focus specifically on this limit, even if in practice the $x$ values may be small, but we do provide results for this limit. We also do not put a restriction on whether the skewness parameter $\xi$ is larger or smaller than $x$. The main purpose is to find all the possible angular dependences for the general case.  

This paper is organized as follows. In Section~\ref{sec:GTMD-def} we recall the operator definition of the gluon-gluon correlation matrix within the framework of Quantum Chromodynamics (QCD) and provide its parametrization in terms of sixteen, leading twist\footnote{Note that we use the term leading twist to mean leading power in the collinear expansion in inverse powers of the hard scale. Just like for TMDs the nonlocal operators of the GTMDs are not of definite twist in the local OPE sense.} GTMDs using a Lorentz basis of symmetric traceless tensors built from the gluon momentum $k_\sT$ and the momentum transfer $\Delta_\sT$. Moreover, we provide the TMDs and GPDs that derive from the GTMDs, where the TMDs are obtained in the forward limit and the GPDs through transverse momentum integrals. 
In Section~\ref{sec:GPD} a similar parametrization, this time directly expressed in terms of GPDs, is presented. The details of the calculation of the cross section for the gluon channel of the exclusive process $e\, N \to e\, Q\, \overline Q\, N$ can be found in Section~\ref{sec:cs-details}. The resulting angular modulations of the cross section are given in Section~\ref{sec:SF-GTMD} as integrals of the GTMDs over $k_\sT$, and in Section~\ref{sec:SF-GPD} in the collinear limit, in terms of GPDs. Section~\ref{sec:conclusions} contains our summary and conclusions. In Appendix~\ref{app-1} we provide the correspondences between our GTMDs and the ones available in the literature employing different notations. In Appendix~\ref{app-2} we show the expression of the square of the gluon-gluon correlator, contributing to the unpolarized cross section, before integration over $k_\sT$. This intermediate result should be helpful in reproducing the results presented in the main text. In Appendix~\ref{app-3} we provide the expressions for all leading twist GPDs in terms of integrals of GTMDs, which complement the expressions given in Section~\ref{sec:GTMD-def}. Finally, in Appendix~\ref{app-4} we include some more results on the small-$x$ limit for the dipole gauge link configuration that is the one related to the Wilson loop correlator and appearing in the process under consideration. We comment on the elliptic parts of GTMDs and on the small-$x$ limit of the expressions given in Appendix~\ref{app-3} that has been considered in the literature. 

\section{The gluon-gluon correlator for GTMDs}
\label{sec:GTMD-def}

\begin{figure}[t]
    \includegraphics[width=0.7\linewidth, keepaspectratio, trim={5cm 21cm 3cm 4cm},clip]{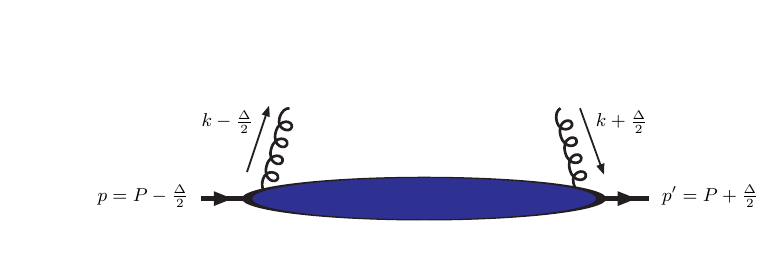}
\caption{The off-forward gluon-gluon correlator.} 
\label{fig:corr}
\end{figure}

The gluon-gluon GTMD correlator describes the nucleon-to-gluon transition~\cite{Lorce:2013pza}, as pictorially represented in Fig.~\ref{fig:corr}, where we have defined the momentum transfer 
\begin{align}
\Delta = p^\prime - p = k_g^\prime - k_g \,,   
\end{align}
and the average nucleon and gluon momenta
\begin{align}
P = \frac{p + p^\prime}{2}\,, \qquad   k = \frac{k_g + k_g^\prime}{2}\,,
\end{align}
with $p\, (p^\prime)$ and $k_g\, (k_g^\prime)$ being the initial (final) nucleon and gluon momenta, respectively. By imposing the on-shell condition for the nucleon, namely  $p^2= p^{\prime 2} = M_N^2$, with $M_N$ being the nucleon mass, the following relations must hold, 
\begin{align}
P\cdot \Delta = 0\,, \qquad \Delta^2 = 4 (M_N^2 - P^2)\,.   
\label{eq:on-shell-N}
\end{align}
Furthermore, we introduce two light-like vectors $n$ and $\overline n$, with $n\cdot \overline n=1$, so that one can perform a Sudakov decomposition of any four-vector $v$,  $v^\mu = v^+ n^\mu + v^- \overline n^\mu + v_\sT^\mu$. Hence, the light-cone components are defined as $v^+ \equiv v \cdot \overline n$ and $v^-\equiv v\cdot n$, while perpendicular vectors $v_\sT$ always refer to the components of $v$ orthogonal to both $n$ and $\overline n$, with $v_\sT^2 = - \bm v_\sT^2$. In a generic frame where $P$ has no transverse component, we can therefore write
\begin{align}
P^\mu & = P^+ n^\mu + \frac{P^2}{2 P^+}\,\overline n^\mu \,, \nonumber \\
k^\mu & = xP^+n^\mu + \frac{k^2 + \bm k_\sT^2}{2 x P^+}\, \overline n^\mu + k^\mu_\sT\,,\nonumber \\
\Delta^\mu & = -2 \xi P^+n^\mu +  \xi\, \frac{P^2}{P^+}\,\overline n^\mu +  \Delta_\sT^\mu\,,
\end{align}
where we have introduced the average light-cone momentum fraction $x = k^+/P^+$, the skewness parameter $\xi$ defined by $2\xi \equiv -\Delta^+/P^+$ and imposed the orthogonality condition $P\cdot \Delta =0$ given in Eq.~\eqref{eq:on-shell-N}. Moreover, from the relations $t \equiv \Delta^2 = -4 \xi^2 P^2 - \bm \Delta_\sT^2$ and $\Delta^2 = 4 (M_N^2 - P^2)$ in Eq.~\eqref{eq:on-shell-N}, one obtains
\begin{align}
P^2 =  \frac{M_N^2 + \frac{1}{4}\,\bm \Delta_\sT^2}{1-\xi^2}\,, \qquad\qquad  \Delta^2 = -  \frac{4 \xi^2 M_N^2 + \bm \Delta^2_\sT}{1-\xi^2}\,.
\end{align}

We can now provide a definition of the gluon-gluon correlator in terms of QCD operators, which is also manifestly covariant: it is given by the  Fourier transform of a nonlocal, off-forward hadronic matrix element of two gluon field tensors between two nucleon states, 
\begin{equation}
\Gamma_g^{[{\cal U}, {\cal U}^\prime]\mu\nu}(x,\bm k_\sT, \xi, \bm \Delta_\sT)
= \frac{\overline n_\rho\,\overline n_\sigma}{x\, (P^+)^2}
{\int}\frac{\d( \lambda {\cdot}P)\,\d^2 \lambda_\sT}{(2\pi)^3}\
e^{i k\cdot \lambda}\,
\left \langle p^\prime \left |\,\tr\left [\,F^{\mu\rho}\left (-\frac{\lambda}{2} \right )\, {\cal U}_{\left [-\frac{\lambda}{2},\frac{\lambda}{2} \right ]}\,
F^{\nu\sigma}\left (\frac{\lambda}{2} \right )\, {\cal U}^\prime_{\left [\frac{\lambda}{2},-\frac{\lambda}{2} \right ]} \right ]\,
\,\right |p \right \rangle\,\bigg\rfloor_{\text{LF}}\,,
\label{eq:corr}
\end{equation}
where the nonlocality is restricted to the  light-front LF ($\lambda \cdot \overline n \equiv \lambda^+ =0$). 
The process dependent gauge links ${\cal U}_{[-\frac{\lambda}{2},\frac{\lambda}{2}]}$ and ${\cal U}^\prime_{[-\frac{\lambda}{2},\frac{\lambda}{2}]}$ are needed to preserve gauge invariance.

The above off-forward matrix element is a $2\times 2$ matrix in helicity space, meaning that it can be decomposed into helicity flip and non-flip components:
\begin{align}
\Gamma_{\lambda' \lambda} = \langle p',\lambda' | O | p,\lambda \rangle = \left[\Gamma_U {\bm 1}+ \vec{\Gamma}\cdot \vec{\bm \sigma} \right]_{\lambda' \lambda}, \label{eq:Gamma-lambda}  
\end{align}
where we will later on use the notation that $\Gamma_3 = \Gamma_L$, $\Gamma_1=\Gamma_{T}^x$, and $\Gamma_2=\Gamma_{T}^y$. Despite what this notation suggests, it is important to emphasize that this decomposition does not correspond to the polarization states of the incoming nucleon. 
Only when the amplitude squared is considered, one is dealing with the spin decomposition of the incoming nucleon:
\begin{align}
\sum_{\lambda'} \Gamma^*_{\lambda'' \lambda'} \Gamma_{\lambda' \lambda} & = \sum_{\lambda'}\langle p,\lambda'' | O^\dagger | p',\lambda' \rangle \langle p',\lambda' | O | p,\lambda \rangle  
= \left[\left(\Gamma_U\Gamma_U^*+\vec{\Gamma}\cdot \vec{\Gamma}^* \right){\bm 1}+ 
\left(\vec{\Gamma}\Gamma_U^* + \Gamma_U \vec{\Gamma}^* + i \left(\vec{\Gamma} \times \vec{\Gamma}^* \right) \right) \cdot \vec{\bm \sigma}  \right]_{\lambda'' \lambda}
\label{productdensity}
\end{align}
To project out the polarization states of the nucleon one takes the trace with the spin density matrix $\frac{1}{2} \left[{\bm 1}+ 
\vec{S} \cdot \vec{\bm \sigma}  \right]$, where $\vec{S}$ denotes the polarization 3-vector. 
This means that for an unpolarized incoming nucleon one has contributions from not just $\Gamma_U\Gamma_U^*$, but also from $\Gamma_L\Gamma_L^*$ and $\Gamma_T\Gamma_T^*$, and for example for a nucleon transversely polarized in the $x$ direction not just from 
$\Gamma_T^x \Gamma_{U}^*$ and $\Gamma_U \Gamma_{T}^{x*}$, but also from $\Gamma_T^y \Gamma_{L}^*$ and so on. Only in the forward limit one can discuss the polarization states of the nucleon on the amplitude level like usually done for TMDs, such that for an unpolarized nucleon only $\Gamma_U$ contributes, for a longitudinally or transversely polarized nucleon only $\Gamma_L$ or  $\Gamma_T$, respectively, such that additional terms will have to be proportional to $\xi$ and/or $\Delta_T$. But for the amplitude squared such terms can remain even in the forward limit. We will see this explicitly below. This is also the reason why one can become sensitive to e.g.\ the Sivers function in exclusive scattering with an unpolarized nucleon \cite{Boussarie:2019vmk}, whereas it does not involve helicity flip states for the incoming proton, it does involve helicity flip matrix elements on the amplitude level. One thus probes the Sivers function squared. This is also the reason why one can access quark helicity distributions in elastic scattering off unpolarized protons, which is commonly used as input for the proton spin sum rule studies. 

Omitting gauge links, the correlator $\Gamma_U$ can be parametrized\footnote{The independent tensor structures correspond to different polarization states of the gluons. They are not all orthogonal, as there are two independent transverse vectors that are in general not orthogonal in the two-dimensional transverse space. However, the tensor structures do not mix under Lorentz transformations, which is the reason for employing the symmetric traceless tensors of definite rank, $a_\sT^{\mu\nu}$, $a_\sT^{\mu\nu\rho}$, and $a_\sT^{\mu\nu\rho \sigma}$, cf.\ Refs.~\cite{Buffing:2013kca,Boer:2016xqr}.} in terms of the complex-valued GTMDs ${\cal F}_i^g$, with $i=1$-$4$~\cite{Boer:2018vdi},
\begin{align}
\Gamma_U^{\mu\nu}(x,\bm k_\sT ,\xi, \bm \Delta_\sT) & =  \frac{1}{2}\,\bigg \{-g_\sT^{\mu\nu}\,{\cal F}_1^g (x,\bm k_\sT^2, \xi, \bm \Delta^2_\sT, \bm k_\sT \cdot \bm \Delta_\sT) + \frac{k_\sT^{\mu\nu}}{M_N^2} \, {\cal F}_2^g (x,\bm k_\sT^2, \xi, \bm \Delta_\sT^2, \bm k_\sT \cdot \bm \Delta_\sT)\,  \nonumber \\  
    & \qquad \qquad\qquad  +\frac{\Delta_\sT^{\mu\nu}}{M_N^2}\, {\cal F}_3^g (x,\bm k_\sT^2, \xi, \bm \Delta_\sT^2, \bm k_\sT \cdot \bm \Delta_\sT) \, + \, \frac{k_\sT^{[\mu} \Delta_\sT^{\nu ]}}{M_N^2} \,  {\cal F}_4^g (x,\bm k_\sT^2, \xi, \bm \Delta_\sT^2, \bm k_\sT \cdot \bm \Delta_\sT)  \bigg \}\,,
\label{eq:GTMDs}
\end{align}
where the symmetric transverse projector $g_\sT^{\mu\nu}$ is given by 
\begin{align}
g^{\mu\nu}_{\sT} = g^{\mu\nu} -  n^{\mu} \, \overline n^{\nu} - n^{\nu}\,\overline n^{\mu}\, ,
\end{align}
with $g_\sT^{11} = g_\sT^{22}=-1$, the symmetric traceless tensors $k_\sT^{\mu\nu}$ and $\Delta_\sT^{\mu\nu}$ are defined as $a_\sT^{\mu\nu} = a_\sT^{\mu} a_\sT^{\nu} + \frac{1}{2}\, \bm a_\sT^2 \, g^{\mu\nu}_{\sT}$, while the square brackets denote antisymmetrization of the indices: $a^{[\mu }\, b^{\nu]} = a^\mu b^\nu \,-\, a^\nu b^\mu$.  The GTMDs ${\cal F}_i^g$ are related to the leading-twist GPDs and TMDs for unpolarized hadrons by integrating over $\bm k_\sT$ and by setting $\Delta=0$, respectively. More explicitly, in the forward limit one obtains the TMDs for an unpolarized hadron:
\begin{align}
 \lim_{\Delta\to 0}\, {\cal F}_1^g = f_1^g\,, \qquad    \lim_{\Delta\to 0}\, {\cal F}_2^g = h_1^{\perp\,g}\,,\qquad \lim_{\Delta\to 0}\, {\cal F}_3^g =\lim_{\Delta\to 0}\, {\cal F}_4^g =0\,,
\end{align}
with $f_1^g$ and $h_1^{\perp\,g}$ being the unpolarized and linearly polarized gluon TMDs, respectively. Moreover, the relation between the GTMD ${\cal F}_1^g$ and the GPDs $H^g$ and $E^g$ reads~\cite{Lorce:2013pza}
\begin {align}
\int \d^2 k_\sT\, {\cal F}_1^g  = \sqrt{1-\xi^2}\,\left [ H^g \, -\, \frac{\xi^2}{1-\xi^2}\,E^g \right ]\,.
\label{eq:F1-GPDH}
\end{align}
Strictly speaking, the following relation is a tree level results, which requires a renormalization prescription beyond tree level, just like $f_1(x;\mu)$ is not simply equal to $\int \d^2 k_\sT\, f_1(x,k_\sT; \zeta, \mu)$. Similarly, one has
\begin{align}
\int \d^2 k_\sT \,\left [ \frac{k_\sT^{\mu\nu}}{M_N^2}\, {\cal F}_2^g \,+ \frac{\Delta_\sT^{\mu\nu}}{M_N^2}\, {\cal F}_3^g\right ] & =  \frac{\Delta_\sT^{\mu\nu}}{M_N^2}\int \d^2 k_\sT\, \left [ \frac{2 (\bm k_\sT \cdot \bm \Delta_\sT)^2 - \bm k_\sT^2 \,\bm \Delta_\sT^2}{\bm \Delta_\sT^4}\,{\cal F}_2^g \,+ \, {\cal F}_3^g\right ] \nonumber \\
&  = \frac{\Delta_\sT^{\mu\nu}}{M_N^2}\,\frac{1}{2\,\sqrt{1-\xi^2}}\,\left [ -2\,\widetilde{H}^g_\sT - E^g_\sT + \xi\,\widetilde{E}_\sT^g\right ] \,.
\label{eq:F23-GPDH}   
\end{align}

To discuss our decomposition of $\Gamma_L$ in terms of GTMDs, we define the antisymmetric transverse projector 
\begin{align}
\epsilon_\sT^{\mu\nu} = \epsilon^{\mu\nu\rho \sigma}\, n_\rho\, \overline  n_\sigma \equiv \epsilon^{-+\mu\nu}, 
\end{align}
with $\epsilon_\sT^{12} = - \epsilon_\sT^{21} = +1$, such that 
\begin{align}
\Gamma_L^{\mu\nu}(x,\bm k_\sT ,\xi, \bm \Delta_\sT) & =  \frac{1}{2} 
\left \{ i \,\epsilon_\sT^{\mu\nu} \, {\cal G}_1^g (x,\bm k_\sT^2, \xi, \bm \Delta_\sT^2, \bm k_\sT \cdot \bm \Delta_\sT) \,+\,  \frac{\epsilon_{\sT \, \alpha}^{\{ \mu }\,  k_\sT^{ \nu\}\alpha }}{2M_N^2} \, {\cal G}_2^g (x,\bm k_\sT^2, \xi, \bm \Delta_\sT^2, \bm k_\sT \cdot \bm \Delta_\sT)\right .  \nonumber   \\
&\qquad \,+\,\left . \frac{\epsilon_{\sT\, \alpha}^{\{ \mu } \, \Delta_\sT^{ \nu\}\, \alpha }}{2M_N^2} \, {\cal G}_3^g (x,\bm k_\sT^2, \xi, \bm \Delta_\sT^2, \bm k_\sT \cdot \bm \Delta_\sT) \, - \,i\,g_\sT^{\mu\nu} \, \frac{ \epsilon_\sT^{k_\sT\, \Delta_\sT }}{M_N^2} \,  {\cal G}_4^g (x,\bm k_\sT^2, \xi, \bm \Delta_\sT^2, \bm k_\sT \cdot \bm \Delta_\sT)\right \} \,, 
\end{align} 
where we have used the notations $\epsilon_{\sT}^{ab} = \epsilon_\sT^{\mu\nu}a_\mu b_\nu$ and $a^{\{\mu }\, b^{\nu\}} = a^\mu b^\nu \,+\, a^\nu b^\mu$. In the forward limit, one obtains the TMDs for a longitudinally polarized hadron:
\begin{align}
\lim_{\Delta\to 0}{\cal G}_1^g = g_{1\sL}^g\,, \qquad \lim_{\Delta\to 0} {\cal G}_2^g = h_{1\sL}^{\perp\,g}\,,\qquad  \lim_{\Delta\to 0}{\cal G}_3^g = \lim_{\Delta\to 0}{\cal G}_4^g = 0\,. 
\end{align}
Furthermore, in this case the relation to GPDs is~\cite{Lorce:2013pza}, 
\begin {align}
\int \d^2 k_\sT\, {\cal G}_1^g  & = \sqrt{1-\xi^2}\,\left [ \widetilde{H}^g \, - \, \frac{\xi^2}{1-\xi^2}\,\widetilde{E}^g \right ]\,, \nonumber \\
\int \d^2 k_\sT \,\left [ \frac{k_\sT^{\nu\alpha}}{M_N^2}\, {\cal G}_2^g \,+ \frac{\Delta_\sT^{\nu\alpha}}{M_N^2}\, {\cal G}_3^g\right ] & = \frac{\Delta_\sT^{\nu\alpha}}{M_N^2}\int \d^2 k_\sT\, \left [ \frac{2 (\bm k_\sT \cdot \bm \Delta_\sT)^2 - \bm k_\sT^2 \,\bm \Delta_\sT^2}{\bm \Delta_\sT^4}\,{\cal G}_2^g \,+ \, {\cal G}_3^g\right ] \nonumber \\
& = \,i\, \frac{\Delta_\sT^{\nu\alpha}}{M_N^2}\, \frac{1}{2 \sqrt{1-\xi^2}}\, \left [-\xi \, E^g_\sT + \widetilde{E}_\sT^g \right ]\,. 
\label{eq:GPDH}
\end{align}

Finally, for the correlator $\Gamma_T^i$, where $i=x,y$, we consider the following GTMD parametrization
\begin{align}
    \Gamma_T^{\mu\nu \, i }(x,\bm k_\sT ,\xi, \bm \Delta_\sT) = & \frac{1}{2}\,\bigg \{g^{\mu\nu}_\sT\,
    \frac{ \epsilon_\sT^{k_\sT i}}{M_N}\,  {\cal H}_1^g (x,\bm k_\sT^2, \xi, \bm \Delta_\sT^2, \bm k_\sT \cdot \bm \Delta_\sT) \, + \, i\, \epsilon_\sT^{\mu\nu}\,
    \frac{k_\sT^i }{M_N}\, {\cal H}_2^g (x,\bm k_\sT^2, \xi, \bm \Delta_\sT^2, \bm k_\sT \cdot \bm \Delta_\sT)  \nonumber \\
    &  \, - \,\frac{\epsilon_\sT^{k_\sT \{ \mu}g_\sT^{\nu \}i}\, + \,      \epsilon_\sT^{ i\{ \mu } k_\sT^{\nu \}}}{4M_N} \, {\cal H}_3^g (x,\bm k_\sT^2, \xi, \bm \Delta_\sT^2, \bm k_\sT \cdot \bm \Delta_\sT) \, - \, \frac{\epsilon_{\sT\,\alpha}^{\{ \mu} k_\sT^{\nu \} \alpha i}}{2M_N^3}\, {\cal H}_4^g (x,\bm k_\sT^2, \xi, \bm \Delta_\sT^2, \bm k_\sT \cdot \bm \Delta_\sT)\, \nonumber \\
   &   + \,i\,g^{\mu\nu}_\sT\,
    \frac{ \epsilon_\sT^{\Delta_\sT i}}{M_N}\,  {\cal H}_5^g (x,\bm k_\sT^2, \xi, \bm \Delta_\sT^2, \bm k_\sT \cdot \bm \Delta_\sT) \, + \,\epsilon_\sT^{\mu\nu}\,
    \frac{\Delta_\sT^i}{M_N}\, {\cal H}_6^g (x,\bm k_\sT^2, \xi, \bm \Delta_\sT^2, \bm k_\sT \cdot \bm \Delta_\sT)  \nonumber \\
    &  \left .  \, - \,i\,\frac{\epsilon_\sT^{\Delta_\sT \{ \mu}g_\sT^{\nu \}i}\, + \,
      \epsilon_\sT^{ i \{ \mu } \Delta_\sT^{\nu \}}}{4M_N} \, {\cal H}_7^g (x,\bm k_\sT^2, \xi, \bm \Delta_\sT^2, \bm k_\sT \cdot \bm \Delta_\sT) \, - \, i\, \frac{\epsilon_{\sT\,\alpha}^{\{ \mu} \Delta_\sT^{\nu \} \alpha i}}{2M_N^3}\, {\cal H}_8^g (x,\bm k_\sT^2, \xi, \bm \Delta_\sT^2, \bm k_\sT \cdot \bm \Delta_\sT) \right \}\,,
      \end{align}
where we have used the notation $a_\sT^{\mu\nu\rho}  =  a_\sT^\mu a_\sT^\nu a_\sT^\rho  \,{+}\,\frac{1}{4}\, \bm a_\sT^2 (g_\sT^{\mu\nu} a_\sT^\rho + g_\sT^{\mu\rho} a_\sT^{\nu} + g^{\nu\rho} a_\sT^{\mu})$. The factors of $i$ in the parametrizations of $\Gamma_{U/L/T}$ could in principle be absorbed in the GTMDs, which are generally complex functions, but as presented here all GTMDs satisfy $F_i^*(k,\Delta) = F_i(k,-\Delta)$, which follows from hermiticity: $\Gamma^{\mu \nu}(k,\Delta)^\dagger = \Gamma^{\nu \mu}(k,-\Delta)$.

The adopted notation for the GTMDs is linked to the nucleon helicity states {\it at the amplitude level} (${\cal F}$ for $\Gamma_U$~\cite{Boer:2018vdi}, ${\cal G}$ for $\Gamma_L$, and ${\cal H}$ for $\Gamma_T$), rather than to polarization states of the incoming nucleon (which relates the helicities of the in-states in the amplitude and the conjugate amplitude) or of the gluons (which is determined by the $\mu\nu$ indices). This choice makes it clear which contributions from helicity difference (${\cal G}$) and helicity flip (${\cal H}$) matrix elements can be accessed with unpolarized nucleon beams (in other words, the $\Gamma_L\Gamma_L^*$ and $\Gamma_T\Gamma_T^*$ contributions in Eq.\ (\ref{productdensity})), besides the helicity sum contributions (${\cal F}$). For unpolarized gluons (obtained by contracting $\Gamma^{\mu\nu}$ with $-g_{\sT\mu\nu}$), the following four functions remain: ${\cal F}_1$, ${\cal G}_4$, ${\cal H}_1$ and ${\cal H}_5$, corresponding to (linear combinations of) $F_{1,1}$, $F_{1,2}$, $F_{1,3}$ and $F_{1,4}$ of Ref.~\cite{Meissner:2009ww}. Similarly, for circularly polarized gluons (obtained by contracting $\Gamma^{\mu\nu}$ with $-i\epsilon_{\sT\mu\nu}$), only four functions remain. The other eight GTMDs correspond to linearly polarized gluons. For convenience, in Appendix~\ref{GTMDrelations} a comparison of our GTMDs to the ones of Ref.~\cite{Meissner:2009ww} and Ref.~\cite{Lorce:2013pza} is given. 

In the forward limit, from $\Gamma_T$ one obtains the TMDs for a transversely polarized hadron:
\begin{align}
\lim_{\Delta\to 0}{\cal H}_1^g & = f_{1\sT}^{\perp\, g}\,, \qquad \lim_{\Delta\to 0}{\cal H}_2^g  = g_{1\sT}^{g}\,, \qquad \lim_{\Delta\to 0} {\cal H}_3^g  = h_{1}^{g}\,,\qquad \lim_{\Delta\to 0}{\cal H}_4^g  = h_{1\sT}^{\perp\, g}\,,\qquad \lim_{\Delta\to 0}{\cal H}_{5,6,7,8}^g  =0\,.
\end{align}
On the other hand, for the connection to GPDs, in analogy with Eqs.~(4.58), (4.88) and (4.89) of Ref.~\cite{Lorce:2013pza} we find, 
\begin{align}
\int \d^2 k_\sT \left [\frac{k_\sT^\mu}{M_N} \, {\cal H}_1^g + i\,\frac{\Delta_\sT^\mu}{M_N}\, {\cal H}_5^g \right ] & =\frac{\Delta^\mu_\sT}{M_N} \int \d^2 k_\sT \,\bigg [ \frac{\bm k_\sT \cdot \bm \Delta_\sT}{\bm \Delta_\sT^2} \, {\cal H}_1^g  \,+\,  i\, {\cal H}_5^g \bigg ]   = i\, \frac{\Delta_\sT^\mu}{M_N}\,\frac{1}{2\sqrt{1-\xi^2}}\, E^g \,,\nonumber \\
\int \d^2 k_\sT \left [\frac{k_\sT^\mu}{M_N} \,i\, {\cal H}_2^g + \frac{\Delta^\mu_\sT}{M_N}\, {\cal H}_6^g \right ] & = \frac{\Delta_\sT^\mu}{M_N}  \int \d^2 k_\sT \,\bigg [\frac{\bm k_\sT \cdot \bm \Delta_\sT}{\bm \Delta_\sT^2} \,i\, {\cal H}_2^g \,+\,   {\cal H}_6^g \bigg ]   = i\, \frac{\Delta_\sT^\mu}{M_N} \,\frac{\xi}{2\sqrt{1-\xi^2}}\, \widetilde{E}^g\,, \nonumber \\
\int \d^2 k_\sT \left [\frac{k_\sT^\mu}{M_N} \, {\cal H}_3^g + i\,\frac{\Delta^\mu_\sT}{M_N}\, {\cal H}_7^g \right ] & =\frac{\Delta_\sT^\mu}{M_N}  \int \d^2 k_\sT \,\bigg [\frac{\bm k_\sT \cdot \bm \Delta_\sT}{\bm \Delta_\sT^2} \, {\cal H}_3^g \,+\, i\,  {\cal H}_7^g \bigg ]  \nonumber \\
& = -\,\frac{\Delta_\sT^\mu}{M_N} \,\frac{i}{\sqrt{1-\xi^2}}\left[ -(1-\xi^2)H^g_\sT + \xi^2\,E^g_\sT - \xi\, \widetilde{E}^g_\sT  - \frac{\bm \Delta_\sT^2}{4 M_N^2}\,\widetilde{H}^g_\sT \right ]\,,  \nonumber \\
\int \d^2 k_\sT \left [\frac{k_\sT^{\mu\nu\rho}}{M_N^3} \, {\cal H}_4^g + i\, \frac{\Delta_\sT^{\mu\nu\rho}}{M_N^3}\, {\cal H}_8^g \right ] & = \frac{\Delta^{\mu\nu\rho}_\sT}{M_N^3} \int \d^2 k_\sT \,\bigg [ \frac{4 \, (\bm k_\sT \cdot \bm \Delta_\sT)^2 - 3\,  \bm k_\sT^2\,  \bm \Delta_\sT^2 \, }{(\bm \Delta_\sT^2)^3} \,(\bm k_\sT \cdot \bm \Delta_\sT )\,  {\cal H}_4^g  \,+\, i\,  {\cal H}_8^g \bigg ]  \nonumber \\ & = -\frac{\Delta_\sT^{\mu\nu\rho}}{M_N^3}\,\frac{i}{2\sqrt{1-\xi^2}}\, \widetilde{H}^g_\sT \,.
 \label{eq:SiVE}
\end{align}

With these amplitudes, one can obtain the amplitude squared and subsequently the cross section expressions. The amplitude squared before transverse momentum integration can be found in Appendix \ref{amplitudesquared}. 
After integration over $k_{1\sT}$ and $k_{2\sT}$, the amplitude squared for an unpolarized nucleon becomes
\begin{align}
\left. \Gamma^{\mu\nu}  \Gamma^{* \rho\sigma} \right|_{S=0} &
 = \frac{1}{4}  \left \{g_\sT^{\mu\nu}  g_\sT^{\rho\sigma} \left[{\cal F}_1^0 {\cal F}_1^{0} + \frac{\bm \Delta_\sT^2}{M_N^2}\left({\cal H}_1^\phi + i {\cal H}_5^0\right)\left({\cal H}_1^{\phi} - i {\cal H}_5^{0}\right)\right] \right. \nonumber \\ 
 & \left. + \frac{\Delta_\sT^{\mu\nu} \Delta_\sT^{\rho\sigma}}{M_N^4} \left({\cal F}_2^{2\phi} + {\cal F}_3^0\right)\left({\cal F}_2^{2\phi} + {\cal F}_3^{0}\right)\right. \nonumber \\
 & \left. + \frac{g_\sT^{\mu\nu}  \Delta_\sT^{\rho\sigma}}{M_N^2} \left[- {\cal F}_1^0 \left({\cal F}_2^{2\phi} + {\cal F}_3^{0}\right) \right. \left.- \left({\cal H}_1^\phi + i {\cal H}_5^0\right)\left({\cal H}_3^{\phi} - i {\cal H}_7^{0}\right) + \frac{\bm \Delta_\sT^2}{2M_N^2}\left({\cal H}_1^\phi + i {\cal H}_5^0\right) \left({\cal H}_4^{3\phi} - i {\cal H}_8^{0}\right)\right] \right. \nonumber \\
 & \left. + \frac{\Delta_\sT^{\mu\nu} g_\sT^{\rho\sigma}}{M_N^2} \left[-\left({\cal F}_2^{2\phi} + {\cal F}_3^0\right) {\cal F}_1^{0} \right.  \left. -\left({\cal H}_3^\phi + i {\cal H}_7^0\right)\left({\cal H}_1^{\phi} - i {\cal H}_5^{0}\right) + \frac{\bm \Delta_\sT^2}{2M_N^2} \left({\cal H}_4^{3\phi} + i {\cal H}_8^{0}\right)\left( {\cal H}_1^{\phi} - i {\cal H}_5^{0}\right) \right] \right. \nonumber \\
 & \left. + \epsilon_\sT^{\mu\nu} \epsilon_\sT^{\rho\sigma} \left[{\cal G}_1^0 {\cal G}_1^{0} + \frac{\bm \Delta_\sT^2}{M_N^2}  \left(i {\cal H}_2^\phi + {\cal H}_6^0\right) \left(- i {\cal H}_2^{\phi} + {\cal H}_6^{0}\right)\right] \right. \nonumber \\
 & \left. + \frac{\epsilon_{\sT\alpha}^{\{\mu} \Delta_\sT^{\nu\}\alpha} \epsilon_{\sT\beta}^{\{\rho} \Delta_\sT^{\sigma\}\beta}}{4M_N^4} \left({\cal G}_2^{2\phi} + {\cal G}_3^0\right) \left({\cal G}_2^{2\phi} + {\cal G}_3^{0}\right)\right. \nonumber \\
 & \left. + \frac{\epsilon_\sT^{\mu\nu} \epsilon_{\sT \alpha}^{\{\rho} \Delta_\sT^{\sigma\}\alpha}}{2M_N^2} \left[i {\cal G}_1^0\left({\cal G}_2^{2\phi} + {\cal G}_3^{0}\right) \right. 
  \left. - \left(i {\cal H}_2^\phi + {\cal H}_6^0\right)\left({\cal H}_3^{\phi} - i {\cal H}_7^{0}\right) - \frac{\bm \Delta_\sT^2}{2M_N^2} \left(i {\cal H}_2^\phi + {\cal H}_6^0\right)\left({\cal H}_4^{3\phi} - i {\cal H}_8^{0}\right) \right] \right. \nonumber \\
 &  + \frac{\epsilon_{\sT \alpha}^{\{\mu} \Delta_\sT^{\nu\}\alpha} \epsilon_\sT^{\rho\sigma}}{2M_N^2} \left[-i \left({\cal G}_2^{2\phi} + {\cal G}_3^0\right) {\cal G}_1^{0} \right. \left. - \left({\cal H}_3^\phi + i {\cal H}_7^0\right) \left(- i {\cal H}_2^{\phi} + {\cal H}_6^{0}\right) \right .\nonumber \\
 & \hspace*{7cm}\left . - \frac{\bm \Delta_\sT^2}{2M_N^2} \left({\cal H}_4^{3\phi} + i {\cal H}_8^0\right) \left(- i {\cal H}_2^{\phi} + {\cal H}_6^{0}\right) \right] \nonumber \\
 &\left. - \left(g_\sT^{\mu\nu} g_\sT^{\rho\sigma}-g_\sT^{\mu\sigma} g_\sT^{\nu\rho}-g_\sT^{\mu\rho} g_\sT^{\nu\sigma}\right)\frac{\bm\Delta_\sT^2}{4M_N^2}\left[\left({\cal H}_3^\phi+i {\cal H}_7^0\right)\left({\cal H}_3^{\phi} -i {\cal H}_7^{0}\right) \right. \left. + \frac{\bm\Delta_\sT^4}{4M_N^4} \left({\cal H}_4^{3\phi} + i {\cal H}_8^0\right)\left({\cal H}_4^{3\phi} - i {\cal H}_8^{0}\right)\right] \right. \nonumber \\
 &\left. - \frac{\Delta_\sT^{\mu\nu\rho\sigma}}{M_N^4} \left[\left({\cal H}_3^\phi + i {\cal H}_7^0\right)\left({\cal H}_4^{3\phi} - i {\cal H}_8^{0}\right) + \left({\cal H}_4^{3\phi} + i {\cal H}_8^{0}\right)  \left({\cal H}_3^{\phi} - i {\cal H}_7^{0}\right)\right]\right\},
 \end{align}
with the symmetric traceless tensor $a_\sT^{\mu \nu \rho \sigma} \equiv \;a_\sT^\mu a_\sT^\nu  a_\sT^\rho  a_\sT^\sigma + \frac{1}{6} \bm{a}_\sT^2 \left( g_\sT^{\mu \nu} a_\sT^{\rho \sigma} + g_\sT^{\mu \rho} a_\sT^{\nu \sigma} + g_\sT^{\mu \sigma} a_\sT^{\nu \rho} + g_\sT^{\nu \rho} a_\sT^{\mu \sigma} + g_\sT^{\nu \sigma} a_\sT^{\mu \rho} + g_\sT^{\rho \sigma} a_\sT^{\mu \nu} \right)$ $ - \frac{1}{8} \bm{a}_\sT^4 \left( g_\sT^{\mu \nu} g_\sT^{\rho \sigma} + g_\sT^{\mu \rho} g_\sT^{\nu \sigma} + g_\sT^{\mu \sigma} g_\sT^{\nu \rho} \right)$,
and where we have introduced the following weighted integrals for a generic GTMD ${\cal X}^g$
\begin{align}
    {\cal X}^0 & =\int\d^2k_\sT \, {\cal X}^g \,, 
    \nonumber \\
          {\cal X}^{\phi} & =\int\d^2k_{\sT}\frac{\bm k_{\sT} \cdot \bm \Delta_\sT}{\bm \Delta_\sT^2}   {\cal X}^g\,, \nonumber\\
    {\cal X}^{2\phi} & = \int\d^2k_{\sT}\frac{2(\bm k_{\sT} \cdot \bm \Delta_\sT)^2\,-\,\bm k_{\sT}^2\bm \Delta_\sT^2}{\bm \Delta_\sT^4}   {\cal X}^g \, , 
    \nonumber  \\
    {\cal X}^{3\phi} & = \int\d^2k_{\sT}\frac{4(\bm k_{\sT} \cdot \bm \Delta_\sT)^3\,-\,3(\bm k_{\sT} \cdot \bm \Delta_\sT)\bm k_{\sT}^2\bm \Delta_\sT^2}{\bm \Delta_\sT^6}\,   {\cal X}^{g}\,.
    \label{eq:integrals-GTMD}
\end{align}
Here inverse powers of $\bm \Delta_\sT^2$ are included for dimensional reasons, which however does not mean that there is a problem in the forward limit, as the functions will appear multiplied by factors of ${\bm \Delta_\sT^2}/{M_N^2}$. Also, for later purposes we note that what is usually referred to as the elliptic part of a GTMD \cite{Hatta:2016dxp,Zhou:2016rnt}, i.e.\ the $\cos 2(\phi_k - \phi_\Delta)$ part, enters in ${\cal X}^{2\phi}$. 

\section{The gluon-gluon correlator for GPDs}
\label{sec:GPD}
The gluon-gluon correlator can be parametrized in terms of GPDs as follows
\begin{align}
\Gamma^{\mu\nu}_{\text{GPD}} = \frac{1}{4 P^+} \,   \overline u (p^\prime, \lambda^\prime)\,G^{\mu\nu}\,u(p,\lambda)\,,
\label{eq:Gamma-GPD-1}
\end{align}
where $\lambda$ and $\lambda^\prime$ denote the helicities of the nucleons in the initial and final state, respectively, whereas 
\begin{align}
G^{\mu\nu} & =  
 -g_\sT^{\mu\nu} \left [ \, \gamma^+\,H^g (x,\xi,t)\,+\, \frac{i \sigma^{+\alpha}\Delta_\alpha}{2 M_N}\,E^g(x,\xi,t) \right ]\, - i \epsilon_\sT^{\mu\nu} \left [ \, \gamma^+\gamma^5\,\widetilde H^g (x,\xi,t)\,+\, \frac{\Delta^+\gamma^5}{2 M_N}\,\widetilde E^g(x,\xi,t) \right ] \nonumber \\
& \qquad \quad  \,-\,\frac{\Delta_\sT^{\mu\nu}}{2 M_N^2}\, \left [ \frac{P^+}{M_N} \,  2\, \widetilde{H}^g_\sT \,+\, \gamma^+ E_\sT^g \right ]\, -\, \frac{1}{2 M_N}\,(\Delta_\sT^{\{\mu}\gamma_\sT^{\nu \}}  - g_\sT ^{\mu\nu} \gamma^\alpha \Delta_{\sT\, \alpha} )\, \left ( \gamma^+ H^g_\sT - \frac{P^+}{M_N}\, \widetilde{E}^g_\sT - \frac{\Delta^+}{2 M_N}\, E^g_\sT \right ) \,. 
\label{eq:G-GPD}
\end{align}

We adopt the following explicit expressions of the light-cone helicity spinors in the usual Dirac representation~\cite{Brodsky:1997de} for the incoming nucleon, see also Eqs.~(28) and (29) in Ref.~\cite{Meissner:2007rx},
\begin{align}
u(p,+) & = \frac{1}{\sqrt{2\sqrt{2}\,p^+}} \begin{pmatrix} 
             \sqrt{2}\, p^+\,+\, M_N\\
               -\frac{\Delta_\sT^1}{2}\, - i \, \frac{\Delta^2_\sT}{2}\\
               \sqrt{2}\,p^+ - M_N\\
               -\frac{\Delta_\sT^1}{2} - i \, \frac{\Delta^2_\sT}{2}
            \end{pmatrix}\,, \qquad \qquad u(p,-) = \frac{1}{\sqrt{2\sqrt{2}\,p^+}}\, \begin{pmatrix} 
               \frac{\Delta_\sT^1}{2}\, - i \, \frac{\Delta^2_\sT}{2}\\
              \sqrt{2}\, p^+\,+\, M_N\\  
                -\frac{\Delta_\sT^1}{2} + i \, \frac{\Delta^2_\sT}{2}\\
               -\sqrt{2}\,p^+ + M_N
            \end{pmatrix}\,, 
            \label{eq:spinors}
\end{align}
while the corresponding ones for the outgoing nucleon can be obtained by replacing $p\to p^\prime$ and $\bm \Delta_\sT \to -\bm \Delta_\sT$. The following relations are fulfilled, 
\begin{align}
\overline u (p^\prime,\lambda^\prime)\, u (p,\lambda) & =  \sqrt{1-\xi^2}\, \frac{1}{1-\xi^2} \left [2 \, M_N\, \delta_{\lambda^\prime\lambda}\,-\,(\lambda\,\Delta_\sT^1\,+\,i\,\Delta_\sT^2)\,\delta_{-\lambda^\prime\lambda} \right ],   \nonumber\\
\overline u (p^\prime,\lambda^\prime)\,\gamma^+\, u (p,\lambda) & = 2\,\sqrt{1-\xi^2}\,P^+\, \delta_{\lambda^\prime\lambda}\,,   \nonumber\\
\overline u (p^\prime,\lambda^\prime)\, \frac{i\, \sigma^{+\alpha} \,\Delta_\alpha}{2 M_N}\, u (p,\lambda) & = -  2\,\sqrt{1-\xi^2}\,\frac{1}{1-\xi^2}\,P^+\, \left [ \xi^2 \,\delta_{\lambda^\prime\lambda}  \,+ \, \frac{1}{2\,M_N}\, (\lambda\, \Delta_\sT^1 + i \, \Delta_\sT^2) \,  \delta_{-\lambda^\prime \lambda}\right ]\,, \nonumber \\
\overline u (p^\prime,\lambda^\prime)\,\gamma^5\, u (p,\lambda) & = \sqrt{1-\xi^2}\, \frac{1}{1-\xi^2} \left [ 2\, \lambda\, \xi\, M_N \,\delta_{\lambda^\prime\lambda}\,-\,(\Delta_\sT^1\,+\,i\,\lambda\,\Delta_\sT^2)\,\delta_{-\lambda^\prime\lambda} \right ]\,,   \nonumber\\
\overline u (p^\prime,\lambda^\prime)\,\gamma^+\gamma^5\, u (p,\lambda) & = 2\, \lambda\,\sqrt{1-\xi^2}\,P^+\, \delta_{\lambda^\prime\lambda}, \nonumber\\
\overline u (p^\prime,\lambda^\prime)\,\gamma^1\, u (p,\lambda) & = \sqrt{1-\xi^2}\,\frac{1}{1-\xi^2}\,\left [ (\xi\,\Delta_\sT^1- i \lambda\,\Delta_\sT^2) \,\delta_{\lambda^\prime\lambda}\,+\,2\,\xi\, \lambda\,M_N\,\delta_{-\lambda^\prime\lambda} \right ]\,, \nonumber\\
\overline u (p^\prime,\lambda^\prime)\,\gamma^2\, u (p,\lambda) & = \sqrt{1-\xi^2}\,\frac{1}{1-\xi^2}\,\left [  (i\,\lambda\,\Delta_\sT^1 + \xi\,\Delta_\sT^2) \,\delta_{\lambda^\prime\lambda}\,+\,2\,i\,\xi\, M_N\,\delta_{-\lambda^\prime\lambda} \right ]\,, \nonumber\\
\overline u (p^\prime,\lambda^\prime)\,\gamma^1\,\gamma^+\, u (p,\lambda) & = -\,2\,\lambda\,\sqrt{1-\xi^2}\,P^+\,\delta_{-\lambda^\prime\lambda},\nonumber\\
\overline u (p^\prime,\lambda^\prime)\,\gamma^2\,\gamma^+\, u (p,\lambda) & = -\,2\,i\,\sqrt{1-\xi^2}\,P^+\,\delta_{-\lambda^\prime\lambda}\,,
\end{align}
which allow us to split the correlator in Eqs.~\eqref{eq:Gamma-GPD-1}-\eqref{eq:G-GPD} into helicity flip and non-flip parts. In analogy with Eq.~\eqref{eq:Gamma-lambda}, we can write
\begin{align}
\Gamma_\text{GPD}^{\mu\nu} = \Gamma_{U \,\text{GPD}}^{\mu\nu}\, {\bm 1}+ \vec{\Gamma}^{\mu\nu}_\text{GPD}\cdot \vec{\bm \sigma} \,,
\label{eq:gamma-GPD}
\end{align}
with
\begin{align}
  \Gamma^{\mu\nu}_{U \,\text{GPD}} & =   \frac{1}{2}\,\sqrt{1-\xi^2}\,\left \{  -g_{\sT}^{\mu\nu} \left [H^g \, -\, \frac{\xi^2}{1-\xi^2}\,E^g \right ] \, + \, \frac{\Delta_\sT^{\mu\nu}}{M_N^2}\, \frac{1}{1-\xi^2}\,\left [ -\widetilde{H}^g_\sT - \frac{1}{2}\,E^g_\sT +\frac{1}{2} \,\xi\,\widetilde{E}_\sT^g\right ]\right \} \,, \nonumber\\
   \Gamma^{\mu\nu}_{L \,\text{GPD}} & =   \frac{1}{2}\,\sqrt{1-\xi^2}\,\left \{  -i \,\epsilon_\sT^{\mu\nu} \left [\widetilde{H}^g \, -\, \frac{\xi^2}{1-\xi^2}\,\widetilde{E}^g \right ] \, +i \, \frac{\epsilon^{\{ \mu}_{\sT \,\alpha}\,\Delta_\sT^{\nu\}\alpha}}{2 M_N^2}\,\frac{1}{1-\xi^2}\, \frac{1}{2}\left [-\xi \, E^g_\sT + \widetilde{E}_\sT^g \right ]\right \} \,, \nonumber \\
   \Gamma^{i\;\mu\nu}_{T \,\text{GPD}} & = \frac{i}{2}\,\sqrt{1-\xi^2}\,\left\{-g_\sT^{\mu\nu}\,\frac{\epsilon_\sT^{\Delta_\sT\,i}}{2M_N}\,\frac{E^g}{1-\xi^2}\,-\,\epsilon_\sT^{\mu\nu}\,\frac{\Delta_T^i}{2M_N}\,\frac{\xi}{1-\xi^2}\,\widetilde{E}^g\,-\,\frac{\epsilon_{\sT\alpha}^{\{\mu}\Delta_\sT^{\nu\}\alpha i}}{4M_N^3}\,\frac{\bm \Delta_\sT^2}{4M_N^2}\,\frac{\widetilde{H}_\sT^g}{1-\xi^2} \right. \nonumber \\ 
   & \qquad \qquad \qquad \qquad \,+\, \left . \frac{\epsilon_\sT^{\Delta_\sT\{\mu}g_\sT^{\nu\}i}\,+\,\epsilon_\sT^{i\{\mu}\Delta_\sT^{\nu\}}}{4M_N}\,\left[H_\sT^g\,+\,\frac{\xi}{1-\xi^2}\,\widetilde{E}_\sT^g\,-\,\frac{\xi^2}{1-\xi^2}\,E_\sT^g \,+ \, \frac{\bm \Delta_\sT^2}{4M_N^2}\,\frac{\widetilde{H}_\sT^g}{1-\xi^2}  \right]   \right\}\,. 
   \label{eq:gamma-GPD-2}
\end{align}

The square of the GPD correlator can also be calculated without resorting to any specific representation of the Dirac spinors. 
If we denote by $S$ the polarization vector of the initial nucleon, with $S\cdot p =0$ and $S^2=-1$, then we have
\begin{align}   
\Gamma^{\mu\nu}_{\text{GPD}}\,\Gamma^{*\,\rho\sigma}_{{\text{GPD}}} & \equiv \sum_{\lambda^\prime} \,\left [ 
\overline u (p^\prime, \lambda^\prime)\,G^{\mu\nu}\,u(p,\lambda) \right ] \left [ 
\overline u (p^\prime, \lambda^\prime)\,G^{\rho\sigma}\,u(p,\lambda) \right ] ^*  = \text{Tr}\left [ \left ( \frac{1 + \gamma^5 S \!\!\! / }{2} \right ) (p \!\!\!/ + M_N) \, \widetilde G^{\,\rho\sigma} \, (p^\prime \!\!\! \!\!/ + M_N) \, G^{\mu\nu} \right ]\,,
\end{align}
where we have summed over the helicities of the final nucleon and we have introduced
\begin{align}
\widetilde{G}^{\rho\sigma} & = \gamma^0\, G^{\rho\sigma \dagger}\,\gamma^0\,.    
\end{align}

In particular, for an unpolarized nucleon ($S=0$), we find:
\begin{align}
\left. \Gamma^{\mu\nu}_{\text{GPD}} \,\Gamma^{*\,\rho\sigma}_{{\text{GPD}}} \right|_{S=0}& 
= \frac{1}{4}\,(1-\xi^2)\, \Biggl \{ g_\sT^{\mu\nu}\,g_\sT^{\rho\sigma}\,  
\left [\left (H^g\,-\,\frac{\xi^2}{1-\xi^2}\,E^g \right )\,\left(H^g\,-\,\frac{\xi^2}{1-\xi^2}\,E^g \right )\,+\, \frac{\bm \Delta_\sT^2}{4M_N^2} \, \frac{1}{(1-\xi^2)^2}\, E^g\, E^g \right] \nonumber \\
&+\,\frac{\Delta_\sT^{\mu\nu}\,\Delta_\sT^{\rho\sigma}}{M_N^4}\,\frac{1}{(1-\xi^2)^2}\,\left(-\,\widetilde{H}^g_{\sT}\,-\frac{1}{2}\,E_{\sT}^g\,+\frac{\xi}{2}\,\widetilde{E}^g_{\sT}\right)\,\left(-\,\widetilde{H}^g_{\sT}\,-\frac{1}{2}\,E_{\sT}^g\,+\frac{\xi}{2}\,\widetilde{E}^g_{\sT}\right) \nonumber \\
&+\,\frac{g_\sT^{\mu\nu}\,\Delta_\sT^{\rho\sigma}}{M_N^2}\,\left[-\,\frac{1}{1-\xi^2}\,\left(H^g\,-\,\frac{\xi^2}{1-\xi^2}\,E^g\right)\,\left(-\,\widetilde{H}^g_{\sT}\,-\frac{1}{2}\,E_{\sT}^g\,+\frac{\xi}{2}\,\widetilde{E}^g_{\sT}\right) \right. \nonumber \\
& \qquad\qquad  \left. +\,\frac{1}{2}\,\frac{E^g}{1-\xi^2}\,\left(-\,H_{\sT}^g\,-\,\frac{\xi}{1-\xi^2}\,\widetilde{E}_{\sT}^g\,+\,\frac{\xi^2}{1-\xi^2}\,E_{\sT}^g\,-\,\frac{\bm \Delta_\sT^2}{2M_N^2}\,\frac{\widetilde{H}_{\sT}^g}{1-\xi^2}\right)\right] \nonumber \\
&+\,\frac{\Delta_\sT^{\mu\nu}\,g_\sT^{\rho\sigma}}{M_N^2}\,\left[-\,\frac{1}{1-\xi^2}\,\left(-\,\widetilde{H}^g_{\sT}\,-\frac{1}{2}\,E_{\sT}^g\,+\frac{\xi}{2}\,\widetilde{E}^g_{\sT}\right)\,\left(H^g\,-\,\frac{\xi^2}{1-\xi^2}\,E^g\right) \right. \nonumber \\
&\left. \qquad \qquad  +\,\frac{1}{2}\,\left(-\,H_{\sT}^g\,-\,\frac{\xi}{1-\xi^2}\,\widetilde{E}_{\sT}^g\,+\,\frac{\xi^2}{1-\xi^2}\,E_{\sT}^g\,-\,\frac{\bm \Delta_\sT^2}{2M_N^2}\,\frac{\widetilde{H}_{\sT}^g}{1-\xi^2}\right)\,\frac{E^g}{1-\xi^2}\right] \nonumber \\
&+\,\epsilon_\sT^{\mu\nu}\,\epsilon_\sT^{\rho\sigma}\,\left[\left(\widetilde{H}^g\,-\,\frac{\xi^2}{1-\xi^2}\,\widetilde{E}^g\right)\,\left(\widetilde{H}^g\,-\,\frac{\xi^2}{1-\xi^2}\,\widetilde{E}^g\right)\,+\,\frac{\bm \Delta_\sT^2}{4M_N^2}\,\frac{\xi^2}{(1-\xi^2)^2}\,\widetilde{E}^g\,\widetilde{E}^g\right]\nonumber \\
&+\,\frac{\epsilon_{\sT\alpha}^{\{\mu}\,\Delta_\sT^{\nu\}\alpha}\,\epsilon_{\sT\beta}^{\{\rho}\,\Delta_\sT^{\sigma\}\beta}}{4M_N^4}\,\frac{1}{4(1-\xi^2)^2}\,\left(-\,\xi\,E_{\sT}^g\,+\,\widetilde{E}_{\sT}^g\right)\,\left(-\,\xi\,E_{\sT}^g\,+\,\widetilde{E}_{\sT}^g\right)\nonumber \\
&+\,\frac{\epsilon_\sT^{\mu\nu}\,\epsilon_{\sT\beta}^{\{\rho}\,\Delta_\sT^{\sigma\}\beta}}{2M_N^2}\,\left[\frac{1}{2(1-\xi^2)}\,\left(\widetilde{H}^g\,-\,\frac{\xi^2}{1-\xi^2}\,\widetilde{E}^g\right)\,\left(-\xi\,E_{\sT}^g\,+\,\widetilde{E}_{\sT}^g\right) \right. \nonumber \\
&\qquad \qquad \left. +\,\frac{\xi}{2(1-\xi^2)}\,\widetilde{E}^g\,\left(-\,H_{\sT}^g\,-\,\frac{\xi}{1-\xi^2}\,\widetilde{E}_{\sT}^g\,+\,\frac{\xi^2}{1-\xi^2}\,E_{\sT}^g\right)\right]\nonumber \\
&+\,\frac{\epsilon_{\sT\alpha}^{\{\mu}\,\Delta_\sT^{\nu\}\alpha}\,\epsilon_\sT^{\rho\sigma}}{2M_N^2}\,\left[\frac{1}{2(1-\xi^2)}\,\left(-\xi\,E_{\sT}^g\,+\,\widetilde{E}_{\sT}^g\right)\,\left(\widetilde{H}^g\,-\,\frac{\xi^2}{1-\xi^2}\,\widetilde{E}^g\right) \right. \nonumber \\
& \qquad \qquad\, \left. +\,\left(-\,H_{\sT}^g\,-\,\frac{\xi}{1-\xi^2}\,\widetilde{E}_{\sT}^g\,+\,\frac{\xi^2}{1-\xi^2}\,E_{\sT}^g\right)\,\frac{\xi}{2(1-\xi^2)}\,\widetilde{E}^g \right]\nonumber \\
&-\,\left(g_\sT^{\mu\nu}\,g_\sT^{\rho\sigma}-g_\sT^{\mu\sigma}\,g_\sT^{\nu\rho}-g_\sT^{\mu\rho}\,g_\sT^{\nu\sigma}\right)\,\frac{\bm \Delta_\sT^2}{4M_N^2}\,\left[\left(-\,H_{\sT}^g\,-\,\frac{\xi}{1-\xi^2}\,\widetilde{E}_{\sT}^g\,+\,\frac{\xi^2}{1-\xi^2}\,E_{\sT}^g\,-\,\frac{\bm \Delta_\sT^2}{4M_N^2}\,\frac{\widetilde{H}_{\sT}^g}{1-\xi^2}\right)\right. \nonumber \\
&\qquad \qquad \left. \times\,\left(-\,H_{\sT}^g\,-\,\frac{\xi}{1-\xi^2}\,\widetilde{E}_{\sT}^g\,+\,\frac{\xi^2}{1-\xi^2}\,E_{\sT}^g\,-\,\frac{\bm \Delta_\sT^2}{4M_N^2}\,\frac{\widetilde{H}_{\sT}^g}{1-\xi^2}\right)\,+\,\frac{\bm \Delta_\sT^4}{16M_N^4}\,\frac{\widetilde{H}^g_{\sT}\,\widetilde{H}^g_{\sT}}{(1-\xi^2)^2}\right]\nonumber \\
&-\,\frac{\Delta_\sT^{\mu\nu\rho\sigma}}{M_N^4}\,\left[\left(-\,H_{\sT}^g\,-\,\frac{\xi}{1-\xi^2}\,\widetilde{E}_{\sT}^g\,+\,\frac{\xi^2}{1-\xi^2}\,E_{\sT}^g\,-\,\frac{\bm \Delta_\sT^2}{4M_N^2}\,\frac{\widetilde{H}_{\sT}^g}{1-\xi^2}\right)\,\frac{\widetilde{H}^g_{\sT}}{2(1-\xi^2)}\right. \nonumber \\
& \qquad \qquad \left. +\,\frac{\widetilde{H}^g_{\sT}}{2(1-\xi^2)}\,\left(-\,H_{\sT}^g\,-\,\frac{\xi}{1-\xi^2}\,\widetilde{E}_{\sT}^g\,+\,\frac{\xi^2}{1-\xi^2}\,E_{\sT}^g\,-\,\frac{\bm \Delta_\sT^2}{4M_N^2}\,\frac{\widetilde{H}_{\sT}^g}{1-\xi^2}\right) \right]\,. 
\end{align}
This result will be useful for the calculation of the cross section for the process $e\,N\to e\,Q\,\overline Q\,N$ in the GPD approach, which will be presented in Section~\ref{sec:SF-GPD}. We point out that the above equation could be obtained in a much faster way from the decomposition in Eqs.~\eqref{eq:gamma-GPD}-\eqref{eq:gamma-GPD-2}, namely
\begin{align}
\left. \Gamma_\text{GPD}^{\mu\nu} \,\Gamma_\text{GPD}^{*\,\rho\sigma} \right|_{S=0} = \Gamma_{U \,\text{GPD}}^{\mu\nu}\, \Gamma_{U \,\text{GPD}}^{*\, \rho\sigma}\,+\, \Gamma_{L \, \text{GPD}}^{\mu\nu}\, \Gamma_{L\, \text{GPD}}^{*\, \rho\sigma}\,+\, \vec{\Gamma}_{T\, \text{GPD}}^{\mu\nu} \cdot \vec{\Gamma}_{T\, \text{GPD}}^{*\, \rho\sigma}\,, 
\end{align}
that is the analogous of Eq.~\eqref{eq:amp-sq-S0} for GPDs.

\section{Cross section for the electroproduction of  heavy quark-antiquark pairs}
\label{sec:cs-details}

\begin{figure}
\centering
\includegraphics[width=1\linewidth, keepaspectratio, trim={4cm 21cm 0cm 1cm},clip]{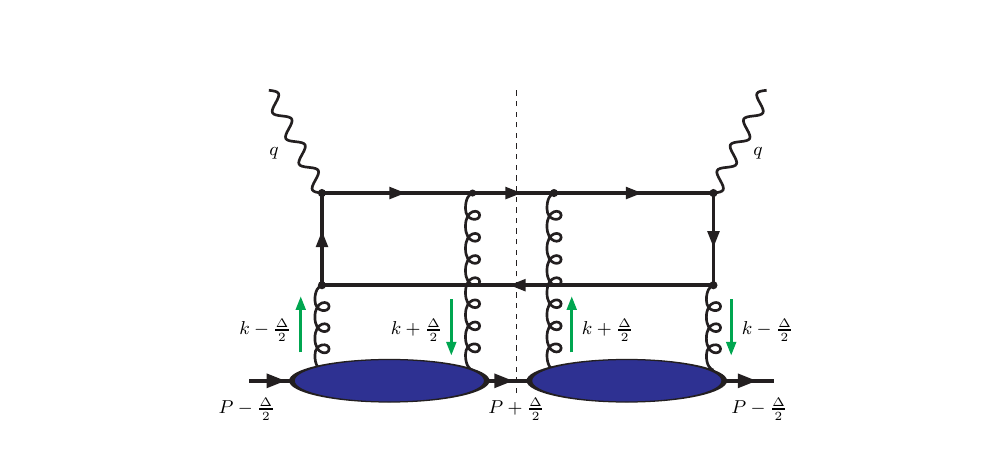} 
\caption{Representative cut diagram for the process $\gamma^*\,N \to Q\,\overline Q\,N$.}
\label{fig:diagram-GTMD}
\end{figure}
\begin{figure}[t]
    \includegraphics[width=1\linewidth, keepaspectratio, trim={0cm 22cm 0cm 3cm},clip]{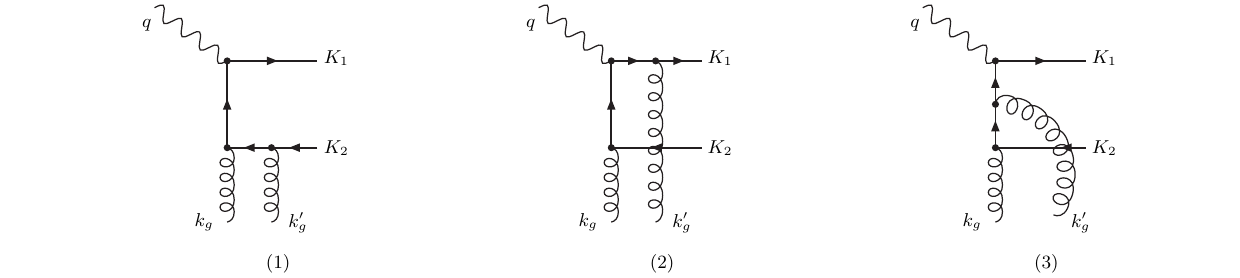}
     \includegraphics[width=1\linewidth, keepaspectratio, trim={0cm 22cm 0cm 2.5cm},clip]{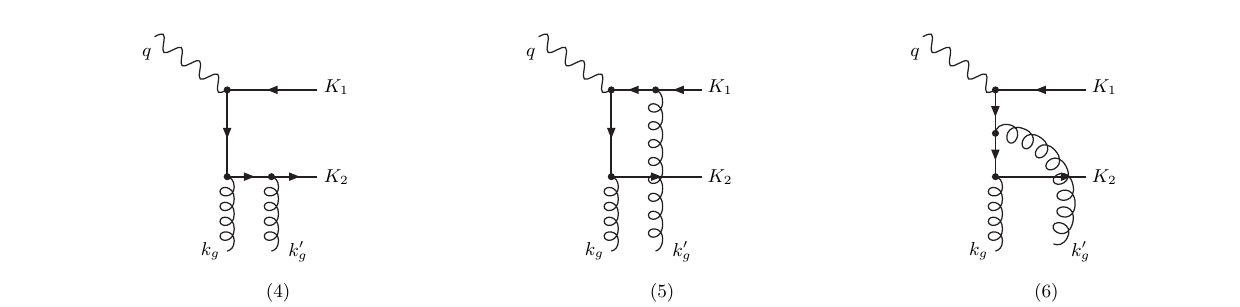}
\caption{Feynman diagrams contributing to the partonic scattering process $\gamma^* g \to Q \overline Q g$.} 
\label{eq:feynman}
\end{figure}

We consider the electron-nucleon exclusive reaction
\begin{align}
e(\ell) \,+\, N(p) \to e(\ell^\prime) \, + \, Q(K_1)\, +\,  \overline Q(K_2)\, + \, N(p^\prime) \,\,,
\end{align}
where the four-momenta of the particles are given within brackets. The dominant channel is the virtual photon-gluon fusion subprocess
\begin{align}
\gamma^*(q) \,+\, g(k_g) \to Q(K_1)\,+\, \overline Q (K_2)  \, + g(k_g^\prime)\,,
\label{eq:subprocess}
\end{align}
as described by a representative cut diagram in Fig.~\ref{fig:diagram-GTMD}. The corresponding amplitude 
can be written as
\begin{align}
{\cal A} (q,k_g; K_1, K_{2},k_g^\prime)\equiv \varepsilon_{\lambda_\gamma\, \alpha}(q) \, \varepsilon_{\lambda_g\,\mu}(k_g) \, \varepsilon^*_{\lambda^\prime_g\,\nu}(k_g^\prime)\,{\cal A}^{\alpha\mu\nu} (q,k_g; K_1,K_{2},k_g^\prime) \,,
\end{align}
with 
\begin{align}
{\cal A}^{\alpha\mu\nu} (q,k_g; K_1,K_{2},k_g^\prime)= \overline u (K_1)\,O^{\alpha\mu\nu}(q,k_g; K_1,K_2,k_g^\prime)\, v(K_2)\,, 
\end{align}
where we have introduced the operator $\cal O^{\alpha\mu\nu}$, which depends neither on the polarization vectors of the external photon and gluons nor on the (anti)quark Dirac spinors. It is given by 
\begin{align}
{\cal O}^{\alpha\mu\nu}(q,k_g;  K_1,K_2,k_g^\prime)  = \sum_{m=1}^6 {\cal C}_{m} \,O^{\alpha\mu\nu}_{m}(q,k_g;  K_1,K_2,k_g^\prime) \,,
\label{eq:O}
\end{align}
where each term can be calculated from the Feynman diagrams depicted in Fig.~\ref{eq:feynman}. From the first three ones we find
\begin{align}
 O^{\alpha\mu\nu}_{\text{1}}(q,k_g;  K_1,K_2,k_g^\prime) & = e e_Q\,  g_s^2 \,      \gamma^\alpha \,\frac{\Kos - \qs \, + M_Q }{(K_1  -q)^2 -M_Q^2}\, \gamma^\mu \, \frac{-\Kts  - \ks_g^\prime  + M_Q}{(K_2 + k_g^\prime)^2  - M_Q^2}\,\gamma^\nu \,,\nonumber \\
  O^{\alpha\mu\nu}_{\text{2}} (q,k_g;  K_1,K_2,k_g^\prime)& = e e_Q\, g_s^2 \,   \gamma^\nu \,\frac{\Kos + \ks_g^\prime + M_Q }{(K_1 + k_g^\prime)^2 -M_Q^2}\, \gamma^\alpha \, \frac{ \ks_g - \Kts + M_Q}{(k_g -K_2)^2  - M_Q^2}\,\gamma^\mu \,,\nonumber \\
  O^{\alpha\mu\nu}_{\text{3}} (q,k_g;  K_1,K_2,k_g^\prime)& = e e_Q\, g_s^2 \,  \gamma^\alpha\,\frac{\Kos -\qs + M_Q }{(K_1 -q )^2 -M_Q^2}\, \gamma^\nu \, \frac{ \ks_g - \Kts + M_Q}{(k_g  -K_2)^2  - M_Q^2}\,\gamma^\mu \,,
  \label{eq:O1-3}
  \end{align}
  with $M_Q$ being the mass of the (anti)quark and $e_Q$ its fractional electric charge in units of the proton charge $e$. The other contributions can be obtained by means of crossing relations, 
  \begin{align}
  O^{\alpha\mu\nu}_{\text{4}}(q,k_g;  K_1,K_2,k_g^\prime) & = e e_Q\, g_s^2  \, \gamma^\nu\,\frac{\Kos + \ks_g^\prime + M_Q }{(K_1 + k_g^\prime )^2 -M_Q^2}\, \gamma^\mu \, \frac{\qs -\Kts + M_Q }{(q-K_2)^2-M_Q^2}\,\gamma^\alpha\,,\nonumber \\   
    O^{\alpha\mu\nu}_{\text{5}}(q,k_g;  K_1,K_2,k_g^\prime) & = e e_Q\, g_s^2 \,  \gamma^\mu\,\frac{\Kos - \ks_g   + M_Q }{(K_1 - k_g )^2 -M_Q^2}\, \gamma^\alpha \, \frac{-\Kts - \ks^\prime_g  + M_Q }{(K_2 + k_g^\prime )^2-M_Q^2}\,\gamma^\nu \,,\nonumber \\ 
O^{\alpha\mu\nu}_{\text{6}}(q,k_g;  K_1,K_2,k_g^\prime) & = e e_Q\, g_s^2  \,  \gamma^\mu\,\frac{\Kos - \ks_g  + M_Q }{(K_1 - k_g )^2 -M_Q^2}\, \gamma^\nu \, \frac{\qs - \Kts + M_Q }{(q-K_2)^2-M_Q^2}\,\gamma^\alpha \,.
\label{eq:O4-6}
\end{align}
The color factors in Eq.~\eqref{eq:O} read
\begin{align}
{\cal C}_{1} = {\cal C}_{5}={\cal C}_{6}= t^{a} t^b\,, \qquad {\cal C}_{2} = {\cal C}_{3}={\cal C}_{4} = t^{b} t^a\,,
\end{align}
with $t^a$ being the $SU(3)$ generators in the fundamental representation, normalized such that Tr$(t^at^b) = \delta^{ab}/2$.  By projecting out the color-singlet contributions to the amplitude by means  of the $SU(3)$ Clebsch-Gordan coefficients,
\begin{equation}
\langle 3i;\bar{3}j\vert 1\rangle = \frac{\delta^{ij}}{\sqrt{N_c}}\,,
\label{eq:cs}
\end{equation}
with $N_c$ being the number of colors, we are able to derive the color-singlet color factors
\begin{align}
{\cal C}_1 ={\cal C}_5 = {\cal C}_6= \sum_{i,j}\,\langle 3i;\bar{3}j\vert 1\rangle \, (t^a t^b)_{ij} \,, \qquad {\cal C}_2 ={\cal C}_3 ={\cal C}_4 =\sum_{i,j}\,\langle 3i;\bar{3}j\vert 1\rangle \, (t^b t^a)_{ij} \,, 
\label{eq:Ci0}
\end{align}
where the sum is taken over the colors of the outgoing quark and antiquark. Hence, it turns out that all the color factors are equal to each other
\begin{equation}
{\cal C}_m =  \frac{1}{2 \sqrt{N_c}}\, \delta^{ab}\, , \qquad m = 1, \hdots, 6~.
\end{equation}

We now choose a reference frame in such a way that both the virtual photon exchanged in the reaction and the incoming nucleon move along the $\hat z$-axis. Azimuthal angles are measured w.r.t.\ to the lepton scattering plane, such that $\phi_\ell = \phi_\ell^\prime =0$.  In terms of two light-cone vectors $n^\mu$ and $\overline{n}^\mu$, with $n \cdot \overline{n}=1$, we have
\begin{align}
P^\mu & = P^+n^\mu + \frac{M_N^2+ \frac{1}{4}\,\bm \Delta_\sT^2}{2 (1-\xi^2)P^+}\,\overline{n}^\mu\,,\nonumber \\
\Delta^\mu & = - 2 \xi\, P^+ n^\mu 
+ \frac{\xi}{1-\xi^2}\, 
\frac{M_N^2+ \frac{1}{4}\,\bm \Delta_\sT^2}{P^+}\, \overline{n}^\mu + \Delta_\sT^\mu\, ,\nonumber \\
p^\mu & = (1+\xi)\,P^+ n^\mu \,+\,\frac{M_N^2 + \frac{1}{4}\, \bm \Delta_\sT^2}{2 (1+\xi)\, P^+}\, \overline n^\mu \, -\, \frac{\Delta_\sT^\mu}{2}\,, \nonumber \\
q^\mu & = - \xB (1+\xi) \,  P^+ n^\mu + \frac{Q^2}{2 \xB (1+\xi)\,P^+}\,\overline {n}^\mu \,, \nonumber \\
\ell^\mu  & = \frac{1-y}{y}\,\xB (1+\xi) P^+ n^\mu  + \frac{1}{y}\,\frac{Q^2}{2\xB (1+\xi)P^+ }\,\overline{n}^\mu
+ \frac{\sqrt{1-y}}{y}\,Q\,\hat\ell_\perp^\mu\, ,\nonumber
\\
\ell^{\prime \mu}& =  \frac{1}{y}\,\xB (1+\xi)\,P^+n^\mu + \frac{1-y}{y}\,\frac{Q^2}{2\xB(1+\xi)\, P^+}\,\overline{n}^\mu
+ \frac{\sqrt{1-y}}{y}\,Q\,\hat\ell^\mu_\perp\,,
\label{eq:sudakov}
\end{align}
where  $Q^2 = -q^2 \equiv -(\ell-\ell^\prime)^2$. Furthermore, we have introduced the Bjorken and inelasticity variables 
\begin{align}
\xB = \frac{Q^2}{2 p\cdot q}\,, \qquad\qquad   y = \frac{p\cdot q}{p\cdot \ell}\,.   
\end{align}
If we define the total invariant mass squared $s = (\ell + p)^2$, the following relations hold: $s = 2\,p\cdot q/y = Q^2/\xB y$. Moreover, the invariant mass squared of the virtual photon-target, namely $W^2 =(q+p)^2$, can be expressed  in terms of the other invariants: $W^2= Q^2(1-\xB)/\xB = (1-\xB)ys$. 

Similarly, the momenta of the final quark-antiquark pair can be parametrized as 
\begin{align}   
K_1^\mu
&= \frac{M_{Q}^2 + \bm K_{{1}\sT}^2}{2z_1 \,p\cdot q}\,(1+\xi) P^+ n^\mu + z_1 \,\frac{p\cdot q}{(1+\xi)P^+}\,\overline{n}^\mu     + K_{{1}\sT}^\mu\,, \nonumber \\
K_2^\mu
&= \frac{M_{Q}^2 + \bm K_{{2}\sT}^2}{2z_2 \,p\cdot q}\,(1+\xi) P^+ n^\mu + z_2 \,\frac{p\cdot q}{(1+\xi) P^+}\,\overline{n}^\mu     + K_{{2}\sT}^\mu\,,
\label{eq:jetmom1}
\end{align}
with $z_i= K_i\cdot p/ q \cdot p$  and $K_{i \sT}^2 = -\bm K_{i\,\sT}^2$. Finally, the Sudakov decomposition of the gluon average momentum $k$ reads 
\begin{align}
k^\mu & = x\, P^+ n^\mu +  \frac{k^2 +  \bm k_\sT^2}{2 x \, P^+}\, \overline{n}^\mu  + k_\sT^\mu \,, 
\label{eq:average-k}
\end{align}
from which we obtain 
\begin{align}
k_g^\mu  & = k^\mu - \frac{1}{2}\,\Delta^\mu = (x+ \xi)\, P^+ n^\mu +  \frac{\bm k_\sT^2 + \frac{1}{4}\, \bm \Delta_\sT^2-  \bm k_\sT \cdot \bm \Delta_\sT}{2 (x+\xi)  \, P^+}\, \overline{n}^\mu  + \left ( k_\sT- \frac{1}{2}\,\Delta_\sT\right)^\mu \,, \nonumber \\
k_g^{\prime\,\mu}  & = k^\mu + \frac{1}{2}\, \Delta^\mu = (x- \xi)\, P^+ n^\mu +  \frac{\bm k_\sT^2 + \frac{1}{4}\, \bm \Delta_\sT^2 +   \bm k_\sT \cdot \bm \Delta_\sT}{2 (x-\xi)  \, P^+}\, \overline{n}^\mu  + \left ( k_\sT+ \frac{1}{2}\,\Delta_\sT\right)^\mu \,,
\end{align}
where $k_g^-$ and $k_g^{\prime\, - }$ have been fixed so that gluons are exactly on shell. 

In order to apply a framework based on GTMD factorization, we consider only the kinematic region in which the component of the quarkonium momentum  transverse w.r.t.\ the lepton plane, that is $p_{\psi \sT}$, is small compared to the virtuality of the photon $Q$ and to the mass of the quarkonium $M_{\psi}$.
The differential cross section can be written as 
\begin{align}
\d\sigma & = \frac{1}{2 s}\,\frac{\d^3 \ell'}{(2\pi)^3\,2 E_e^{\prime}}\, \frac{\d^3 K_{1}}{(2\pi)^3\,2 E_{1}}\,\frac{\d^3 K_{2}}{(2\pi)^3\,2 E_{2}}\,\frac{\d^3 p^\prime}{(2\pi)^3\,2 E^\prime}
{\int}\d \widetilde x_1\, \d \widetilde x_2\, \d^2 k_{1\sT}\, \d^2 k_{2\sT}\,(2\pi)^4\,  \delta^4(q  - K_1 - K_2 - \Delta ) \nonumber \\
&\qquad  \times \frac{1}{Q^4}\, 
L^{\alpha \beta}(\ell,q) \,  \Gamma_{g}^{\mu\nu}(x_1, \bm k_{1\sT}, \xi, \bm \Delta_\sT)\,  \Gamma_{g}^{\star\rho\sigma}(x_2{,} \bm k_{2\sT}, 
\xi, \bm \Delta_\sT)\,
\, {\cal A}_{\alpha\mu\nu}(x_1, \xi) \,{\cal A}^{\star}_{\beta\rho\sigma}(x_2, \xi) \, , 
\label{CrossSec}
\end{align}
where we have introduced the notation
\begin{align}
\d \widetilde x_i = \frac{\d x_i}{(x_i + \xi - i \epsilon)(x_i - \xi + i \epsilon)} \, ,
\qquad \qquad i = 1, 2\,, 
\end{align}
with $x_i$ and $ k_{i\sT}$ being defined through the Sudakov decomposition of the average gluon momentum, see Eq.~\eqref{eq:average-k}, the index $i=1,2$ referring to the quantities entering the amplitude and its complex conjugated, respectively.

The scattering amplitude ${\cal A}$ for the process $\gamma^*g\to Q\, \overline Q\, g$ has been calculated in the previous section, the gluon correlator $ \Gamma_g$ is given in Eq.~\eqref{eq:GTMDs} and the leptonic tensor $L(\ell, q)$ is defined as
\begin{equation}
L^{\alpha\beta}(\ell, q) = e^2 \left [-g^{\alpha\beta}\,Q^2 + 2\,( 
 \ell^{\alpha}\ell'^\beta + \ell^{\beta}\ell'^\alpha) + 2 i \lambda_e \epsilon^{\alpha\beta\rho \sigma} \ell_\rho q_\sigma\right ]\, ,
\end{equation}
with $e$ being the electric charge of the electron and $\lambda_e = \pm 1$ twice its helicity. By substituting in the previous equation the Sudakov decompositions of $\ell$ and $\ell^\prime$ in Eq.~\eqref{eq:sudakov}, the leptonic tensor can be alternatively written as 
\begin{align}
L^{\alpha\beta} (\ell, q) 
& = e^2\frac{Q^2}{y^2}\, \bigg  \{ -[ 1+(1-y)^2 ]\, g_\sT^{\alpha\beta} \,+ \, 4 (1-y)\,  \epsilon^\alpha_L \epsilon^\beta_L  \,+\, 4 (1-y) 
\left ( \hat \ell_\sT^\alpha  \hat \ell_\sT ^\beta   + \frac{1}{2}\, g_\sT^{\alpha\beta}\right )  \nonumber \\
& \qquad \qquad \qquad  + 2 (2-y)\, \sqrt{1-y} \, ( \epsilon_L ^{\alpha}\,  \hat \ell_\sT^{\beta}  +  \epsilon_L ^{\beta}\,  \hat  \ell_\sT^{\alpha} )  + 2 i \lambda_e y  (2-y)\,\epsilon_\sT^{\alpha\beta} + 4 i \lambda_e y \sqrt{1-y}\, \epsilon_L^{[\alpha}\, \epsilon_\sT^{\beta]\sigma}\, \hat \ell_{\sT\sigma} \bigg \} \,,
\end{align}
where  the transverse projector $g_\sT^{\alpha\beta}$ is given by 
\begin{align}
g_\sT^{\alpha\beta} & \equiv g^{\alpha\beta} - n^\alpha \, \overline{n}^\beta - n^\beta \, \overline{n}^\alpha \,,  
\end{align}
and we have introduced the longitudinal polarization vector of the exchanged virtual photon, 
\begin{align}
\epsilon_L^\alpha (q)= \frac{1}{Q}\left (  q^\alpha + \frac{Q^2}{p\cdot q} \, p^\alpha \right ) \,,
\end{align}
which fulfills the relations $\epsilon^2_L (q)= 1$ and $\epsilon_L^\alpha(q)\, q_\alpha = 0$. 

By rewriting the phase-space elements in  Eq.~(\ref{CrossSec}) as
\begin{align}
\frac{\d^3 \ell'}{(2\pi)^3\, 2 E_e^{\prime}} 
& = \frac{1}{16 \pi^2} \,s y\,\d \xB\,\d y\, ,\nonumber \\
\frac{\d^3 K_{1}}{(2\pi)^3\, 2 E_{1}} \, \frac{\d^3 K_{2}}{(2\pi)^3\, 2 E_{2}}& = \frac{1}{4 (2\pi)^6}\,  \frac{\d z_1}{z_1}\,\frac{\d z_2}{z_2}\,\d^2  K_{1\sT}\,\d^2 K_{2\sT} = \frac{1}{4 (2\pi)^6}\,  \frac{\d z_1}{z_1}\,\frac{\d z_2}{z_2}\,\d^2 (K_{1 \sT} + K_{2\sT})\,\d^2 K_{\perp}\,,\nonumber \\
\frac{\d^3 p^\prime}{(2\pi)^3\, 2 E^\prime} & = \frac{1}{2 (2\pi)^3}\,  \frac{\d \xi}{\xi}\,\d^2 \bm \Delta_{\sT}\,,
\end{align}
and the momentum-conserving $\delta$-function as
\begin{equation}
\label{DeltaFunc}
\delta^4(q- K_1-K_2  - \Delta)
= \frac{1}{y \,s}\, \delta\bigg(\xi - \frac{\xB}{2} -\frac{M_{1\perp}^2}{2\, y \,z_1\,s}  -\frac{M_{2\perp}^2}{2\, y \,z_2\,s}\bigg)
\,\delta \left (1-z_1-z_2 \right ) 
\,\delta^2\left (\bm K_{1\sT} + \bm K_{2\sT} + \bm \Delta_\sT\right )\,,
\end{equation}
we find that $z_2$ is fixed to the value $z_2=1-z$, with $z\equiv z_1$, $\bm K_{1\sT} + \bm K_{2\sT} = -\bm \Delta_\sT\,$, and the Bjorken variable is related to the skewness by
\begin{align}
 \xB = \frac{2\xi}{1+\xi}\, \frac{z(1-z)Q^2}{z(1-z)Q^2 + M^2_\perp}  \,. 
\end{align}
Hence the cross section takes the form
\begin{align}
\frac{\d\sigma}
{\d \xB\, \d y\,\d z\, \d^2 K_\perp\, \d^2\bm{\Delta}_{\sT}} & = \frac{\alpha}{16\, (2\pi)^6 \,Q^2} \, \frac{1}{M_\perp^2 + Q^2 z(1-z)}\left \{  - \frac{1+ (1-y)^2}{y}\,g_\sT^{\alpha\beta} \,+ \, \frac{4(1-y)}{y} \, \left [\epsilon_L^\alpha\epsilon_L^\beta +  \hat \ell_\sT^\alpha  \hat \ell_\sT ^\beta   + \frac{1}{2}\, g_\sT^{\alpha\beta}    \right ] \right .  \nonumber \\
& \qquad \left .\!\! +\,  2\, \frac{(2-y)}{y}\, \sqrt{1-y} \, \right .  ( \epsilon_L ^{\alpha}\,  \hat \ell_\sT^{\beta}  +  \epsilon_L ^{\beta}\,  \hat  \ell_\sT^{\alpha} ) + 2 i \lambda_e (2-y)\,\epsilon_\sT^{\alpha\beta} + 4 i \lambda_e \sqrt{1-y}\, \epsilon_L^{[\alpha}\, \epsilon_\sT^{\beta]\sigma}\, \hat \ell_{\sT\sigma} \bigg \} \nonumber \\
& \qquad \times \int  \d \widetilde x_1\, \d \widetilde x_2\, \d^2 k_{1\sT}\,  \d^2 k_{2\sT} \,\Gamma_{g}^{\mu\nu}(x_1 {,} \bm k_{1\sT}, \xi,  \bm \Delta_\sT)\,  \Gamma_{g}^{\star\rho\sigma}(x_2 {,} \bm k_{2\sT},\xi , \bm \Delta_\sT) {\cal A}_{\alpha\mu\nu} (x_1 {,}  \xi) \,{\cal A}^{\star}_{\beta\rho\sigma} (x_2 {,} \xi) \, .
\label{eq:cs}
\end{align}

\label{eq:xiresult}

Coming back to the expression of the amplitudes, the denominators in Eqs.~\eqref{eq:O1-3}-\eqref{eq:O4-6} can be written as
\begin{align}
(K_1-q)^2 - M_Q^2 & =  -\frac{z (1-z)\,Q^2 + {M_\perp^2}}{z}\,, \nonumber \\
(K_2 + k_g^\prime)^2 -M_Q^2 & = \frac{x-\xi}{2\,\xi (1+\xi)} \,\frac{z(1-z) Q^2 +M_\perp^2}{z} 
\simeq \frac{x-\xi}{2\,\xi } \,\frac{z(1-z) Q^2 +M_\perp^2}{z}  \,,\nonumber \\
 (K_1 + k_g^\prime)^2 -M_Q^2 & = \frac{x-\xi}{2\xi (1+\xi)}\, \frac{z(1-z) Q^2 +M_\perp^2}{1-z}\simeq \frac{x-\xi}{2\xi}\, \frac{z(1-z) Q^2 +M_\perp^2}{1-z}\,,\nonumber \\
  (k_g - K_2)^2 -M_Q^2 & = -\frac{x+\xi}{2\xi (1+\xi)}\, \frac{z(1-z) Q^2 +M_\perp^2}{z} \simeq  -\frac{x+\xi}{2\xi}\, \frac{z(1-z) Q^2 +M_\perp^2}{z}\,,\nonumber \\
  (q-K_2)^2 - M_Q^2 & =  -\frac{z (1-z)\,Q^2 + {M_\perp^2}}{1-z}\,, \nonumber \\
  (K_1-k_g)^2 - M_Q^2 & = - \frac{x+\xi}{2\,\xi (1+\xi)} \,\frac{z(1-z) Q^2 +M_\perp^2}{1-z}
 \simeq  - \frac{x+\xi}{2\,\xi} \,\frac{z(1-z) Q^2 +M_\perp^2}{1-z} \,. 
\end{align}
Hence, if we denote by $\phi_\perp$, $\phi_\Delta$ and $\phi_\ell$ the azimuthal angles of $K_\perp$, $\Delta_\sT$ and the scattered electron, respectively, the angular structure of the cross section can be written as 
\begin{align}
\frac{\d\sigma}
{\d \xB\, \d y\,\d z\, \d^2 K_\perp\, \d^2\bm{\Delta}_{\sT}} & = \frac{\alpha}{\pi\,y \,Q^2}\left [ \frac{1+(1-y)^2}{2}\, \d \sigma_{U, T} (\phi_\Delta, \phi_\perp)\,+\, (1-y)\,\d \sigma_{U,L} (\phi_\Delta, \phi_\perp) \, + \,(1-y)\,\d\sigma_{U,TT} (\phi_\ell, \phi_\Delta, \phi_\perp) \right . \nonumber \\
& \qquad \qquad \qquad \left . +\, (2-y) \sqrt{1-y}\,\d \sigma_{U,LT} (\phi_\ell, \phi_\Delta, \phi_\perp) \right ]\,,
\end{align}
where the subscripts denote the polarization of the proton and the exchanged virtual photon, respectively, and where
\begin{align}
\d\sigma_{U,T} & = \frac{\alpha\, \alpha_s^2\, e^2_Q}{8\, \pi^2 N_c}\,
\frac{\bm K_\perp^2}{z(1-z)[z(1-z)Q^2 +M_\perp^2]^3} \left [ \, F^0_{U,T}  \,+\,F^{\,\cos 2(\phi_\Delta - \phi_\perp)}_{U,T}\cos 2(\phi_\Delta - \phi_\perp) \right . \nonumber \\
& \hspace*{8cm}\,+\, \left . F^{\,\cos 4 (\phi_\Delta - \phi_\perp)}_{U,T} \cos 4 (\phi_\Delta - \phi_\perp)\right ]   \,,\nonumber \\
\d\sigma_{U,L} & = \frac{2\, \alpha\,\alpha_s^2\, e^2_Q}{\pi^2 \,N_c}\, \frac{Q^2 \,z(1-z)}{[z(1-z)Q^2 +M_\perp^2]^3}\left [ \, F^0_{U,L}  \,+\,F^{\,\cos 2(\phi_\Delta - \phi_\perp)}_{U,L}\cos 2(\phi_\Delta - \phi_\perp) \,+\,F^{\,\cos 4 (\phi_\Delta - \phi_\perp)}_{U,L} \cos 4 (\phi_\Delta - \phi_\perp)
\right ]  \,,\nonumber \\
\d\sigma_{U,TT} & =  \frac{\alpha\,\alpha_s^2\, e^2_Q}{4\, \pi^2 \,N_c} \,\frac{\bm K_\perp^2}{[z(1-z)Q^2 +M_\perp^2]^3}\left [ F_{U,TT}^{\cos 2 (\phi_\perp-\phi_\ell)} \, \cos 2(\phi_\perp-\phi_\ell) \,+\, F_{U,TT}^{\cos 2 (\phi_\Delta-\phi_\ell)} \, \cos 2(\phi_\Delta-\phi_\ell)  \right . \nonumber \\
& \hspace*{2.4cm}\left . +\,F^{\,\cos 2(2 \phi_\perp -\phi_\Delta - \phi_\ell)}_{U,TT}\cos 2(2 \phi_\perp-\phi_\Delta - \phi_\ell) \,+\,F^{\,\cos 2( \phi_\perp - 2 \phi_\Delta + \phi_\ell)}_{U,TT}\cos 2(\phi_\perp - 2 \phi_\Delta + \phi_\ell)
\right . \nonumber \\
& \hspace*{2.4cm}\left .  +\,F^{\,\cos 2( 3\phi_\perp -2\phi_\Delta -\phi_\ell)}_{U,TT}\cos 2(3 \phi_\perp - 2\phi_\Delta -\phi_\ell) \right ] 
\,, \nonumber \\
\d\sigma_{U,LT} & = \frac{\alpha\,\alpha_s^2\, e^2_Q}{2\, \pi^2 \,N_c} \,\frac{Q\,K_{\perp}(1-2z)}{[z(1-z)Q^2 +M_\perp^2]^3}\left [ \, F_{U,LT}^{\,\cos (\phi_\perp - \phi_\ell)}\cos (\phi_\perp - \phi_\ell)  \,+\,F_{U,LT}^{\,\cos (3\phi_\perp - 2\phi_{\Delta}-\phi_\ell)}\cos (3\phi_\perp - 2\phi_{\Delta}-\phi_\ell)\right . \nonumber\\
& \hspace*{2.4cm}\left .+\,F_{U,LT}^{\,\cos (5\phi_\perp - 4\phi_{\Delta}-\phi_\ell)}\cos (5\phi_\perp - 4\phi_{\Delta}-\phi_\ell)\,+\,F_{U,LT}^{\,\cos (3\phi_\perp - 4\phi_{\Delta} + \phi_\ell)} \cos (3\phi_\perp - 4\phi_{\Delta} + \phi_\ell) \right. \nonumber \\
& \hspace*{2.4cm}\left . +\,F_{U,LT}^{\,\cos (\phi_\perp - 2\phi_{\Delta} + \phi_\ell)} \cos (\phi_\perp - 2\phi_{\Delta} + \phi_\ell)\right ]\,. 
\label{eq:xss}
\end{align}
The explicit expressions of the structure functions will be given in the next section in terms of GTMDs, and in Section~\ref{sec:SF-GPD} in terms of GPDs. 

\section{Structure functions in the GTMD framework}
\label{sec:SF-GTMD}

If we neglect smearing effects in the transverse momenta of the outgoing quark-antiquark pair, the structure functions for the process $e\,N\to e \,Q\, \overline Q\, N$ can be expressed in terms of the integrals of GTMDs introduced in Eq.~\eqref{eq:integrals-GTMD}, 
which allow us to  define the following Compton form factors for a generic GTMD ${\cal Y}^{n\phi}_i$ (with $n=0...3$ and $i=1...8$),  
\begin{align}\label{compton}
\bm{{\cal Y}}^{n\phi}_{i,1} & = \int_{-1}^{1}\d x\, \left [\frac{1-2\beta}{(x+\xi - i \epsilon)^2} \,+\, \frac{1-2\beta}{(x-\xi + i \epsilon)^2} + \frac{4\beta}{(x+\xi - i \epsilon)(x - \xi +i \epsilon)} \right ] {\cal Y}^{n\phi}_i \,, \nonumber\\
\bm{{\cal Y}}^{n\phi}_{i,2} & = 2 
\int_{-1}^{1}\d x\, \left [\frac{1-\beta}{(x+\xi - i \epsilon)^2} \,+\, \frac{1-\beta}{(x-\xi + i \epsilon)^2} - \frac{1- 2\beta}{(x+\xi - i \epsilon)(x - \xi +i \epsilon)} \right ] {\cal Y}^{n\phi}_i \,, \nonumber \\
\bm{{\cal Y}}^{n\phi}_{i,3} & = \hide{(1-\beta)}\int_{-1}^{1}\d x\, \left [\frac{1}{2(x+\xi - i \epsilon)^2} \,+\, \frac{1}{2(x-\xi + i \epsilon)^2} - \frac{1}{(x+\xi - i \epsilon)(x - \xi +i \epsilon)} \right ] {\cal Y}^{n\phi}_i
\,,\nonumber \\
\bm{{\cal Y}}^{n\phi}_{i,4} & = \hide{(1-\beta)}
\int_{-1}^{1}\d x\, \left [-\,\frac{1}{(x+\xi - i \epsilon)^2} \,+\, \frac{1}{(x-\xi + i \epsilon)^2}\right ]  {\cal Y}^{n\phi}_i \,,
\end{align}
with the following identifications: 
\begin{align*}
    \bm{{\cal F}}^{\phi}_{2}&\equiv\bm{{\cal F}}^{\phi}_{2,3} & \bm{{\cal F}}^{0}_{3}&\equiv\bm{{\cal F}}^{0}_{3,3}\,,\\
    \bm{{\cal G}}^{0}_{1}&\equiv\bm{{\cal G}}^{0}_{1,4} & \bm{{\cal G}}^{2\phi}_{2}&\equiv\bm{{\cal G}}^{2\phi}_{2,3} & \bm{{\cal G}}^{0}_{3}&\equiv\bm{{\cal G}}^{0}_{3,3}\,,\\\bm{{\cal H}}^{\phi}_{3}&\equiv\bm{{\cal H}}^{\phi}_{3,3} & \bm{{\cal H}}^{3\phi}_{4}&\equiv\bm{{\cal H}}^{3\phi}_{4,3} & \bm{{\cal H}}^{0}_{7}&\equiv\bm{{\cal H}}^{0}_{7,3} & \bm{{\cal H}}^{0}_{8}&\equiv\bm{{\cal H}}^{0}_{8,3} & \bm{{\cal H}}^{\phi}_{2}&\equiv\bm{{\cal H}}^{\phi}_{2,4} & \bm{{\cal H}}^{0}_{6}&\equiv\bm{{\cal H}}^{0}_{6,4}\,.
\end{align*}

where, along the lines of Ref.~\cite{Braun:2005rg}, we have introduced the parameter
\begin{align}
\beta = \frac{z(1-z)Q^2\, + \, M_Q^2}{z(1-z)Q^2\, + \, M_\perp^2 } \,,   
\end{align}
which for outgoing light quarks $(M_Q=0)$ coincides with the conventional $\beta$ parameter used in the description of diffractive deep-inelastic scattering. 

The integrands in Eq.~\eqref{compton} potentially contain singularities at $x=\pm\,\xi$ (a similar situation occurs in e.g.\ Ref.~\cite{Ji:2016jgn}).

Double poles in combination with cusps in the GTMDs will require a proper distributional definition, which however is not the objective of this paper. For more on this topic we refer to the GPD literature, for instance Refs.\ \cite{Radyushkin:2011dh,Chall:2026oes}.

The structure functions in Eqs.~\eqref{eq:xss} are thus given by 
\begin{align}
F^0_{U,T} & =   [z^2 + (1-z)^2]\,\left\{ \left \vert  \bm{{\cal F}}_{1, 1}^0  \right \vert^2\,+\,\left \vert  \bm{{\cal G}}_{1}^0  \right \vert^2\,+\,\frac{\bm \Delta_\sT^2}{M_N^2}\,\left\vert{\bm{{\cal H}}}_{1, 1}^\phi\,+\,i\,{\bm{{\cal H}}}_{5, 1}^0\right\vert^2\,+\, \frac{\bm \Delta_\sT^2}{M_N^2}\,\left\vert{i\,\bm{{\cal H}}}_{2}^\phi\,+\,{\bm{{\cal H}}}_{6}^0\right\vert^2\right\} \nonumber \\
&\qquad+\, \frac{M^2_Q}{\bm K_\perp^2}\,\left\{ \,\left \vert \bm{{\cal F}}_{1, 2}^0  \right \vert^2\,+\,\frac{\bm \Delta_\sT^2}{M_N^2}\,\left\vert{\bm{{\cal H}}}_{1, 2}^\phi\,+\,i\,{\bm{{\cal H}}}_{5, 2}^0\right\vert^2\right\} \,\nonumber \\
&\qquad+\,\left \{ [\beta^2 +  (1-\beta)^2]\,[z^2 + (1-z)^2]\,+\,2 \,(1-\beta)^2\,\frac{M^2_Q}{\bm K_\perp^2}\right \}\,\frac{\bm \Delta_\sT^2}{M_N^2}\,\left\{2\,\left\vert\bm{{\cal H}}_3^\phi\,+\,i\,\bm{{\cal H}}_7^0\right\vert^2\right. \nonumber\\
&\qquad\left.+\,\frac{\bm \Delta_\sT^2}{M_N^2}\,\left \vert \bm{{\cal F}}_{2}^{2\phi}\,+\, \bm{{\cal F}}_{3}^{0}\right \vert^2\,+\,\frac{\bm \Delta_\sT^2}{M_N^2}\,\left \vert \bm{{\cal G}}_{2}^{2\phi}\,+\, \bm{{\cal G}}_{3}^{0}\right \vert^2 \,+\,\frac{\bm\Delta_\sT^4}{2M_N^4} \left\vert\bm{{\cal H}}_4^{3\phi}\,+\,i\,\bm{{\cal H}}_8^0\right\vert^2\right\} \,,\nonumber
\end{align}
\begin{align}
F_{U,T}^{\cos{2(\phi_{\Delta} - \phi_{\perp}})} &  = 2\,(1-2\beta)\,[z^2 + (1-z)^2] \,\frac{\bm \Delta_\sT^2}{M_N^2}\,\, \Re\left \{\bm{{\cal F}}^0_{1, 1}\, \left [ \,\bm{{\cal F}}_{2}^{2\phi\,*} \,+ \,\bm{{\cal F}}_{3}^{0\,*}\right ]\,+\,\left(\bm{{\cal H}}_{1, 1}^\phi\,+\,i\,\bm{{\cal H}}_{5, 1}^0\right)\,\left(\bm{{\cal H}}_3^{\phi*}\,+\,i\,\bm{{\cal H}}_7^{0*}\right)\,\right. \nonumber \\
&\qquad\left. -\,\,\frac{\bm \Delta_\sT^2}{2M_N^2}\, \left(\bm{{\cal H}}_{1, 1}^\phi\,+\,i\,\bm{{\cal H}}_{5, 1}^0\right)\,\left(\bm{{\cal H}}_4^{3\phi*}\,+\,i\,\bm{{\cal H}}_8^{0*}\right)\right \} \nonumber \\ 
 &\qquad+\, 4\,\frac{M^2_Q}{\bm K_\perp^2}\, (1-\beta) \,\frac{\bm \Delta_\sT^2}{M_N^2}\, \Re  \left \{\bm{{\cal F}}^0_{1, 2}\, \left [ \,\bm{{\cal F}}_{2}^{2\phi\,*} \,+ \,\bm{{\cal F}}_{3}^{0\,*}\right ]\,+\,\left(\bm{{\cal H}}_{1, 2}^\phi\,+\,i\,\bm{{\cal H}}_{5, 2}^0\right)\,\left(\bm{{\cal H}}_3^{\phi*}\,+\,i\,\bm{{\cal H}}_7^{0*}\right)\,\right. \nonumber \\
&\qquad \left. -\,\,\frac{\bm \Delta_\sT^2}{2M_N^2}\, \left(\bm{{\cal H}}_{1, 2}^\phi\,+\,i\,\bm{{\cal H}}_{5, 2}^0\right)\,\left(\bm{{\cal H}}_4^{3\phi*}\,+\,i\,\bm{{\cal H}}_8^{0*}\right)\right \} \nonumber \\
&\qquad-\,2\,[z^2 + (1-z)^2]\,\,\frac{\bm \Delta_\sT^2}{M_N^2}\,\Re\left\{i\,\bm{{\cal G}}_1^0\,\left(\bm{{\cal G}}_2^{2\phi*}\,+\,\bm{{\cal G}}_3^{0*}\right) \right. 
\left. -\,\left(i\,\bm{{\cal H}}_2^\phi\,+\,\bm{{\cal H}}_6^0\right)\,\left(\bm{{\cal H}}_3^{\phi*}\,-\,i\,\bm{{\cal H}}_7^{0*}\right) \right. \nonumber \\
&\qquad\left.-\,\frac{\bm \Delta_\sT^2}{2M_N^2}\,\left(i\,\bm{{\cal H}}_2^\phi\,+\,\bm{{\cal H}}_6^0\right)\,\left(\bm{{\cal H}}_4^{3\phi*}\,-\,i\,\bm{{\cal H}}_8^{0*}\right)\right\}\,,\nonumber \\
F_{U,T}^{\cos{4(\phi_{\Delta} - \phi_{\perp}})} &  = 2\,z (1-z)(1-\beta)^2\left\{ [z^2 + (1-z)^2]\,\frac{Q^2}{\bm K_\perp^2} - 2\, \frac{M_Q^2}{\bm K_\perp^2}\right\}\,\frac{\bm \Delta_\sT^4}{M_N^4}\,\left\{\left \vert \bm{{\cal F}}_{2}^{2\phi}\,+\,\bm{{\cal F}}_{3}^{0}\right \vert^2\,+\,\left \vert \bm{{\cal G}}_{2}^{2\phi}\,+\, \bm{{\cal G}}_{3}^{0}\right \vert^2\right.\nonumber \\
&\qquad\left. +\,2\,\Re\left\{\left(\bm{{\cal H}}_3^\phi\,+\,i\,\bm{{\cal H}}_7^0\right)\,\left(\bm{{\cal H}}_4^{3\phi*}\,-\,i\,\bm{{\cal H}}_8^{0*}\right)\right\}\right\} \,,
\end{align}
\begin{align}
F^0_{U,L} & = \left \vert  
 \bm{{\cal F}}_{1, 2}^0  \right \vert ^2\,+\,\frac{\bm \Delta_\sT^2}{M_N^2}\,\left\vert{\bm{{\cal H}}}_{1, 2}^\phi\,+\,i\,{\bm{{\cal H}}}_{5, 2}^0\right\vert^2  \,+\,(1-\beta)^2\,\frac{\bm \Delta_\sT^2}{M_N^2}\left\vert\bm{{\cal H}}_3^\phi\,+\,i\,\bm{{\cal H}}_7^0\right\vert^2 \,,\nonumber \\
 &\qquad+\, \frac{1}{2}\,(1-\beta)^2\,\frac{\bm \Delta_\sT^4}{M_N^4}\,\left\{\left \vert \bm{{\cal F}}_{2}^{2\phi}  \,+ \, \bm{{\cal F}}^{0}_{3} \right \vert ^2\,+\,\left \vert \bm{{\cal G}}_{2}^{2\phi}\,+\, \bm{{\cal G}}_{3}^{0}\right \vert^2\right\}\nonumber \\
 &\qquad+\,(1-\beta)^2\,\frac{\bm\Delta_\sT^6}{4M_N^6} \left\vert\bm{{\cal H}}_4^{3\phi}\,+\,i\,\bm{{\cal H}}_8^0\right\vert^2 \,,\nonumber \\
 F^{\,\cos 2(\phi_\Delta - \phi_\perp)}_{U,L} & = (1-\beta)\,\frac{\bm \Delta_\sT^2}{M_N^2}\,\Re \left \{\bm{{\cal F}}^0_{1, 2} \,\left [ \bm{{\cal F}}_{2}^{2\phi\,*} \,+\,\bm{{\cal F}}_{3}^{0\,*}\right ]\,+\,\left(\bm{{\cal H}}_{1, 2}^\phi\,+\,i\,\bm{{\cal H}}_{5, 2}^0\right)\,\left(\bm{{\cal H}}_3^{\phi*}\,-\,i\,\bm{{\cal H}}_7^{0*}\right)\,,\right. \nonumber \\
&\qquad\left. -\,\,\frac{\bm \Delta_\sT^2}{2M_N^2}\, \left(\bm{{\cal H}}_{1, 2}^\phi\,+\,i\,\bm{{\cal H}}_{5, 2}^0\right)\,\left(\bm{{\cal H}}_4^{3\phi*}\,-\,i\,\bm{{\cal H}}_8^{0*}\right)\right \}\,,\nonumber \\
  F^{\,\cos 4 (\phi_\Delta - \phi_\perp)}_{U,L}
  & =-\,\frac{1}{2}\,(1-\beta)^2\, \frac{\bm \Delta_\sT^4}{M_N^4}\,\left\{\left \vert \bm{{\cal F}}_{2}^{2\phi}  \,+\,  \bm{{\cal F}}^{0}_{3} \right \vert ^2\,+\,\left \vert \bm{{\cal G}}_{2}^{2\phi}\,+\, \bm{{\cal G}}_{3}^{0}\right \vert^2\,+\,4\,\Re\left\{\left(\bm{{\cal H}}_3^\phi\,+\,i\,\bm{{\cal H}}_7^0\right)\,\left(\bm{{\cal H}}_4^{3\phi*}\,-\,i\,\bm{{\cal H}}_8^{0*}\right)\right\}\right\}\,,
\label{eq:gamma-cs-UU}
\end{align}
\begin{align}
F_{U,TT}^{\,\cos 2 (\phi_\perp - \phi_\ell)} & = -\,\left \vert \bm{{\cal F}}_{1, 1}^0  \right \vert ^2+\,\left \vert  \bm{{\cal G}}_{1}^0  \right \vert^2\,+\,\frac{\bm \Delta_\sT^2}{M_N^2}\,\left\{-\,\left\vert{\bm{{\cal H}}}_{1, 1}^\phi\,+\,i\,{\bm{{\cal H}}}_{5, 1}^0\right\vert^2\,+\,\left\vert{\bm{{\cal H}}}_{2}^\phi\,+\,i\,{\bm{{\cal H}}}_{6}^0\right\vert^2\,+\,4\,\beta^3\,\left\vert\bm{{\cal H}}_3^\phi\,+\,i\,\bm{{\cal H}}_7^0\right\vert^2\right\}\nonumber\\
&\qquad- \, 2\,\beta(1-\beta)\,\frac{\bm \Delta_\sT^4}{M_N^4}\, \left\{\left \vert  \bm{{\cal F}}_{2}^{2\phi} \,+ \, \bm{{\cal F}}_{3}^{0}  \right \vert ^2\,-\,\left \vert \bm{{\cal G}}_{2}^{2\phi}\,+\, \bm{{\cal G}}_{3}^{0}\right \vert^2\right\}\,+\,\beta^3\,\frac{\bm \Delta_\sT^6}{M_N^6}\,\left\vert\bm{{\cal H}}_4^{3\phi}\,+\,i\,\bm{{\cal H}}_8^0\right\vert^2
 \,,\nonumber \\
 F_{U,TT}^{\,\cos 2 (\phi_\Delta - \phi_\ell)} & = 2 \beta \, \frac{\bm \Delta_\sT^2}{M_N^2}\,\Re \left \{\bm{{\cal F}}_{1, 1}^0 \, \left [ \bm{{\cal F}}_{2}^{\phi\, *} \,+\, \bm{{\cal F}}_{3}^{0 \,*}  \right ]\,+\,\left(\bm{{\cal H}}_{1, 1}^\phi\,+\,i\,\bm{{\cal H}}_{5, 1}^0\right)\,\left(\bm{{\cal H}}_3^{\phi*}\,-\,i\,\bm{{\cal H}}_7^{0*}\right)\right. \nonumber \\
&\qquad \left. -\,i\,\bm{{\cal G}}_1^0\,\left(\bm{{\cal G}}_2^{2\phi*}\,+\,\bm{{\cal G}}_3^{0*}\right)\,+\,\left(i\,\bm{{\cal H}}_2^\phi\,+\,\bm{{\cal H}}_6^0\right)\,\left(\bm{{\cal H}}_3^{\phi*}\,-\,i\,\bm{{\cal H}}_7^{0*}\right) \right. \nonumber \\ 
&\qquad \left. -\,\,\frac{\bm \Delta_\sT^2}{2M_N^2}\, \left(\bm{{\cal H}}_{1, 1}^\phi\,+\,i\,\bm{{\cal H}}_{5, 1}^0\right)\,\left(\bm{{\cal H}}_4^{3\phi*}\,-\,i\,\bm{{\cal H}}_8^{0*}\right)\,+\,\frac{\bm \Delta_\sT^2}{2M_N^2}\,\left(i\,\bm{{\cal H}}_2^\phi\,+\,\bm{{\cal H}}_6^0\right)\,\left(\bm{{\cal H}}_4^{3\phi*}\,-\,i\,\bm{{\cal H}}_8^{0*}\right)\right \}\,, \nonumber \\ 
 F_{U,TT}^{\,\cos 2 (2\phi_\perp - \phi_\Delta - \phi_\ell)} & =- 2 \,(1-\beta)\,\frac{\bm \Delta_\sT^2}{M_N^2}\, \Re \left \{ \bm{{\cal F}}_{1, 1}^0  \, \left [ \bm{{\cal F}}_{2}^{\phi\, *} \,+\, \bm{{\cal F}}_{3}^{0 \,*}  \right ] \,+\,\left(\bm{{\cal H}}_{1, 1}^\phi\,+\,i\,\bm{{\cal H}}_{5, 1}^0\right)\,\left(\bm{{\cal H}}_3^{\phi*}\,-\,i\,\bm{{\cal H}}_7^{0*}\right)\,\right. \nonumber \\
&\qquad \left. +\,i\,\bm{{\cal G}}_1^0\,\left(\bm{{\cal G}}_2^{2\phi*}\,+\,\bm{{\cal G}}_3^{0*}\right)\,-\,\left(i\,\bm{{\cal H}}_2^\phi\,+\,\bm{{\cal H}}_6^0\right)\,\left(\bm{{\cal H}}_3^{\phi*}\,-\,i\,\bm{{\cal H}}_7^{0*}\right) \right. \nonumber \\ 
&\qquad \left. -\,\frac{\bm \Delta_\sT^2}{2M_N^2}\, \left(\bm{{\cal H}}_{1, 1}^\phi\,+\,i\,\bm{{\cal H}}_{5, 1}^0\right)\,\left(\bm{{\cal H}}_4^{3\phi*}\,-\,i\,\bm{{\cal H}}_8^{0*}\right)\,-\,\frac{\bm \Delta_\sT^2}{2M_N^2}\,\left(i\,\bm{{\cal H}}_2^\phi\,+\,\bm{{\cal H}}_6^0\right)\,\left(\bm{{\cal H}}_4^{3\phi*}\,-\,i\,\bm{{\cal H}}_8^{0*}\right)
\right \}\,,\nonumber
\end{align}
\begin{align}
F_{U,TT}^{\,\cos 2 (\phi_\perp - 2\phi_\Delta + \phi_\ell)}
 & = \beta^2\,\frac{\bm \Delta_\sT^4}{M_N^4}\, \left\{ \left \vert \bm{{\cal F}}_2^{2\phi} \,  + \, \bm{{\cal F}}_3^0 \right \vert^2\,+\,\left \vert \bm{{\cal G}}_{2}^{2\phi}\,+\, \bm{{\cal G}}_{3}^{0}\right \vert^2\,+\,2\,\Re\left\{\left(\bm{{\cal H}}_3^\phi\,+\,i\,\bm{{\cal H}}_7^0\right)\,\left(\bm{{\cal H}}_4^{3\phi*}\,-\,i\,\bm{{\cal H}}_8^{0*}\right)\right\}\right\}\,,\nonumber \\
 F_{U,TT}^{\,\cos 2 (3\phi_\perp - 2\phi_\Delta -\phi_\ell)} & = 
 (1-\beta)^2\,\frac{\bm \Delta_\sT^4}{M_N^4}\,\left\{\left \vert \bm{{\cal F}}_2^{2\phi} \,  + \, \bm{{\cal F}}_{3}^0 \right \vert^2\,+\,\left \vert \bm{{\cal G}}_{2}^{2\phi}\,+\, \bm{{\cal G}}_{3}^{0}\right \vert^2\,+\,2\,\,\Re\left\{\left(\bm{{\cal H}}_3^\phi\,+\,i\,\bm{{\cal H}}_7^0\right)\,\left(\bm{{\cal H}}_4^{3\phi*}\,-\,i\,\bm{{\cal H}}_8^{0*}\right)\right\}\right\}\,. 
\end{align}

\begin{align}
F_{U,LT}^{\,\cos (\phi_\perp - \phi_\ell)} & = 
 \bm{{\cal F}}_{1, 1}^0\,\bm{{\cal F}}_{1, 2}^{0\;*}\,+\,\,\frac{\bm \Delta_\sT^2}{M_N^2}\,\left({\bm{{\cal H}}}_{1, 1}^\phi\,+\,i\,{\bm{{\cal H}}}_{5, 1}^0\right)\,\left({\bm{{\cal H}}}_{1, 2}^{\phi*}\,-\,i\,{\bm{{\cal H}}}_{5, 2}^{0*}\right) \nonumber \\
 &\qquad-\,(1-\beta)(1-2\beta)\,\frac{\bm \Delta_\sT^2}{M_N^2}\,\left\{\left\vert\bm{{\cal H}}_3^\phi\,+\,i\,\bm{{\cal H}}_7^0\right\vert^2\,+\,\frac{\bm \Delta_\sT^2}{2M_N^2}\left \vert \bm{{\cal F}}_{2}^{\phi}  \,+\, \bm{{\cal F}}^{0}_{3} \right \vert ^2\right.\nonumber \\
 &\qquad\left.+\,\frac{\bm \Delta_\sT^2}{2M_N^2}\,\left \vert \bm{{\cal G}}_{2}^{2\phi}\,+\, \bm{{\cal G}}_{3}^{0}\right \vert^2\,+\,\frac{\bm\Delta_\sT^4}{4M_N^4} \left\vert\bm{{\cal H}}_4^{3\phi}\,-\,i\,\bm{{\cal H}}_8^0\right\vert^2\right\} \,, \nonumber \\
F_{U,LT}^{\,\cos (3\phi_\perp - 2\phi_{\Delta}-\phi_\ell)} & =  -\,\frac{1}{2}\,(1-\beta)\,\frac{\bm \Delta_\sT^2}{M_N^2}\, \Re\left \{\bm{{\cal F}}^0_{1, 1}\, \left [ \,\bm{{\cal F}}_{2}^{2\phi\,*} \,+ \,\bm{{\cal F}}_{3}^{0\,*}\right ]\,+\,\left(\bm{{\cal H}}_{1, 1}^\phi\,+\,i\,\bm{{\cal H}}_{5, 1}^0\right)\,\left(\bm{{\cal H}}_3^{\phi*}\,-\,i\,\bm{{\cal H}}_7^{0*}\right)\,\right. \nonumber \\
&\qquad\left. +\,\bm{{\cal F}}^0_{1, 2}\, \left [ \,\bm{{\cal F}}_{2}^{2\phi\,*} \,+ \,\bm{{\cal F}}_{3}^{0\,*}\right ]\,+\,\left(\bm{{\cal H}}_{1, 2}^\phi\,+\,i\,\bm{{\cal H}}_{5, 2}^0\right)\,\left(\bm{{\cal H}}_3^{\phi*}\,-\,i\,\bm{{\cal H}}_7^{0*}\right)\right. \nonumber\\
& \qquad+\,i\,\bm{{\cal G}}_1^0\,\left(\bm{{\cal G}}_2^{2\phi*}\,+\,\bm{{\cal G}}_3^{0*}\right)\,-\,\left(i\,\bm{{\cal H}}_2^\phi\,+\,\bm{{\cal H}}_6^0\right)\,\left(\bm{{\cal H}}_3^{\phi*}\,-\,i\,\bm{{\cal H}}_7^{0*}\right)\nonumber
\end{align}
\begin{align}
& \qquad-\,\frac{\bm \Delta_\sT^2}{2M_N^2}\, \left(\bm{{\cal H}}_{1, 1}^\phi\,+\,i\,\bm{{\cal H}}_{5, 1}^0\right)\,\left(\bm{{\cal H}}_4^{3\phi*}\,-\,i\,\bm{{\cal H}}_8^{0*}\right)\, -\,\,\frac{\bm \Delta_\sT^2}{2M_N^2}\, \left(\bm{{\cal H}}_{1, 2}^\phi\,+\,i\,\bm{{\cal H}}_{5, 2}^0\right)\,\left(\bm{{\cal H}}_4^{3\phi*}\,-\,i\,\bm{{\cal H}}_8^{0*}\right)\nonumber \\
&\qquad\left.-\,\frac{\bm \Delta_\sT^2}{2M_N^2}\,\left(i\,\bm{{\cal H}}_2^\phi\,+\,\bm{{\cal H}}_6^0\right)\,\left(\bm{{\cal H}}_4^{3\phi*}\,-\,i\,\bm{{\cal H}}_8^{0*}\right)\right\} \,,\nonumber \\
F_{U,LT}^{\,\cos (\phi_\perp - 2\phi_{\Delta}+\phi_\ell)} & = \frac{1}{2}\,(1-\beta)\,\frac{\bm \Delta_\sT^2}{M_N^2}\,\, \Re\left \{-\,\bm{{\cal F}}^0_{1, 1}\, \left [ \,\bm{{\cal F}}_{2}^{2\phi\,*} \,+\,\bm{{\cal F}}_{3}^{0\,*}\right ]\,-\,\left(\bm{{\cal H}}_{1, 1}^\phi\,+\,i\,\bm{{\cal H}}_{5, 1}^0\right)\,\left(\bm{{\cal H}}_3^{\phi*}\,-\,i\,\bm{{\cal H}}_7^{0*}\right)\,\right. \nonumber \\
&\qquad +\,i\,\bm{{\cal G}}_1^0\,\left(\bm{{\cal G}}_2^{2\phi*}\,+\,\bm{{\cal G}}_3^{0*}\right)\,-\,\left(i\,\bm{{\cal H}}_2^\phi\,+\,\bm{{\cal H}}_6^0\right)\,\left(\bm{{\cal H}}_3^{\phi*}\,-\,i\,\bm{{\cal H}}_7^{0*}\right) \nonumber \\
&\qquad\left. +\,\,\frac{\bm \Delta_\sT^2}{2M_N^2}\, \left(\bm{{\cal H}}_{1, 1}^\phi\,+\,i\,\bm{{\cal H}}_{5, 1}^0\right)\,\left(\bm{{\cal H}}_4^{3\phi*}\,-\,i\,\bm{{\cal H}}_8^{0*}\right)\,-\,\frac{\bm \Delta_\sT^2}{2M_N^2}\,\left(i\,\bm{{\cal H}}_2^\phi\,+\,\bm{{\cal H}}_6^0\right)\,\left(\bm{{\cal H}}_4^{3\phi*}\,-\,i\,\bm{{\cal H}}_8^{0*}\right)\right \}  \nonumber\\
&\qquad+\,\frac{1}{2}\,\beta\,\frac{\bm \Delta_\sT^2}{M_N^2}\,\, \Re\left \{\bm{{\cal F}}^0_{1, 2}\, \left [ \,\bm{{\cal F}}_{2}^{2\phi\,*} \,+ \,\bm{{\cal F}}_{3}^{0\,*}\right ]\,+\,\left(\bm{{\cal H}}_{1, 2}^\phi\,+\,i\,\bm{{\cal H}}_{5, 2}^0\right)\,\left(\bm{{\cal H}}_3^{\phi*}\,-\,i\,\bm{{\cal H}}_7^{0*}\right)\,\right. \nonumber \\
&\qquad\left. -\,\frac{\bm \Delta_\sT^2}{2M_N^2}\, \left(\bm{{\cal H}}_{1, 2}^\phi\,+\,i\,\bm{{\cal H}}_{5, 2}^0\right)\,\left(\bm{{\cal H}}_4^{3\phi*}\,-\,i\,\bm{{\cal H}}_8^{0*}\right)\right \} \,,  \nonumber\\
F_{U,LT}^{\,\cos (5\phi_\perp - 4\phi_{\Delta}-\phi_\ell)} & =-\,\frac{1}{2}\,(1-\beta)^2\,\frac{\bm \Delta_\sT^4}{M_N^4}\,\left\{\left \vert \bm{{\cal F}}_{2}^{2\phi}  \,+ \, \bm{{\cal F}}^{0}_{3} \right \vert ^2\,-\,\left \vert \bm{{\cal G}}_{2}^{2\phi}\,+\, \bm{{\cal G}}_{3}^{0}\right \vert^2\,-\,\Re\left\{\left(\bm{{\cal H}}_3^\phi\,+\,i\,\bm{{\cal H}}_7^0\right)\,\left(\bm{{\cal H}}_4^{3\phi*}\,-\,i\,\bm{{\cal H}}_8^{0*}\right)\right\}\right\} \,, \nonumber\\
F_{U,LT}^{\,\cos (3\phi_\perp - 4\phi_{\Delta} + \phi_\ell)} & = \frac{1}{2}\,\beta(1-\beta)\,\frac{\bm \Delta_\sT^4}{M_N^4}\,\left\{\left \vert \bm{{\cal F}}_{2}^{2\phi}  \,+ \, \bm{{\cal F}}^{0}_{3} \right \vert ^2\,-\,\left \vert \bm{{\cal G}}_{2}^{2\phi}\,+\, \bm{{\cal G}}_{3}^{0}\right \vert^2\,-\,\Re\left\{\left(\bm{{\cal H}}_3^\phi\,+\,i\,\bm{{\cal H}}_7^0\right)\,\left(\bm{{\cal H}}_4^{3\phi*}\,-\,i\,\bm{{\cal H}}_8^{0*}\right)\right\}\right\}\,.
\end{align}
We note that the GTMDs ${\cal F}^g_4$ and ${\cal G}_4^g$ do not appear in any of the above structure functions, at least at leading order in the collinear expansion that we are considering. This is due to the fact that GTMDs depend in general on $\bm k_\sT\cdot \bm \Delta_\sT$. Furthermore, ${\cal F}^g_4$ and ${\cal G}_4^g$ are multiplied by a factor proportional to $\sin(\phi_\Delta-\phi_\sT)$, with $\phi_\sT$ being the azimuthal angle of the gluon average momentum $k_\sT$, and therefore their integral over $k_\sT$ is zero\footnote{The functions ${\cal F}_4$ and ${\cal G}_4$ do not show up in our expressions but they would show up beyond leading power, like in the twist-3 asymmetries considered in Ref.~\cite{Ji:2016jgn}.}.  This is in contrast to the cross section for $\gamma\gamma \to Q \,\overline Q$ in ultraperipheral collisions of two nuclei in Ref.~\cite{Boer:2024cnw}, where a contribution from ${\cal F}_4^\gamma$ is present, although convoluted with another ${\cal F}_4^\gamma$. 

In general, every structure function receives contributions from different GTMDs, which complicates their phenomelogical extraction. In order to single them out, one needs to look at different angular modulations in different kinematic regions. For instance, from $F_{U,L}^0$ at $\vert \bm \Delta_\sT\vert  \ll M_N$ one can extract the unpolarized gluon GTMD ${\cal F}_1^g$ (or rather its integral over $x$ and $\bm k_\sT$) and then from $F^0_{U,T}$ one can gather information on the distribution of circularly polarized gluons inside a longitudinally polarized nucleon ${\cal G}^g_1$.

We end this section with some comments on the small-$x$ limit. Like for TMDs, at vanishing $x$ and $\xi$ some GTMDs for the dipole gauge link configuration ($\Gamma^{[+,-]}$), which enters in the process under consideration, become related. In Ref.~\cite{Boer:2018vdi} it was shown that (see Appendix~\ref{app-4} for more details):
\begin{equation}
    \lim_{x,\xi \to 0} \mathcal{F}_1 = \lim_{x,\xi \to 0} \mathcal{F}_2^{(1)} = -4 \lim_{x,\xi \to 0} \mathcal{F}_3^{(1)} = -2 \lim_{x,\xi \to 0} \mathcal{F}_4^{(1)} ,
    \label{relationbetweenFs}
\end{equation}
which employs the shorthand notation: $\mathcal{F}_i^{(n)} \equiv [(\bm{k}^2 - \bm{\Delta}^2/4)/(2M_N^2)]^n \,\mathcal{F}_i$. Because of this relation the $\bm{k}_\sT\cdot \bm{\Delta}_\sT$ dependence of the four functions become related, in particular, their so-called elliptic part, i.e.\ their $\cos 2 (\phi_k - \phi_\Delta)$ dependence. Therefore, at small $x$ one can discuss ``the'' elliptic Wigner function (ignoring the elliptic parts of the ${\cal H}$ GTMDs) and in several small-$x$ analyses \cite{Hatta:2016dxp,Zhou:2016rnt,Hagiwara:2017fye,ReinkePelicer:2018gyh,Mantysaari:2019csc} the $(\phi_{\Delta} - \phi_{\perp})$-dependence of the cross section due to this elliptic Wigner function is studied. It might be expected that this is associated with the elliptic part of ${\cal F}_1$, but as can be seen, 
no ${\cal F}_{1}^{2\phi}$ enters in any of the above expressions (since no $\bm{{\cal F}}_{1,i}^{2\phi}$ appears), not even in $F_{U,T/L}^{\cos{2(\phi_{\Delta} - \phi_{\perp}})}$. However, due to the equalities in Eq.\ (\ref{relationbetweenFs}) for the dipole gauge link gluon GTMDs at small-$x$, ${{\cal F}}_{1}^{2\phi}$ becomes related to the elliptic part of ${\cal F}_2$, which {\it does} enter in the expressions. The fact that this refers to ${{\cal F}}_{2}^{2\phi}$ is relevant for the integral relation between a GPD and ``the'' elliptic GTMD given in Ref.~\cite{Hatta:2017cte}. This is addressed further in Appendix~\ref{app-4}. Moreover, in Appendix~\ref{app-3} we provide leading order expressions for the GPDs in terms of integrals over GTMDs, which also show that ${{\cal F}}_{1}^{2\phi}$ does not enter in any of them.

Besides the contributions from various ${\cal F}$-type GTMDs to the $(\phi_{\Delta} - \phi_{\perp})$-dependence of the cross section, we note that there are many contributions from ${\cal G}$-type and ${\cal H}$-type GTMDs, even at leading power. In the small-$x$ limit, the ${\cal G}$-type functions vanish, but the ${\cal H}$-type GTMDs do not, as we discuss further in Appendix~\ref{app-4}. Needless to say, it will be a very difficult task to isolate the elliptic contribution ${\cal F}_{2}^{2\phi}$ from all the other effects and this will considerably complicate the analysis of the $(\phi_{\Delta} - \phi_{\perp})$-dependence, such as in Ref.~\cite{Linek:2023kga}.

\section{Structure functions in the GPD framework}
\label{sec:SF-GPD}
For a generic GPD $Y$ we define the following Compton form factors,
\begin{align}
{\bm Y}_{1} & = \int_{-1}^{1}\d x\, \left [\frac{1-2\beta}{(x+\xi - i \epsilon)^2} \,+\, \frac{1-2\beta}{(x-\xi + i \epsilon)^2} + \frac{4\beta}{(x+\xi - i \epsilon)(x - \xi +i \epsilon)} \right ] Y \,, \nonumber\\
{\bm Y}_{2} & = 2 
\int_{-1}^{1}\d x\, \left [\frac{1-\beta}{(x+\xi - i \epsilon)^2} \,+\, \frac{1-\beta}{(x-\xi + i \epsilon)^2} - \frac{1- 2\beta}{(x+\xi - i \epsilon)(x - \xi +i \epsilon)} \right ] Y \,, \nonumber \\
{\bm Y}_{3} & = \hide{(1-\beta)}\int_{-1}^{1}\d x\, \left [\frac{1}{2(x+\xi - i \epsilon)^2} \,+\, \frac{1}{2(x-\xi + i \epsilon)^2} - \frac{1}{(x+\xi - i \epsilon)(x - \xi +i \epsilon)} \right ] Y
\,,\nonumber \\
{\bm Y}_{4} & = \hide{(1-\beta)}
\int_{-1}^{1}\d x\, \left [-\,\frac{1}{(x+\xi - i \epsilon)^2} \,+\, \frac{1}{(x-\xi + i \epsilon)^2}\right ]  Y \,,
\end{align}
in terms of which the structure functions in Eq.~\eqref{eq:xss} can be written as 
\begin{align}
F_{U,T}^0  & =   (1-\xi^2)\,[z^2 + (1-z)^2]\,\left\{ \left ({\bm H}_1^g\,-\,\frac{\xi^2}{1-\xi^2}\,{\bm{E}}_1^g  \right )^2\,+\,\frac{1}{(1-\xi^2)^2}\,\frac{\bm \Delta_\sT^2}{4M_N^2}\,\left({\bm{E}}_1^{g2}\,+\,\xi^2\,{\bm{\widetilde E}}_4^{g2}\right) \right. \nonumber \\
& \qquad  \left. +\,\left ({\bm{\widetilde H}}_4^g\,-\,\frac{\xi^2}{1-\xi^2}\,{\bm{\widetilde E}}_4^g\right)^2\right\}\,+\, (1-\xi^2)\,\frac{M^2_Q}{\bm K_\perp^2}\,\left\{ \,\left ({\bm H}_2^g\,-\,\frac{\xi^2}{1-\xi^2}\,{\bm{E}}_2^g\right)^2\,+\frac{1}{(1-\xi^2)^2}\,\frac{\bm \Delta_\sT^2}{4M_N^2}\,{\bm{E}}_2^{g2}\right\} \,\nonumber \\
&\qquad +\,\left \{ [\beta^2 +  (1-\beta)^2]\,[z^2 + (1-z)^2]\,+\,2 \,(1-\beta)^2\,\frac{M^2_Q}{\bm K_\perp^2}\right \}\nonumber \\
&\qquad \times\,\frac{\bm \Delta_\sT^2}{M_N^2}\left\{2\,(1-\xi^2)\,\left(-\,{\bm H}_{\bm{\sT} 3}^g\,-\,\frac{\xi}{1-\xi^2}\,\widetilde{{\bm{E}}}_{\bm{\sT} 3}^g\,+\,\frac{\xi^2}{1-\xi^2}\,{\bm{E}}_{\bm{\sT} 3}^g\,-\,\frac{\bm \Delta_\sT^2}{4M_N^2}\,\frac{\widetilde{{\bm H}}_{\bm{\sT} 3}^g}{1-\xi^2}\right)^2 \right. \nonumber \\
&\qquad  \left .+\,\frac{1}{1-\xi^2}\,\frac{\bm \Delta_\sT^2}{M_N^2}\,\left\{\left (-\,\widetilde{\bm H}^g_{\bm{\sT} 3}\,-\frac{1}{2}\,{\bm{E}}_{\bm{\sT} 3}^g\,+\frac{\xi}{2}\,\widetilde{{\bm{E}}}^g_{\bm{\sT}3}\right)^2\,+\,\frac{1}{4}\,\left ( -\,\xi\,{\bm{E}}_{\bm{\sT} 3}^g\,+\,\widetilde{{\bm{E}}}_{\bm{\sT} 3}^g\right )^2\right\} +\,\frac{\bm\Delta_\sT^4}{8M_N^4}\frac{\widetilde{{\bm H}}^{g2}_{\bm{\sT} 3}}{(1-\xi^2)}\right\}\,, \nonumber\\
F_{U,T}^{\cos{2(\phi_{\Delta} - \phi_{\perp}})} &  = 2\,(1-2\beta)\,[z^2 + (1-z)^2] \,\frac{\bm \Delta_\sT^2}{M_N^2}\,\, \left \{\left({\bm H}_1^g\,-\,\frac{\xi^2}{1-\xi^2}\,{\bm{E}}_1^g\right)\,\left(-\,\widetilde{{\bm{H}}}^g_{\bm{\sT} 3}\,-\frac{1}{2}\,{\bm{E}}_{\bm{\sT} 3}^g\,+\frac{\xi}{2}\,\widetilde{{\bm{E}}}^g_{\bm{\sT} 3}\right)\, \right. \nonumber \\
& \left. -\,\frac{1}{2}\,{\bm{E}}^g_1\,\left(-\,{\bm H}_{\bm{\bm{\sT}} 3}^g\,-\,\frac{\xi}{1-\xi^2}\,\widetilde{{\bm{E}}}_{\bm{\sT} 3}^g\,+\,\frac{\xi^2}{1-\xi^2}\,{\bm{E}}_{\bm{\sT} 3}^g\,-\,\frac{\bm \Delta_\sT^2}{2M_N^2}\,\frac{\widetilde{{\bm{H}}}_{\bm{\sT} 3}^g}{1-\xi^2}\right)\right \} \nonumber \\
&+\, 4\,\frac{M^2_Q}{\bm K_\perp^2}\, (1-\beta) \,\frac{\bm \Delta_\sT^2}{M_N^2}\,  \left \{\left({\bm H}_2^g\,-\,\frac{\xi^2}{1-\xi^2}\,{\bm{E}}_2^g\right)\,\left(-\,\widetilde{{\bm{H}}}^g_{\bm{\sT} 3}\,-\frac{1}{2}\,{\bm{E}}_{\bm{\sT} 3}^g\,+\frac{\xi}{2}\,\widetilde{{\bm{E}}}^g_{\bm{\sT} 3}\right)\, \right. \nonumber \\
& \left. -\,\frac{1}{2}\,{\bm{E}}^g_2\,\left(-\,{\bm H}_{\bm{\sT} 3}^g\,-\,\frac{\xi}{1-\xi^2}\,\widetilde{{\bm{E}}}_{\bm{\sT} 3}^g\,+\,\frac{\xi^2}{1-\xi^2}\,{\bm{E}}_{\bm{\sT} 3}^g\,-\,\frac{\bm \Delta_\sT^2}{2M_N^2}\,\frac{\widetilde{{\bm{H}}}_{\bm{\sT} 3}^g}{1-\xi^2}\right)\right \} \nonumber \\
&-\,2\,[z^2 + (1-z)^2]\,\frac{\bm \Delta_\sT^2}{M_N^2}\,\left\{\frac{1}{2}\,\left(\widetilde{{\bm{H}}}_4^g\,-\,\frac{\xi^2}{1-\xi^2}\,\widetilde{{\bm{E}}}_4^g\right)\,\left(-\xi\,{\bm{E}}_{\bm{\sT} 3}^g\,+\,\widetilde{{\bm{E}}}_{\bm{\sT} 3}^g\right) \right. \nonumber \\
&\left. +\,\frac{\xi}{2}\,\widetilde{{\bm{E}}}_4^g\,\left(-\,{\bm H}_{\bm{\sT} 3}^g\,-\,\frac{\xi}{1-\xi^2}\,\widetilde{{\bm{E}}}_{\bm{\sT} 3}^g\,+\,\frac{\xi^2}{1-\xi^2}\,{\bm{E}}_{\bm{\sT} 3}^g\right)\right\}\,,\nonumber\\
F_{U,T}^{\cos{4(\phi_{\Delta} - \phi_{\perp}})} &  = 2\,z (1-z)(1-\beta)^2\left\{ [z^2 + (1-z)^2]\,\frac{Q^2}{\bm K_\perp^2} - 2\, \frac{M_Q^2}{\bm K_\perp^2}\right\}\,\frac{1}{1-\xi^2} \,\frac{\bm \Delta_\sT^4}{M_N^4} \nonumber \\
& \times\,\left\{\left(-\,\widetilde{{\bm{H}}}^g_{\bm{\sT} 3}\,-\frac{1}{2}\,{\bm{E}}_{\bm{\sT} 3}^g\,+\frac{\xi}{2}\,\widetilde{{\bm{E}}}^g_{\bm{\sT} 3}\right)^2\,+\,\frac{1}{4}\,\left(-\,\xi\,{\bm{E}}_{\bm{\sT} 3}^g\,+\,\widetilde{{\bm{E}}}_{\bm{\sT} 3}^g\right)^2\right. \nonumber \\
&\left.+\,(1-\xi^2)\,\left(-\,{\bm H}_{\bm{\sT} 3}^g\,-\,\frac{\xi}{1-\xi^2}\,\widetilde{{\bm{E}}}_{\bm{\sT} 3}^g\,+\,\frac{\xi^2}{1-\xi^2}\,{\bm{E}}_{\bm{\sT} 3}^g\,-\,\frac{\bm \Delta_\sT^2}{4M_N^2}\,\frac{\widetilde{{\bm{H}}}_{\bm{\sT} 3}^g}{1-\xi^2}\right)\,\widetilde{{\bm{H}}}^g_{\bm{\sT} 3}\right\} \,,
\end{align}
\begin{align}
    F^0_{U,L} & = (1-\xi^2)\,\left({\bm H}_2^g\,-\,\frac{\xi^2}{1-\xi^2}\,{\bm{E}}_2^g \right ) ^2 \nonumber \\
&\qquad +\,\frac{\bm \Delta_\sT^2}{M_N^2}\,\left\{ \, \frac{1}{4(1-\xi^2)}\, {\bm{E}}_2^{g2}\,+\,(1-\beta)^2\,(1-\xi^2)\,\left(-\,{\bm H}_{\bm{\sT} 3}^g\,-\,\frac{\xi}{1-\xi^2}\,\widetilde{{\bm{E}}}_{\bm{\sT} 3}^g\,+\,\frac{\xi^2}{1-\xi^2}\,{\bm{E}}_{\bm{\sT} 3}^g\right)^2\right\} \nonumber \\
& \qquad +  \, \frac{1}{2}\,(1-\beta)^2\,\frac{1}{1-\xi^2}\,\frac{\bm \Delta_\sT^4}{M_N^4}\,\left\{\left(-\,\widetilde{{\bm{H}}}^g_{\bm{\sT} 3}\,-\frac{1}{2}\,{\bm{E}}_{\bm{\sT} 3}^g\,+\frac{\xi}{2}\,\widetilde{{\bm{E}}}^g_{\bm{\sT} 3}\right) ^2\,+\,\frac{1}{4}\,\left(-\,\xi\,{\bm{E}}_{\bm{\sT} 3}^g\,+\,\widetilde{{\bm{E}}}_{\bm{\sT} 3}^g\right)^2\right\}\nonumber \\
 &\qquad +(1-\beta)^2\,\frac{1}{1-\xi^2}\,\frac{\bm \Delta_\sT^6}{16M_N^6}\,\widetilde{{\bm{H}}}^{g2}_{\bm{\sT} 3} \,, \nonumber 
  \end{align}
 \begin{align}
 F^{\,\cos 2(\phi_\Delta - \phi_\perp)}_{U,L} & = (1-\beta)\,\frac{\bm \Delta_\sT^2}{M_N^2}\, \left \{\left({\bm H}_2^g\,-\,\frac{\xi^2}{1-\xi^2}\,{\bm{E}}_2^g\right)\,\left(-\,\widetilde{{\bm{H}}}^g_{\bm{\sT} 3} -\,\frac{1}{2}\,{\bm{E}}_{\bm{\sT} 3}^g\,+\frac{\xi}{2}\,\widetilde{{\bm{E}}}^g_{\bm{\sT} 3}\right)\right.\nonumber \\
 & \qquad \left. -\,\frac{1}{2}\,{\bm{E}}_2^g\,\left(-\,{\bm H}_{\bm{\sT} 3}^g\,-\,\frac{\xi}{1-\xi^2}\,\widetilde{{\bm{E}}}_{\bm{\sT} 3}^g\,+\,\frac{\xi^2}{1-\xi^2}\,{\bm{E}}_{\bm{\sT} 3}^g\,-\,\frac{\bm \Delta_\sT^2}{2M_N^2}\,\frac{\widetilde{{\bm{H}}}_{\bm{\sT} 3}^g}{1-\xi^2}\right)\right \}\,,\nonumber \\
F^{\,\cos 4 (\phi_\Delta - \phi_\perp)}_{U,L}
& =-\frac{1}{2}\,(1-\beta)^2\, \frac{1}{1-\xi^2}\,\frac{\bm \Delta_\sT^4}{M_N^4}\,\left\{\left(-\,\widetilde{{\bm{H}}}^g_{\bm{\sT} 3}\,-\frac{1}{2}\,{\bm{E}}_{\bm{\sT} 3}^g\,+\frac{\xi}{2}\,\widetilde{{\bm{E}}}^g_{\bm{\sT} 3}\right) ^2\,-\,\frac{1}{4}\,\left(-\,\xi\,{\bm{E}}_{\bm{\sT} 3}^g\,+\,\widetilde{{\bm{E}}}_{\bm{\sT} 3}^g\right)^2\right\}\nonumber \\
& \qquad -\,(1-\beta)^2\, \frac{\bm \Delta_\sT^4}{M_N^4}\,\left(-\,{\bm H}_{\bm{\sT} 3}^g\,-\,\frac{\xi}{1-\xi^2}\,\widetilde{{\bm{E}}}_{\bm{\sT} 3}^g\,+\,\frac{\xi^2}{1-\xi^2}\,{\bm{E}}_{\bm{\sT} 3}^g\,-\,\frac{\bm \Delta_\sT^2}{4M_N^2}\,\frac{\widetilde{{\bm{H}}}_{\bm{\sT} 3}^g}{1-\xi^2}\right)\,\widetilde{{\bm{H}}}^g_{\bm{\sT} 3}\,, 
\end{align}
\begin{align}
F_{U,TT}^{\,\cos 2 (\phi_\perp - \phi_\ell)} & = (1-\xi^2)\,\left\{-\,\left ({\bm H}_1^g\,-\,\frac{\xi^2}{1-\xi^2}\,{\bm{E}}_1^g \right )^2\,+\,\left(\widetilde{{\bm{H}}}_4^g\,-\,\frac{\xi^2}{1-\xi^2}\,\widetilde{{\bm{E}}}_4^g\right)^2 \right\}  \nonumber \\
& \qquad +\, \frac{\bm \Delta_\sT^2}{4M_N^2} \, \frac{1}{1-\xi^2}\,\left( \xi^2\,\widetilde{{\bm{E}}}_4^{g2}\,-\,{\bm{E}}_1^{g2}\right)\nonumber \\
&+\,4\,\beta^3\,(1-\xi^2)\,\frac{\bm \Delta_\sT^2}{M_N^2}\,\left(-\,{\bm H}_{\bm{\sT} 4}^g\,-\,\frac{\xi}{1-\xi^2}\,\widetilde{{\bm{E}}}_{\bm{\sT} 3}^g\,+\,\frac{\xi^2}{1-\xi^2}\,{\bm{E}}_{\bm{\sT} 3}^g\,-\,\frac{\bm \Delta_\sT^2}{4M_N^2}\,\frac{\widetilde{{\bm{H}}}_{\bm{\sT} 3}^g}{1-\xi^2}\right)^2\nonumber \\
& \qquad - \, 2\,\beta(1-\beta)\,\frac{1}{1-\xi^2}\,\frac{\bm \Delta_\sT^4}{M_N^4}\,\left\{\left(-\,\widetilde{{\bm{H}}}^g_{\bm{\sT} 3}\,-\frac{1}{2}\,{\bm{E}}_{\bm{\sT} 3}^g\,+\frac{\xi}{2}\,\widetilde{{\bm{E}}}^g_{\bm{\sT} 3}\right) ^2\,-\,\frac{1}{4}\,\left(-\,\xi\,{\bm{E}}_{\bm{\sT} 3}^g\,+\,\widetilde{{\bm{E}}}_{\bm{\sT} 3}^g\right)^2\right\}  \nonumber \\
& \qquad +\,\beta^3\,\frac{1}{1-\xi^2}\,\frac{\bm \Delta_\sT^6}{4 M_N^6}\,\widetilde{{\bm{H}}}^{g2}_{\bm{\sT} 3} \,,\nonumber \\
 F_{U,TT}^{\,\cos 2 (\phi_\Delta - \phi_\ell)} & = 2 \beta \, \frac{\bm \Delta_\sT^2}{M_N^2}\, \left \{\left({\bm H}_1^g\,-\,\frac{\xi^2}{1-\xi^2}\,{\bm{E}}_1^g\right)\,\left(-\,\widetilde{{\bm{H}}}^g_{\bm{\sT} 3}\,-\frac{1}{2}\,{\bm{E}}_{\bm{\sT} 3}^g\,+\frac{\xi}{2}\,\widetilde{{\bm{E}}}^g_{\bm{\sT} 3}\right)\right.\nonumber \\
 & \qquad \left. -\,\frac{1}{2}\,\left(\widetilde{{\bm{H}}}_4^g\,-\,\frac{\xi^2}{1-\xi^2}\,\widetilde{{\bm{E}}}_4^g\right)\,\left(-\xi\,{\bm{E}}_{\bm{\sT} 3}^g\,+\,\widetilde{{\bm{E}}}_{\bm{\sT} 3}^g\right) \right.\nonumber \\
 & \qquad \left. -\,\frac{1}{2}\,{\bm{E}}^g_1\,\left(-\,{\bm H}_{\bm{\sT} 3}^g\,-\,\frac{\xi}{1-\xi^2}\,\widetilde{{\bm{E}}}_{\bm{\sT} 3}^g\,+\,\frac{\xi^2}{1-\xi^2}\,{\bm{E}}_{\bm{\sT} 3}^g\,-\,\frac{\bm \Delta_\sT^2}{2M_N^2}\,\frac{\widetilde{{\bm{H}}}_{\bm{\sT} 3}^g}{1-\xi^2}\right)\,\right. \nonumber \\
& \qquad \left.-\,\frac{\xi}{2}\,\widetilde{{\bm{E}}}_4^g\,\left(-\,{\bm H}_{\bm{\sT} 3}^g\,-\,\frac{\xi}{1-\xi^2}\,\widetilde{{\bm{E}}}_{\bm{\sT} 3}^g\,+\,\frac{\xi^2}{1-\xi^2}\,{\bm{E}}_{\bm{\sT} 3}^g\right)\right\} \,, \nonumber \\
F_{U,TT}^{\,\cos 2 (2\phi_\perp - \phi_\Delta - \phi_\ell)} & =- 2 \,(1-\beta)\,\frac{\bm \Delta_\sT^2}{M_N^2}\, \left \{\left({\bm H}_1^g\,-\,\frac{\xi^2}{1-\xi^2}\,{\bm{E}}_1^g\right)\,\left(-\,\widetilde{{\bm{H}}}^g_{\bm{\sT} 3}\,-\frac{1}{2}\,{\bm{E}}_{\bm{\sT} 3}^g\,+\frac{\xi}{2}\,\widetilde{{\bm{E}}}^g_{\bm{\sT} 3}\right)\right.\nonumber\\
& \qquad \left.+\,\frac{1}{2}\,\left(\widetilde{{\bm{H}}}_4^g\,-\,\frac{\xi^2}{1-\xi^2}\,\widetilde{{\bm{E}}}_4^g\right)\,\left(-\xi\,{\bm{E}}_{\bm{\sT} 3}^g\,+\,\widetilde{{\bm{E}}}_{\bm{\sT} 3}^g\right) \right.\nonumber \\
 & \qquad \left. -\,\frac{1}{2}\,{\bm{E}}^g_1\,\left(-\,{\bm H}_{\bm{\sT} 3}^g\,-\,\frac{\xi}{1-\xi^2}\,\widetilde{{\bm{E}}}_{\bm{\sT} 3}^g\,+\,\frac{\xi^2}{1-\xi^2}\,{\bm{E}}_{\bm{\sT} 3}^g\,-\,\frac{\bm \Delta_\sT^2}{2M_N^2}\,\frac{\widetilde{{\bm{H}}}_{\bm{\sT} 3}^g}{1-\xi^2}\right)\,\right. \nonumber \\
& \qquad \left.+\,\frac{\xi}{2}\,\widetilde{{\bm{E}}}_4^g\,\left(-\,{\bm H}_{\bm{\sT} 3}^g\,-\,\frac{\xi}{1-\xi^2}\,\widetilde{{\bm{E}}}_{\bm{\sT} 3}^g\,+\,\frac{\xi^2}{1-\xi^2}\,{\bm{E}}_{\bm{\sT} 3}^g\right)\right \}\,,\nonumber \\
F_{U,TT}^{\,\cos 2 (\phi_\perp - 2\phi_\Delta + \phi_\ell)}
 & = \beta^2\,\frac{1}{1-\xi^2}\,\frac{\bm \Delta_\sT^4}{M_N^4}\, \left\{\left(-\,\widetilde{{\bm{H}}}^g_{\bm{\sT} 3}\,-\frac{1}{2}\,{\bm{E}}_{\bm{\sT} 3}^g\,+\frac{\xi}{2}\,\widetilde{{\bm{E}}}^g_{\bm{\sT} 3}\right)^2\,+\,\frac{1}{4}\,\left(-\,\xi\,{\bm{E}}_{\bm{\sT} 3}^g\,+\,\widetilde{{\bm{E}}}_{\bm{\sT} 3}^g\right)^2\right. \nonumber \\
 & \qquad \left. +\,(1-\xi^2)\,\left(-\,{\bm H}_{\bm{\sT} 3}^g\,-\,\frac{\xi}{1-\xi^2}\,\widetilde{{\bm{E}}}_{\bm{\sT} 3}^g\,+\,\frac{\xi^2}{1-\xi^2}\,{\bm{E}}_{\bm{\sT} 3}^g\,-\,\frac{\bm \Delta_\sT^2}{4M_N^2}\,\frac{\widetilde{{\bm{H}}}_{\bm{\sT} 3}^g}{1-\xi^2}\right)\,\widetilde{{\bm{H}}}^g_{\bm{\sT} 3}\right\}\,,\nonumber \\
 F_{U,TT}^{\,\cos 2 (3\phi_\perp - 2\phi_\Delta -\phi_\ell)} & = 
 (1-\beta)^2\,\frac{1}{1-\xi^2}\,\frac{\bm \Delta_\sT^4}{M_N^4}\,\left\{\left(-\,\widetilde{{\bm{H}}}^g_{\bm{\sT} 3}\,-\frac{1}{2}\,{\bm{E}}_{\bm{\sT} 3}^g\,+\frac{\xi}{2}\,\widetilde{{\bm{E}}}^g_{\bm{\sT} 3}\right)^2\,+\,\frac{1}{4}\,\left(-\,\xi\,{\bm{E}}_{\bm{\sT} 3}^g\,+\,\widetilde{{\bm{E}}}_{\bm{\sT} 3}^g\right)^2\right. \nonumber \\
 &\qquad \left. +\,(1-\xi^2)\,\left(-\,{\bm H}_{\bm{\sT} 3}^g\,-\,\frac{\xi}{1-\xi^2}\,\widetilde{{\bm{E}}}_{\bm{\sT} 3}^g\,+\,\frac{\xi^2}{1-\xi^2}\,{\bm{E}}_{\bm{\sT} 3}^g\,-\,\frac{\bm \Delta_\sT^2}{4M_N^2}\,\frac{\widetilde{{\bm{H}}}_{\bm{\sT} 3}^g}{1-\xi^2}\right)\,\widetilde{{\bm{H}}}^g_{\bm{\sT} 3}\right\}\,,
\end{align}
\begin{align}
F_{U,LT}^{\,\cos (\phi_\perp - \phi_\ell)} & = 
 (1-\xi^2)\,\left ({\bm H}_1^g\,-\,\frac{\xi^2}{1-\xi^2}\,{\bm{E}}_1^g \right )\,\left({\bm H}_2^g\,-\,\frac{\xi^2}{1-\xi^2}\,{\bm{E}}_2^g \right ) \,+\,\frac{\bm \Delta_\sT^2}{M_N^2}\,\left\{\frac{1}{4(1-\xi^2)}\, {\bm{E}}_1^g\, {\bm{E}}_2^g \right. \nonumber \\
 & \qquad \left.-\,(1-\beta)\,(1-2\beta)\,(1-\xi^2)\,\left(-\,{\bm H}_{\bm{\sT} 3}^g\,-\,\frac{\xi}{1-\xi^2}\,\widetilde{{\bm{E}}}_{\bm{\sT} 3}^g \,+\,\frac{\xi^2}{1-\xi^2}\,{\bm{E}}_{\bm{\sT} 3}^g\,-\,\frac{\bm \Delta_\sT^2}{4M_N^2}\,\frac{\widetilde{{\bm{H}}}_{\bm{\sT} 3}^g}{1-\xi^2}\right)^2\right\} \nonumber \\
 & \qquad -\,\frac{1}{2}\,(1-\beta)(1-2\beta)\,\frac{1}{1-\xi^2}\,\frac{\bm \Delta_\sT^4}{M_N^4}\,\left\{\left(-\,\widetilde{{\bm{H}}}^g_{\bm{\sT} 3}\,-\frac{1}{2}\,{\bm{E}}_{\bm{\sT} 3}^g\,+\frac{\xi}{2}\,\widetilde{{\bm{E}}}^g_{\bm{\sT} 3}\right)^2\,+\,\frac{1}{4}\,\left(-\,\xi\,{\bm{E}}_{\bm{\sT} 3}^g\,+\,\widetilde{{\bm{E}}}_{\bm{\sT} 3}^g\right)^2\right\} \nonumber\\
 &\qquad -\,(1-\beta)\,(1-2\beta)\,\frac{\bm \Delta_\sT^6}{16M_N^6}\,\frac{\widetilde{{\bm{H}}}^{g2}_{\bm{\sT} 3}}{(1-\xi^2)} \,, \nonumber \\
F_{U,LT}^{\,\cos (3\phi_\perp - 2\phi_{\Delta}-\phi_\ell)} & =  -\,\frac{1}{2}\,(1-\beta)\,\frac{\bm \Delta_\sT^2}{M_N^2}\, \left \{\left({\bm H}_1^g\,-\,\frac{\xi^2}{1-\xi^2}\,{\bm{E}}_1^g\right)\,\left(-\,\widetilde{{\bm{H}}}^g_{\bm{\sT} 3}\,-\frac{1}{2}\,{\bm{E}}_{\bm{\sT} 3}^g\,+\frac{\xi}{2}\,\widetilde{{\bm{E}}}^g_{\bm{\sT} 3}\right) \right. \nonumber \\
& \qquad \left. -\,\frac{1}{2}\,{\bm{E}}_1^g\,\left(-\,{\bm H}_{\bm{\sT} 3}^g\,-\,\frac{\xi}{1-\xi^2}\,\widetilde{{\bm{E}}}_{\bm{\sT} 3}^g\,+\,\frac{\xi^2}{1-\xi^2}\,{\bm{E}}_{\bm{\sT} 3}^g\,-\,\frac{\bm \Delta_\sT^2}{2M_N^2}\,\frac{\widetilde{{\bm{H}}}_{\bm{\sT} 3}^g}{1-\xi^2}\right)\right \} \nonumber\\
& \qquad  -\,\frac{1}{2}\,(1-\beta)\,\,\frac{\bm \Delta_\sT^2}{M_N^2}\, \left \{\left({\bm H}_2^g\,-\,\frac{\xi^2}{1-\xi^2}\,{\bm{E}}_2^g\right)\,\left(-\,\widetilde{{\bm{H}}}^g_{\bm{\sT} 3}\,-\frac{1}{2}\,{\bm{E}}_{\bm{\sT} 3}^g\,+\frac{\xi}{2}\,\widetilde{{\bm{E}}}^g_{\bm{\sT} 3}\right) \right. \nonumber \\
& \qquad \left. -\,\frac{1}{2}\,{\bm{E}}_2^g\,\left(-\,{\bm H}_{\bm{\sT} 3}^g\,-\,\frac{\xi}{1-\xi^2}\,\widetilde{{\bm{E}}}_{\bm{\sT} 3}^g\,+\,\frac{\xi^2}{1-\xi^2}\,{\bm{E}}_{\bm{\sT} 3}^g\,-\,\frac{\bm \Delta_\sT^2}{2M_N^2}\,\frac{\widetilde{{\bm{H}}}_{\bm{\sT} 3}^g}{1-\xi^2}\right)\right \}\nonumber \\
& \qquad -\,\frac{1}{2}\,(1-\beta)\,\frac{\bm \Delta_\sT^2}{M_N^2}\left\{\frac{1}{2}\,\left(\bm{\widetilde{H}}^g\,-\,\frac{\xi^2}{1-\xi^2}\,\bm{\widetilde{E}}^g\right)\,\left(-\xi\,\bm{E}_{\bm{\sT}}^g\,+\,\bm{\widetilde{E}}_{\bm{\sT}}^g\right) \right. \nonumber \\
& \qquad \left. +\,\frac{\xi}{2}\,\bm{\widetilde{E}}^g\,\left(-\,\bm{H}_{\bm{\sT}}^g\,-\,\frac{\xi}{1-\xi^2}\,\bm{\widetilde{E}}_{\bm{\sT}}^g\,+\,\frac{\xi^2}{1-\xi^2}\,\bm{E}_{\bm{\sT}}^g\right) \right\} \,, \nonumber\\
F_{U,LT}^{\,\cos (\phi_\perp - 2\phi_{\Delta}+\phi_\ell)} & = -\,\frac{1}{2}\,(1-\beta)\,\frac{\bm \Delta_\sT^2}{M_N^2}\,\,\left \{\left({\bm H}_1^g\,-\,\frac{\xi^2}{1-\xi^2}\,{\bm{E}}_1^g\right)\,\left(-\,\widetilde{{\bm{H}}}^g_{\bm{\sT} 3}\,-\frac{1}{2}\,{\bm{E}}_{\bm{\sT} 3}^g\,+\frac{\xi}{2}\,\widetilde{{\bm{E}}}^g_{\bm{\sT} 3}\right) \right. \nonumber \\
& \qquad \left. -\,\frac{1}{2}\,{\bm{E}}_1^g\,\left(-\,{\bm H}_{\bm{\sT} 3}^g\,-\,\frac{\xi}{1-\xi^2}\,\widetilde{{\bm{E}}}_{\bm{\sT} 3}^g\,+\,\frac{\xi^2}{1-\xi^2}\,{\bm{E}}_{\bm{\sT} 3}^g\,-\,\frac{\bm \Delta_\sT^2}{2M_N^2}\,\frac{\widetilde{{\bm{H}}}_{\bm{\sT} 3}^g}{1-\xi^2}\right)\right \} \nonumber\\
&\qquad +\frac{1}{2}\,\beta\,\frac{\bm \Delta_\sT^2}{M_N^2}\,\, \left \{\left({\bm H}_2^g\,-\,\frac{\xi^2}{1-\xi^2}\,{\bm{E}}_2^g\right)\,\left(-\,\widetilde{{\bm{H}}}^g_{\bm{\sT} 3}\,-\frac{1}{2}\,{\bm{E}}_{\bm{\sT} 3}^g\,+\frac{\xi}{2}\,\widetilde{{\bm{E}}}^g_{\bm{\sT} 3}\right) \right. \nonumber \\
&\qquad  \left. -\,\frac{1}{2}\,{\bm{E}}_2^g\,\left(-\,{\bm H}_{\bm{\sT} 3}^g\,-\,\frac{\xi}{1-\xi^2}\,\widetilde{{\bm{E}}}_{\bm{\sT} 3}^g\,+\,\frac{\xi^2}{1-\xi^2}\,{\bm{E}}_{\bm{\sT} 3}^g\,-\,\frac{\bm \Delta_\sT^2}{2M_N^2}\,\frac{\widetilde{{\bm{H}}}_{\bm{\sT} 3}^g}{1-\xi^2}\right)\right \} \nonumber\\
& \qquad +\,\frac{1}{2}\,(1-\beta)\,\frac{\bm \Delta_\sT^2}{M_N^2}\left\{\frac{1}{2}\,\left(\widetilde{\bm{H}}^g\,-\,\frac{\xi^2}{1-\xi^2}\,\bm{\widetilde E}^g\right)\,\left(-\xi\,\bm{E}_{\bm{\sT}}^g\,+\,\bm{\widetilde E}_{\bm{\sT}}^g\right) \right. \nonumber \\
& \qquad \left.+\,\frac{\xi}{2}\,\bm{\widetilde E}^g\,\left(-\,\bm{H}_{\bm{\sT}}^g\,-\,\frac{\xi}{1-\xi^2}\,\bm{\widetilde E}_{\bm{\sT}}^g\,+\,\frac{\xi^2}{1-\xi^2}\,\bm{E}_{\bm{\sT}}^g\right) \right\} \,, \nonumber\\
F_{U,LT}^{\,\cos (5\phi_\perp - 4\phi_{\Delta}-\phi_\ell)} & =-\,\frac{1}{2}\,(1-\beta)^2\,\frac{1}{1-\xi^2}\,\frac{\bm \Delta_\sT^4}{M_N^4}\,\left\{\left(-\,\widetilde{{\bm{H}}}^g_{\bm{\sT} 3}\,-\frac{1}{2}\,{\bm{E}}_{\bm{\sT} 3}^g\,+\frac{\xi}{2}\,\widetilde{{\bm{E}}}^g_{\bm{\sT}}\right)^2\,-\,\frac{1}{4}\,\left(-\,\xi\,{\bm{E}}_{\bm{\sT} 3}^g\,+\,\widetilde{{\bm{E}}}_{\bm{\sT} 3}^g\right)^2\right. \nonumber \\
&\qquad \left. -\,(1-\xi^2)\,\left(-\,{\bm H}_{\bm{\sT} 3}^g\,-\,\frac{\xi}{1-\xi^2}\,\widetilde{{\bm{E}}}_{\bm{\sT} 3}^g\,+\,\frac{\xi^2}{1-\xi^2}\,{\bm{E}}_{\bm{\sT} 3}^g\,-\,\frac{\bm \Delta_\sT^2}{4M_N^2}\,\frac{\widetilde{{\bm{H}}}_{\bm{\sT} 3}^g}{1-\xi^2}\right)\,\widetilde{{\bm{H}}}^g_{\bm{\sT} 3}\right\} \,,\nonumber\\
F_{U,LT}^{\,\cos (3\phi_\perp - 4\phi_{\Delta} + \phi_\ell)} & = \frac{1}{2}\,\beta(1-\beta)\,\frac{1}{1-\xi^2}\,\frac{\bm \Delta_\sT^4}{M_N^4}\,\left\{\left(-\,\widetilde{{\bm{H}}}^g_{\bm{\sT} 3}\,-\frac{1}{2}\,{\bm{E}}_{\bm{\sT} 3}^g\,+\frac{\xi}{2}\,\widetilde{{\bm{E}}}^g_{\bm{\sT}}\right)^2\,-\,\frac{1}{4}\,\left(-\,\xi\,{\bm{E}}_{\bm{\sT} 3}^g\,+\,\widetilde{{\bm{E}}}_{\bm{\sT} 3}^g\right)^2\right. \nonumber \\
&\qquad \left. -\,(1-\xi^2)\,\left(-\,{\bm H}_{\bm{\sT} 3}^g\,-\,\frac{\xi}{1-\xi^2}\,\widetilde{{\bm{E}}}_{\bm{\sT} 3}^g\,+\,\frac{\xi^2}{1-\xi^2}\,{\bm{E}}_{\bm{\sT} 3}^g\,-\,\frac{\bm \Delta_\sT^2}{4M_N^2}\,\frac{\widetilde{{\bm{H}}}_{\bm{\sT} 3}^g}{1-\xi^2}\right)\,\widetilde{{\bm{H}}}^g_{\bm{\sT} 3}\right\}\,.
\end{align}

The above results can be obtained either directly by using, in the calculation of the cross section, the decomposition of the gluon-gluon correlator in terms of GPDs, see Eqs.~\eqref{eq:gamma-GPD}-\eqref{eq:gamma-GPD-2}, or by integration over $k_\sT$ of the GTMD expressions presented in the previous section, in combination with Eqs.~\eqref{eq:F1-GPDH}, \eqref{eq:F23-GPDH}, \eqref{eq:GPDH}, \eqref{eq:SiVE}.

\section{Conclusions}
\label{sec:conclusions}
In this paper we have provided expressions for exclusive electroproduction of heavy quark-antiquark pairs off unpolarized nucleons, including all possible azimuthal modulations of the differential cross section. The corresponding structure functions are expressed in terms of integrals of gluon GTMDs. As we restrict to leading order in $\alpha_s$ and in a collinear expansion, these expressions can also be given in terms of GPDs, but the GTMD expressions make it clear which terms in the cross section provide constraints on which particular GTMDs. Furthermore, the expressions extend and complete currently available GPD results \cite{Braun:2005rg,Chall:2026oes,Pang:2026lsr} and are relevant for experimental studies of this exclusive process at the future Electron Ion Collider\footnote{In particular, while finishing this paper Ref.~\cite{Pang:2026lsr}
appeared which partly overlaps with our GPD results.}. 

We have also introduced a decomposition of the gluon-gluon correlation matrix in terms of GTMDs, where the functions are directly related to the ones of Ref.~\cite{Lorce:2013pza}, but now they appear in a convenient expansion of the correlator in a Lorentz basis of symmetric traceless tensors obtained from the partonic momentum $k_\sT$ and the momentum transfer $\Delta_\sT$. For the GTMDs we adopt a convenient and compact notation that relates to the helicity states at the amplitude level, rather than to polarization states of the incoming nucleon or of the gluons. This makes it directly clear which contributions from helicity difference and helicity flip matrix elements can be accessed with unpolarized nucleon beams. This extends the observation of Ref.~\cite{Boussarie:2019vmk} that one can become sensitive to the Sivers function in exclusive scattering with an unpolarized nucleon, or rather to the GTMD that becomes the Sivers TMD in the forward limit. 

\section{Acknowledgments}
The work of C.P.\ is supported by Fondazione
di Sardegna through the project {\it Journey to the center of
the proton}, No.\ F23C25000150007.

\appendix

\section{Other GTMD parameterizations\label{GTMDrelations}} 
\label{app-1}

In this appendix we provide the correspondences between our GTMDs and those of Refs.~\cite{Lorce:2013pza} and \cite{Meissner:2009ww}, where the latter are also employed in Ref.~\cite{Bhattacharya:2018lgm}. We find for the GTMDs for unpolarized gluons:
\begin{align}
{\cal F}_1 & = \frac{1}{\sqrt{1-\xi^2}}F_{1,1} = S_a^+,\nonumber\\
{\cal G}_4 & = \frac{1}{\sqrt{1-\xi^2}} F_{1,4} = - S_b^+,\nonumber\\
{\cal H}_1 & = i \sqrt{1-\xi^2} F_{1,2} -\frac{i\xi}{\sqrt{1-\xi^2}} \frac{\vec{\Delta}_\sT^2}{2M_N^2} F_{1,4} = i P_a^+,\nonumber\\
{\cal H}_5 & = -\frac{1}{2\sqrt{1-\xi^2}}F_{1,1} + \sqrt{1-\xi^2} F_{1,3} + \frac{\xi}{\sqrt{1-\xi^2}} \frac{\bm {k}_\sT \cdot \bm{\Delta}_\sT}{2M_N^2} F_{1,4} = - P_b^+,
\end{align}
with for the leading twist GTMDs of Ref.~\cite{Lorce:2013pza} we use a shorthand notation, where $S_{a/b}^\pm \equiv S_{1,1a/b}^{0,\pm;g}$ and similarly for $P$. Apart from some factors of $i$ and minus signs the correspondence between our GTMDs and those of Ref.~\cite{Lorce:2013pza} is one-to-one, whereas the relations to the functions of Ref.~\cite{Meissner:2009ww,Bhattacharya:2018lgm} is more involved.

The result for ${\cal H}_1$ shows that in the forward limit it corresponds to the real part of $i F_{1,2}$, which in Ref.~\cite{Boussarie:2019vmk} is identified with $- {\rm Im} F_{1,2} =-g_{1,2}= x f_{1\sT}^{\perp\, g}/2$, which differs by a factor $x/2$ from our limit of ${\cal H}_1$. Note that our notation for the GTMDs follows the helicity states of the nucleons, rather than of the gluons, which is just a matter of preference. Our choice corresponds to the parameterizations used in Ref.~\cite{Boer:2016xqr,Boer:2018vdi} for $\Gamma_U$ and emphasizes that contributions associated with helicity flip of the nucleon states on the amplitude level enter in the cross section for unpolarized scattering, which agrees with the main observation of Ref.\ \cite{Boussarie:2019vmk}. Our results show that besides the GTMD that becomes the Sivers TMD in the forward limit, other helicity difference and helicity flip GTMDs contribute as well in the off-forward case.

Similarly, for circularly polarized gluons we obtain for the GTMDs correspondences:
\begin{align}
{\cal F}_4 & = \frac{1}{\sqrt{1-\xi^2}}\,G_{1,1} = S_b^-,\nonumber\\
{\cal G}_1 & = \frac{1}{\sqrt{1-\xi^2}}\,G_{1,4} = S_a^-,\nonumber\\
{\cal H}_2 & = \sqrt{1-\xi^2}\, G_{1,2} -\frac{1}{\sqrt{1-\xi^2}} \frac{\bm{\Delta}_\sT^2}{2M_N^2}\, G_{1,1} = P_a^-,\nonumber\\
{\cal H}_6 & = -i\,\frac{\xi}{2\sqrt{1-\xi^2}}\, G_{1,4} +i\, \sqrt{1-\xi^2}\, G_{1,3} + i\,\frac{1}{\sqrt{1-\xi^2}} \frac{\bm{k}_\sT \cdot \bm{\Delta}_\sT}{2M_N^2} G_{1,1} = i\, P_b^-.
\end{align}

Finally, the identification with the GTMDs of Ref.~\cite{Lorce:2013pza} for linearly polarized gluons: 
\begin{align}
{\cal F}_2 & = -\,2\,D_a^{2+}\,, & \quad {\cal F}_3 & = -\,2\,D_b^{2+}\,, \nonumber\\
{\cal G}_2 & = 2\,i\,D_a^{'2+}\,, & \quad {\cal G}_3 & = 2\,i\,D_b^{'2+}\,, \nonumber\\
{\cal H}_3 & = -\,i\,P_a^{2+}\,, & \quad {\cal H}_7 & = -\,P_b^{2+}\,, \nonumber\\
{\cal H}_4 & = -\,2\,i\,F_a^{2+}\,, & \quad {\cal H}_8 & = -\,2\,F_b^{2+}\,,
\end{align}
where in this case $D_{a/b}^{2+}$ stands for $D_{1,1a/b}^{2,+;g}$, and similarly for $D^{'}$, $P$ and $F$ (where the superscript $2$ is added to differentiate $P_{a/b}^{2+}$ from $P_{a/b}^\pm$ used before). For the identification with the $H_{1,i}$ functions of Ref.~\cite{Meissner:2009ww}, we refer to the appendix of Ref.~\cite{Lorce:2013pza}.

\section{Amplitude squared expressions in terms of GTMDs\label{amplitudesquared}} 
\label{app-2}

With the amplitudes $\Gamma_U$, $\Gamma_L$, and $\Gamma_T^i$ one can obtain the amplitude squared and subsequently the cross section expressions. In this appendix we provide the amplitude squared expressions before transverse momentum integration. This provides an intermediate step that can be helpful for reproduction of the final results. The amplitude squared for unpolarized incoming nucleons is given by:
\begin{align}
\left. \Gamma^{\mu\nu} \,\Gamma^{*\,\rho\sigma} \right|_{S=0} = \Gamma_U^{\mu\nu}\, \Gamma_U^{*\, \rho\sigma}\,+\, \Gamma_L^{\mu\nu}\, \Gamma_L^{*\, \rho\sigma}\,+\, \vec{\Gamma}_T^{\mu\nu} \cdot \vec{\Gamma}_T^{*\, \rho\sigma}\,, 
\label{eq:amp-sq-S0}
\end{align}
with
\begin{align}
\Gamma_U^{\mu\nu}\, \Gamma_U^{*\, \rho\sigma}
 &  = \frac{1}{4}\, \Biggl \{ g_\sT^{\mu\nu}\, g_\sT^{\rho\sigma} \, {\cal F}_1^g \,{\cal F}_1^{g *} \,+\,\frac{k_{1\sT}^{\mu\nu}\,k_{2\sT}^{\rho\sigma}}{M_N^4}\,{\cal F}_2^g \,{\cal F}_2^{g *}\,+\,\frac{\Delta_{\sT}^{\mu\nu}\,\Delta_{\sT}^{\rho\sigma}}{M_N^4}\,{\cal F}_3^g \,{\cal F}_3^{g *}\,+\,\frac{k_{1\sT}^{[\mu}\Delta_{\sT}^{\nu]}\,k_{2\sT}^{[\rho}\Delta_{\sT}^{\sigma]}}{M_N^4}\,{\cal F}_4^g \,{\cal F}_4^{g *} \nonumber \\
& \qquad - \, \frac{g_{\sT}^{\mu\nu}\,k_{1\sT}^{\rho\sigma}}{M_N^2}\,{\cal F}_1^g{\cal F}_2^{g*}\,-\,\frac{k_{\sT}^{\mu\nu}\,g_\sT^{\rho\sigma}}{M_N^2}\,{\cal F}_2^g{\cal F}_1^{g*}\,-\,\frac{g_\sT^{\mu\nu}\,\Delta_\sT^{\rho\sigma}}{M_N^2}\,{\cal F}_1^g{\cal F}_3^{g*}\,+\,\frac{\Delta_\sT^{\mu\nu}\,g_\sT^{\rho\sigma}}{M_N^2}{\cal F}_3^g{\cal F}_1^{g*} \nonumber \\
& \qquad - \, \frac{g_\sT^{\mu\nu}\,k_{2\sT}^{[\rho}\,\Delta_\sT^{\sigma]}}{M_N^2}{\cal F}_1^g{\cal F}_4^{g*}\,-\,\frac{k_{1\sT}^{[\mu}\,\Delta_\sT^{\nu]}\,g_\sT^{\rho\sigma}}{M_N^2}{\cal F}_4^g{\cal F}_1^{g*}\,+\,\frac{k_{1\sT}^{\mu\nu}\,\Delta_\sT^{\rho\sigma}}{M_N^4}\,{\cal F}_2^g{\cal F}_3^{g*}\,+\,\frac{\Delta_\sT^{\mu\nu}\,k_{2\sT}^{\rho\sigma}}{M_N^4}\,{\cal F}_3^g{\cal F}_2^{g*}\nonumber \\
& \qquad +\,\left . \frac{k_{1\sT}^{\mu\nu}\,k_{2\sT}^{[\rho}\,\Delta_{\sT}^{\sigma]}}{M_N^4}\,{\cal F}_2^g{\cal F}_4^{g*}\,+\,\frac{k_{1\sT}^{[\mu}\,\Delta_{\sT}^{\nu]}\,k_{2\sT}^{\rho\sigma}}{M_N^4}\,{\cal F}_4^g{\cal F}_2^{g*}\,+\,\frac{\Delta_\sT^{\mu\nu}\,k_{2\sT}^{[\rho}\,\Delta_{\sT}^{\sigma]}}{M_N^4}\,{\cal F}_3^g{\cal F}_4^{g*}\,+\,\frac{k_{1\sT}^{[\mu}\,\Delta_{\sT}^{\nu]}\,\Delta_{\sT}^{\rho\sigma}}{M_N^4}\,{\cal F}_4^g{\cal F}_3^{g*} \right \}\,,
\end{align}
\begin{align}
\Gamma_L^{\mu\nu}\, \Gamma_L^{*\, \rho\sigma}
&  = \frac{1}{4}\, \left \{ g_\sT^{\mu\nu} g_\sT^{\rho\sigma} \, \frac{1}{M_N^4}\,\epsilon_\sT^{k_{1\sT} \Delta_\sT}\epsilon_\sT^{k_{2\sT} \Delta_\sT}\, {\cal G}_4^g \, {\cal G}_4^{g*} \, +\, \epsilon_{\sT}^{\mu\nu}\, \epsilon_{\sT}^{\rho\sigma} \,  {\cal G}_1^g \, {\cal G}_1^{g*}\, -\,g_\sT^{\mu\nu}\,\epsilon_{\sT}^{\rho\sigma}\,\frac{\epsilon_\sT^{k_{1\sT}\, \Delta_\sT}}{M_N^2}\,{\cal G}_4^g\,{\cal G}_1^{g*} \right . \nonumber \\
& \qquad \,- \,\epsilon_{\sT}^{\mu\nu}\,g_{\sT}^{\rho\sigma}\,\frac{\epsilon_\sT^{k_{2\sT}\, \Delta_\sT}}{M_N^2}\,{\cal G}_1^g\,{\cal G}_4^{g*} \,
+ \, \frac{\epsilon_{\sT\,\alpha}^{\{\mu}\, g_\sT^{\nu \} \beta}\epsilon_{\sT \gamma}^{\{ \rho} \, g_\sT^{\sigma \} \delta}}{4 M_N^4} \left [ k_{1\sT \beta}^\alpha \, {\cal G}_2^g \, + \, \Delta_{\sT \beta}^\alpha \,{\cal G}_3^g \right ] \left [ k_{2\sT \delta}^\gamma \, {\cal G}_2^{g*} \, + \, \Delta_{\sT \delta}^\gamma \, {\cal G}_3^{g*}\right ] \nonumber \\
& \qquad -\,i \,  \frac{\epsilon_{\sT\,\alpha}^{\{\mu}\, g_\sT^{\nu \} \gamma}\, \epsilon_\sT^{\rho\sigma}}{2 M_N^2} \left [ k_{1\sT \gamma}^\alpha \, {\cal G}_2^g \, + \, \Delta_{\sT \gamma}^\alpha \,{\cal G}_3^g \right ]\,{\cal G}_1^{g*}\,+\,i\,\frac{\epsilon_\sT^{\mu\nu} \, \epsilon_{\sT \alpha}^{\{ \rho} \, g_\sT^{\sigma \} \gamma}}{2 M_N^2}\,{\cal G}_1^g\,\left [ k_{2\sT \gamma}^\alpha \, {\cal G}_2^{g*} \, + \, \Delta_{\sT \gamma}^\alpha  \, {\cal G}_3^{g*}\right ] \nonumber \\
& \qquad +\,i \,\left . \frac{\epsilon_{\sT\,\alpha}^{\{\mu}\, g_\sT^{\nu \} \beta}\, \epsilon_\sT^{\rho\sigma}}{2 M_N^2} \left [ k_{1\sT \beta}^\alpha \, {\cal G}_2^g \, + \, \Delta_{\sT \beta}^\alpha \,{\cal G}_3^g \right ]\,\frac{\epsilon_\sT^{k_{2\sT}\, \Delta_\sT}}{M_N^2} \, {\cal G}_4^{g*}\,-\,i\,\frac{\epsilon_\sT^{\mu\nu} \, \epsilon_{\sT \gamma}^{\{ \rho} \, g_\sT^{\sigma \} \gamma}}{2 M_N^2}\,\frac{\epsilon_\sT^{k_{1\sT}\, \Delta_\sT}}{M_N^2} \, {\cal G}_4^{g}\,\left [ k_{2\sT \delta}^\gamma \, {\cal G}_2^{g*} \, + \, \Delta_{\sT \delta}^\gamma  \, {\cal G}_3^{g*}\right ] \right \}\, ,
\end{align}
and 
\begin{align}
\vec{\Gamma}_T^{\mu\nu}\cdot \vec{\Gamma}_T^{*\, \rho\sigma} 
& = \frac{1}{4}\,\left\{g_\sT^{\mu\nu}g_\sT^{\rho\sigma}\,\left(\frac{ \epsilon_\sT^{k_{1\sT} i}}{M_N}\,  {\cal H}_1^g\,+\,i\,\frac{ \epsilon_\sT^{\Delta_\sT i}}{M_N}\,  {\cal H}_5^g\right)\left(\frac{ \epsilon_\sT^{k_{2\sT} i}}{M_N}\,  {\cal H}_1^{g*}\,-\,i\,\frac{ \epsilon_\sT^{\Delta_\sT i}}{M_N}\,  {\cal H}_5^{g*}\right) \right.\nonumber\\
&\qquad+\,g_\sT^{\mu\nu}\epsilon_\sT^{\rho\sigma}\,\left(\frac{ \epsilon_\sT^{k_{1\sT} i}}{M_N}\,  {\cal H}_1^g\,+\,i\,\frac{ \epsilon_\sT^{\Delta_\sT i}}{M_N}\,  {\cal H}_5^g\right)\left(-\,i\,\frac{k_{2\sT}^i}{M_N}\,  {\cal H}_2^{g*}\,+\,\frac{\Delta_\sT^i}{M_N}\,{\cal H}_6^{g*}\right)\nonumber\\
&\qquad+\,\epsilon_\sT^{\mu\nu}g_\sT^{\rho\sigma}\,\left(i\,\frac{k_{1\sT}^i}{M_N}\,  {\cal H}_2^{g}\,+\,\frac{\Delta_\sT^i}{M_N}\,{\cal H}_6^{g}\right)\left(\frac{ \epsilon_\sT^{k_{2\sT} i}}{M_N}\,  {\cal H}_1^{g*}\,-\,i\,\frac{ \epsilon_\sT^{\Delta_\sT i}}{M_N}\,  {\cal H}_5^{g*}\right)\nonumber\\
&\qquad-\,g_\sT^{\mu\nu}\left(\epsilon_{\sT}^{\alpha\{\rho}g_\sT^{\sigma\}i}+\epsilon_{\sT}^{i\{\rho}g_\sT^{\sigma\}\alpha}\right)\,\left(\frac{ \epsilon_\sT^{k_{1\sT} i}}{M_N}\,  {\cal H}_1^g\,+\,i\,\frac{ \epsilon_\sT^{\Delta_\sT i}}{M_N}\,  {\cal H}_5^g\right)\left(\frac{k_{2\sT\alpha}}{M_N}\,  {\cal H}_3^{g*}\,-\,i\,\frac{\Delta_{\sT \alpha}}{M_N}\,  {\cal H}_7^{g*}\right)\nonumber\\
&\qquad-\,\left(\epsilon_{\sT}^{\alpha\{\mu}g_\sT^{\nu\}i}+\epsilon_{\sT}^{i\{\mu}g_\sT^{\nu\}\alpha}\right)g_\sT^{\rho\sigma}\,\left(\frac{k_{1\sT\alpha}}{M_N}\,  {\cal H}_3^{g}\,+\,i\,\frac{\Delta_{\sT \alpha}}{M_N}\,  {\cal H}_7^{g}\right)\left(\frac{ \epsilon_\sT^{k_{2\sT} i}}{M_N}\,  {\cal H}_1^{g*}\,-\,i\,\frac{ \epsilon_\sT^{\Delta_\sT i}}{M_N}\,  {\cal H}_5^{g*}\right)\nonumber\\
&\qquad-\,g_\sT^{\mu\nu}\,\left(\frac{ \epsilon_\sT^{k_{1\sT} i}}{M_N}\,  {\cal H}_1^g\,+\,i\,\frac{ \epsilon_\sT^{\Delta_\sT i}}{M_N}\,  {\cal H}_5^g\right)\left(\frac{\epsilon_{\sT\,\alpha}^{\{ \rho} k_{2\sT}^{\sigma \} \alpha i}}{2M_N^3}\, {\cal H}_4^{g*}\,-\,i\,\frac{\epsilon_{\sT\,\alpha}^{\{ \rho} \Delta_\sT^{\sigma \} \alpha i}}{2M_N^3}\, {\cal H}_8^{g*}\right)\nonumber\\
&\qquad-\,\left(\frac{\epsilon_{\sT\,\alpha}^{\{ \mu} k_{1\sT}^{\nu \} \alpha i}}{2M_N^3}\, {\cal H}_4^{g}\,+\,i\,\frac{\epsilon_{\sT\,\alpha}^{\{ \mu} \Delta_\sT^{\nu \} \alpha i}}{2M_N^3}\, {\cal H}_8^{g}\right)\,g_\sT^{\rho\sigma}\,\left(\frac{ \epsilon_\sT^{k_{2\sT} i}}{M_N}\,  {\cal H}_1^{g*}\,-\,i\,\frac{ \epsilon_\sT^{\Delta_\sT i}}{M_N}\,  {\cal H}_5^{g*}\right)\nonumber\\
&\qquad+\,\epsilon_\sT^{\mu\nu}\epsilon_\sT^{\rho\sigma}\,\left(i\,\frac{k_{1\sT}^i}{M_N}\,  {\cal H}_2^{g}\,+\,\frac{\Delta_\sT^i}{M_N}\,{\cal H}_6^{g}\right)\left(-\,i\,\frac{k_{2\sT}^i}{M_N}\,  {\cal H}_2^{g*}\,+\,\frac{\Delta_\sT^i}{M_N}\,{\cal H}_6^{g*}\right)\nonumber\\
&\qquad-\,\epsilon_\sT^{\mu\nu}\left(\epsilon_{\sT}^{\alpha\{\rho}g_\sT^{\sigma\}i}+\epsilon_{\sT}^{i\{\rho}g_\sT^{\sigma\}\alpha}\right)\,\left(i\,\frac{k_{1\sT}^i}{M_N}\,  {\cal H}_2^{g}\,+\,\frac{\Delta_\sT^i}{M_N}\,{\cal H}_6^{g}\right)\left(\frac{k_{2\sT\alpha}}{M_N}\,  {\cal H}_3^{g*}\,-\,i\,\frac{\Delta_{\sT \alpha}}{M_N}\,  {\cal H}_7^{g*}\right)\nonumber\\
&\qquad-\,\left(\epsilon_{\sT}^{\alpha\{\mu}g_\sT^{\nu\}i}+\epsilon_{\sT}^{i\{\mu}g_\sT^{\nu\}\alpha}\right)\epsilon_\sT^{\rho\sigma}\,\left(\frac{k_{1\sT\alpha}}{M_N}\,  {\cal H}_3^{g}\,+\,i\,\frac{\Delta_{\sT \alpha}}{M_N}\,  {\cal H}_7^{g}\right)\left(-\,i\,\frac{k_{2\sT}^i}{M_N}\,  {\cal H}_2^{g*}\,+\,\frac{\Delta_\sT^i}{M_N}\,{\cal H}_6^{g*}\right)\nonumber\\
&\qquad-\,\epsilon_\sT^{\mu\nu}\,\left(i\,\frac{k_{1\sT}^i}{M_N}\,  {\cal H}_2^{g}\,+\,\frac{\Delta_\sT^i}{M_N}\,{\cal H}_6^{g}\right)\left(\frac{\epsilon_{\sT\,\alpha}^{\{ \rho} k_{2\sT}^{\sigma \} \alpha i}}{2M_N^3}\, {\cal H}_4^{g*}\,-\,i\,\frac{\epsilon_{\sT\,\alpha}^{\{ \rho} \Delta_\sT^{\sigma \} \alpha i}}{2M_N^3}\, {\cal H}_8^{g*}\right)\nonumber\\
&\qquad-\,\left(\frac{\epsilon_{\sT\,\alpha}^{\{ \mu} k_{1\sT}^{\nu \} \alpha i}}{2M_N^3}\, {\cal H}_4^{g}\,+\,i\,\frac{\epsilon_{\sT\,\alpha}^{\{ \mu} \Delta_\sT^{\nu \} \alpha i}}{2M_N^3}\, {\cal H}_8^{g}\right)\epsilon_\sT^{\rho\sigma}\,\left(-\,i\,\frac{k_{2\sT}^i}{M_N}\,  {\cal H}_2^{g*}\,+\,\frac{\Delta_\sT^i}{M_N}\,{\cal H}_6^{g*}\right)\nonumber\\
&\qquad+\,\left(\epsilon_{\sT}^{\alpha\{\mu}g_\sT^{\nu\}i}+\epsilon_{\sT}^{i\{\mu}g_\sT^{\nu\}\alpha}\right)\left(\epsilon_{\sT}^{\beta\{\rho}g_\sT^{\sigma\}i}+\epsilon_{\sT}^{i\{\rho}g_\sT^{\sigma\}\beta}\right)\,\left(\frac{k_{1\sT\alpha}}{M_N}\,  {\cal H}_3^{g}\,+\,i\,\frac{\Delta_{\sT \alpha}}{M_N}\,  {\cal H}_7^{g}\right)\left(\frac{k_{2\sT\beta}}{M_N}\,  {\cal H}_3^{g*}\,-\,i\,\frac{\Delta_{\sT \beta}}{M_N}\,  {\cal H}_7^{g*}\right)\nonumber\\
&\qquad+\,\left(\epsilon_{\sT}^{\alpha\{\mu}g_\sT^{\nu\}i}+\epsilon_{\sT}^{i\{\mu}g_\sT^{\nu\}\alpha}\right)\,\left(\frac{k_{1\sT\alpha}}{M_N}\,  {\cal H}_3^{g}\,+\,i\,\frac{\Delta_{\sT \alpha}}{M_N}\,  {\cal H}_7^{g}\right)\left(\frac{\epsilon_{\sT\,\beta}^{\{ \rho} k_{2\sT}^{\sigma \} \beta i}}{2M_N^3}\, {\cal H}_4^{g*}\,-\,i\,\frac{\epsilon_{\sT\,\beta}^{\{ \rho} \Delta_\sT^{\sigma \} \beta i}}{2M_N^3}\, {\cal H}_8^{g*}\right)\nonumber \\
&\qquad+\,\left(\frac{\epsilon_{\sT\,\alpha}^{\{ \mu} k_{1\sT}^{\nu \} \alpha i}}{2M_N^3}\, {\cal H}_4^{g}\,+\,i\,\frac{\epsilon_{\sT\,\alpha}^{\{ \mu} \Delta_\sT^{\nu \} \alpha i}}{2M_N^3}\, {\cal H}_8^{g}\right)\,\left(\epsilon_{\sT}^{\beta\{\rho}g_\sT^{\sigma\}i}+\epsilon_{\sT}^{i\{\rho}g_\sT^{\sigma\}\beta}\right)\,\left(\frac{k_{2\sT\beta}}{M_N}\,  {\cal H}_3^{g*}\,-\,i\,\frac{\Delta_{\sT \beta}}{M_N}\,  {\cal H}_7^{g*}\right)\nonumber\\
&\qquad\left.+\,\left(\frac{\epsilon_{\sT\,\alpha}^{\{ \mu} k_{1\sT}^{\nu \} \alpha i}}{2M_N^3}\, {\cal H}_4^{g}\,+\,i\,\frac{\epsilon_{\sT\,\alpha}^{\{ \mu} \Delta_\sT^{\nu \} \alpha i}}{2M_N^3}\, {\cal H}_8^{g}\right)\left(\frac{\epsilon_{\sT\,\beta}^{\{ \rho} k_{2\sT}^{\sigma \} \beta i}}{2M_N^3}\, {\cal H}_4^{g*}\,-\,i\,\frac{\epsilon_{\sT\,\beta}^{\{ \rho} \Delta_\sT^{\sigma \} \beta i}}{2M_N^3}\, {\cal H}_8^{g*}\right)\right\}\,.
\end{align}

\section{GPDs in terms of integrals of GTMDs}
\label{app-3}
It is well-known that the collinear PDFs expressed as integrals of TMDs require careful consideration of the ultraviolet divergences, the renormalization scheme and scale dependence, as well as of gauge invariance and the way to deal with rapidity divergences, cf.~Refs.~\cite{Collins:2003fm,Collins:2011zzd}. Similarly, the GPDs can be expressed as integrals of GTMDs, but that comes with the same issues. In this appendix we give the expressions of the GPDs in terms of integrals of GTMDs as one can obtain them simply from their operator definitions, without considering any of these issues. In that sense they should be considered as tree level expressions.

\begin{align}
    H^g=& \int d^2k_\sT\,\left[\frac{{\cal{F}}_1^g}{\sqrt{1-\xi^2}}\,+\,2i\xi^2\left(\frac{k_\sT^\mu \Delta_{\sT\,\mu}}{\bm \Delta_\sT^2}\,{\cal H}_1^g\,-\,i\,{\cal H}_5^g\right)\right]\,, \nonumber\\
    E^g=&2i\sqrt{1-\xi^2}\int d^2k_\sT\,\left[\frac{k_\sT^\mu\Delta_{\sT\,\mu}}{\bm \Delta_\sT^2}\,{\cal H}_1^g\,-\,i\,{\cal H}_5^g\right] \,, \nonumber\\
    \tilde{H}^g=& \int d^2k_\sT\,\left[\frac{{\cal{G}}_1^g}{\sqrt{1-\xi^2}}\,-\,2\xi\left(\frac{k_\sT^\mu \Delta_{\sT\,\mu}}{\bm \Delta_\sT^2}\,{\cal H}_2^g\,+\,i\,{\cal H}_6^g\right)\right] \,, \nonumber\\
    \tilde{E}^g=&-2i\frac{\sqrt{1-\xi^2}}{\xi}\int d^2k_\sT\,\left[\frac{k_\sT^\mu\Delta_{\sT\,\mu}}{\bm \Delta_\sT^2}\,{\cal H}_2^g\,+\,i\,{\cal H}_6^g\right] \,, \nonumber\\
    H_\sT^g=&\frac{i}{\sqrt{1-\xi^2}}\int d^2 k_\sT\,\left[4\xi\,\left(\frac{k_\sT^{\mu\nu} \Delta_{\sT\,\mu\nu}}{\bm \Delta_\sT^4}\,{\cal G}_2^g\,+\,\frac{{{\cal G}_3^g}}{2}\right)\,+\,\frac{k_\sT^\mu \Delta_{\sT\,\mu}}{\bm \Delta_\sT^2}\,{\cal H}_3^g\,-\,i{\cal H}_7^g\,+\,2\frac{\bm \Delta_\sT^2}{M_N^2}\left(\frac{k_\sT^{\mu\nu\rho} \Delta_{\sT\,\mu\nu\rho}}{\bm \Delta_\sT^6}\,{\cal H}_4^g\,-\,\frac{i}{4}\,{\cal H}_8^g\right)\right] \,, \nonumber\\
    E_\sT^g=&\frac{4}{\sqrt{1-\xi^2}}\int d^2 k_\sT\,\left[-\frac{k_\sT^{\mu\nu} \Delta_{\sT\,\mu\nu}}{\bm \Delta_\sT^4}\left({\cal{F}}_2^g\,+\,i\xi {\cal{G}}_2^g\right)\,-\,\frac{1}{2}\left({\cal{F}}_3^g\,+\,i\xi{\cal{G}}_3^g\right)\,+\,4i\left(\frac{k_\sT^{\mu\nu\rho} \Delta_{\sT\,\mu\nu\rho}}{\bm \Delta_\sT^6}\,{\cal H}_4^g\,-\,\frac{i}{4}\,{\cal H}_8^g\right)\right] \,, \nonumber\\   
    \tilde{H}_\sT^g=&-8i\sqrt{1-\xi^2}\int d^2k_\sT \left[\frac{k_\sT^{\mu\nu\rho} \Delta_{\sT\,\mu\nu\rho}}{\bm \Delta_\sT^6}\,{\cal H}_4^g\,-\,\frac{i}{4}\,{\cal H}_8^g\right] \,, \nonumber\\
    \tilde{E}_\sT^g=&\frac{4}{\sqrt{1-\xi^2}}\int d^2 k_\sT\left[4i\xi\,\left(\frac{k_\sT^{\mu\nu\rho} \Delta_{\sT\,\mu\nu\rho}}{\bm \Delta_\sT^6}\,{\cal H}_4^g\,-\,\frac{i}{4}\,{\cal H}_8^g\right)\,-\,\frac{k_\sT^{\mu\nu} \Delta_{\sT\,\mu\nu}}{\bm \Delta_\sT^4}\left(\xi{\cal{F}}_2^g\,+\,i {\cal{G}}_2^g\right)\,-\,\frac{1}{2}\left(\xi{\cal{F}}^g_3\,+\,i{\cal{G}}_3^g\right)\right]\,. \label{GPDintegrals}
\end{align}

\section{GPDs and GTMDs in the small-$x$ limit}
\label{app-4}

In this appendix we provide some further results on the small-$x$ limit for the dipole gauge link correlator $\Gamma^{[+,-]\mu\nu}$. Like for TMDs, in the limit of vanishing $x$ and $\xi$ some GTMDs for the dipole gauge link configuration become related. This is due to the fact that the dipole gauge link correlator can be expressed in terms of the Wilson loop correlator $\Gamma^{[\Box ]}$ as follows: 
\begin{align} 
\Gamma^{[+,-]\mu\nu} & =\frac{2N_c}{\alpha_s} \left(k_\sT^\mu + \frac{\Delta_\sT^\mu}{2}\right)\left(k_\sT^\nu - \frac{\Delta_\sT^\nu}{2}\right) \Gamma^{[\Box ]},
\end{align}
where the parametrization of $\Gamma^{[\Box ]}$ becomes very simple (the factors of $M_N$ are chosen such that the ${\cal E}$-type functions have the same dimension ($-2$) as the other GTMDs): 
\begin{align}
\Gamma^{[\Box ]} = \frac{\alpha_s}{2N_c} \frac{1}{M_N^2} \left[ {\cal E} + \epsilon_\sT^{kS} \frac{{\cal E}_{\sT}}{M_N}  + i \epsilon_\sT^{\Delta S} \frac{{\cal E}_{\sT}^\prime}{M_N}\right].
\end{align}
This implies the relation \cite{Boer:2018vdi}:
\begin{equation}
    \lim_{x,\xi \to 0} \mathcal{F}_1 = \lim_{x,\xi \to 0} \mathcal{F}_2^{(1)} = -4 \lim_{x,\xi \to 0} \mathcal{F}_3^{(1)} = -2 \lim_{x,\xi \to 0} \mathcal{F}_4^{(1)} = {\cal E}^{(1)},
    \label{relationbetweenFs2}
\end{equation}
which employs the shorthand notation: $\mathcal{F}_i^{(n)} \equiv [(\bm{k}^2 - \bm{\Delta}^2/4)/(2M_N^2)]^n \,\mathcal{F}_i$. In the forward limit this reduces to the equality $\lim_{x\to0} f_1 = \lim_{x\to0} h_1^{\perp(1)}$ for dipole gauge link gluon TMDs \cite{Metz:2011wb}. 

A similar TMD relation was obtained for the transversely polarized case in Refs.~\cite{Boer:2015pni,Boer:2022njw}: $\lim_{x\to 0} f_{1T}^\perp = \lim_{x\to0} h_1 = -
\lim_{x\to0} h_{1T}^{\perp(1)}$, but the extension to GTMDs is not simply a proportionality of the GTMDs ${\cal H}_{1/3/4}$. All eight ${\cal H}$ functions are nonzero in the off-forward case at small $x$ and are determined by the two Wilson loop GTMDs ${\cal E}_\sT$ and ${\cal E}_\sT^\prime$. The expressions were first obtained in Ref.~\cite{Benic:2026idy}, which can be translated into our parameterization by means of the identifications in Appendix~\ref{app-1}. For completeness, we provide them here in our notation. Upon identifying $i{\cal E}_\sT/M_N^2 = 2N_c/\alpha_s\,  {\cal S}_{1T}^\perp$ and ${\cal E}_\sT^\prime/M_N^2 = - 2N_c/\alpha_s\, {\cal S}_T$ our expressions reduce to the relations in Ref.~\cite{Benic:2026idy} (their contributions to near-forward elastic proton–proton scattering were studied in Ref.~\cite{Hagiwara:2020mqb}). We find

\begin{align}
 \lim_{x,\xi \to 0} {\cal H}_1 & =- \left(\bm{k}_\sT^2 -\frac{\bm{\Delta}_\sT^2}{4}\right) \frac{{\cal E}_\sT}{M_N^2}\, ,\nonumber\\
\lim_{x,\xi \to 0} {\cal H}_2 & = \left[i \bm{k}_\sT\cdot\bm{\Delta}_\sT \frac{{\cal E}_\sT}{M_N^2} + \bm{\Delta}_\sT^2 \frac{{\cal E}_\sT^\prime}{M_N^2}\right]\, ,\nonumber\\
\lim_{x,\xi \to 0} {\cal H}_3 & = - \left[\left(\bm{k}_\sT^2 +\frac{\bm{\Delta}_\sT^2}{4}\right) \frac{{\cal E}_\sT}{M_N^2} + 2i \bm{k}_\sT\cdot\bm{\Delta}_\sT\frac{{\cal E}_\sT^\prime}{M_N^2}\right]\, ,\nonumber\\
\lim_{x,\xi \to 0} {\cal H}_4 & = -\frac{2M_N^2}{\bm{k}_\sT^2\bm{\Delta}_\sT^2 -4 (\bm{k}_\sT\cdot\bm{\Delta}_\sT)^2} \left[\left(4 (\bm{k}_\sT\cdot\bm{\Delta}_\sT)^2 - \bm{k}_\sT^2 \bm{\Delta}_\sT^2 -\frac{\bm{\Delta}_\sT^4}{4}\right) \frac{{\cal E}_\sT}{M_N^2} + 2i \bm{\Delta}_\sT^2 \bm{k}_\sT\cdot\bm{\Delta}_\sT \frac{{\cal E}_\sT^\prime}{M_N^2}\right]\, ,\nonumber\\
\lim_{x,\xi \to 0} {\cal H}_5 & =- \left(\bm{k}_\sT^2 -\frac{\bm{\Delta}_\sT^2}{4}\right) \frac{{\cal E}_\sT^\prime}{M_N^2}\, ,\nonumber\\
\lim_{x,\xi \to 0} {\cal H}_6 & = \left[\bm{k}_\sT^2  \frac{{\cal E}_\sT}{M_N^2} + i \bm{k}_\sT\cdot\bm{\Delta}_\sT\frac{{\cal E}_\sT^\prime}{M_N^2}\right] \, ,\nonumber\\
\lim_{x,\xi \to 0} {\cal H}_7 & = - \left[\frac{i}{2} \bm{k}_\sT\cdot\bm{\Delta}_\sT \frac{{\cal E}_\sT}{M_N^2} - \left(\bm{k}_\sT^2 +\frac{\bm{\Delta}_\sT^2}{4}\right) \frac{{\cal E}_\sT^\prime}{M_N^2}\right]\, ,\nonumber\\
\lim_{x,\xi \to 0} {\cal H}_8 & = -\frac{M_N^2}{\bm{k}_\sT^2\bm{\Delta}_\sT^2 -4 (\bm{k}_\sT\cdot\bm{\Delta}_\sT)^2} \left[ i\bm{k}_\sT^2 \bm{k}_\sT\cdot\bm{\Delta}_\sT \frac{{\cal E}_\sT}{M_N^2} -  \frac{1}{2} \left(4 (\bm{k}_\sT\cdot\bm{\Delta}_\sT)^2 - \bm{k}_\sT^2 \bm{\Delta}_\sT^2-4\bm{k}_\sT^4\right) \frac{{\cal E}_\sT^\prime}{M_N^2}\right]\, .   
\end{align}
All ${\cal G}$ functions are zero in the small-$x$ limit, also in the off-forward case.

In Eqs.~(11) and (12) of Ref.~\cite{Hatta:2017cte} small-$x$ expressions are given for the gluon GPDs $H^g$ and $E^g_\sT$ (or rather $E_\sT^{\prime g} \equiv E^g_\sT + 2 \tilde{H}^g_T$, see its footnote 2) in terms of $F^g_0$ and $F^g_\epsilon$ that are part of the angular expansion of the small-$x$ dipole gluon GTMD $F_x^g$:
\begin{align} 
F^g_x({\bm k}_\sT,{\bm \Delta}_\sT) & = F^g_0(|k_\sT|,|\Delta_\sT|) + 2 \cos 2(\phi_k - \phi_\Delta) F^g_\epsilon(|k_\sT|,|\Delta_\sT|) + \ldots \,.\label{ellipticexp}
\end{align}
To make the connection with the expressions for $H^g$ and $E_T^g$ given above in Eq.\ (\ref{GPDintegrals}), one needs the small-$x$ identification (through the Wilson loop correlator): ${\cal F}_1^g = {\cal F}_2^{(1)\, g}= -4 {\cal F}_3^{(1) \,g} = (4 N_c M_N^2/\alpha_s) F_x^{(1) \,g}\, $. In this way one finds for zero skewness ($\xi \to 0$):
\begin{align}
    & H^g= \int \d^2k_\sT\,{\cal F}_1^g = \frac{2N_c}{\alpha_s} \int \d^2k_\sT\,k_\sT^2\, F_0 = xH_g\ \text{of Ref.~\cite{Hatta:2017cte}},\nonumber\\
    & E_\sT^g + 2 \tilde{H}_T^g = 4 \int \d^2 k_\sT\,\left[-\frac{k_\sT^{\mu\nu} \Delta_{\sT\,\mu\nu}}{\bm \Delta_\sT^4}{\cal F}_2^g\,-\,\frac{1}{2}{\cal F}_3 \right] = - \frac{8N_c}{\alpha_s} \frac{M_N^2}{\bm \Delta_\sT^2} \int \d^2k_\sT\,k_\sT^2\, F_\epsilon = -\frac{xE_{Tg}^\prime}{2}\ \text{of Ref.~\cite{Hatta:2017cte}},\nonumber   
\end{align}
where in the latter result only ${\cal F}_2^{2\phi}$ contributes, which once again shows that what is referred to as ``the'' elliptic Wigner distribution (the Fourier transform of $F_\epsilon$) is actually associated with ${\cal F}_2$. This is as opposed to $F_0$, which {\it is} associated with ${\cal F}_1$. The elliptic part of ${\cal F}_1$ does not enter in GPDs, nor in any of the angular dependences of the cross section in leading order. So the two functions in Eq.\ (\ref{ellipticexp}) actually refer to two different GTMDs, which (only) in the small-$x$ limit can be treated as the same.

\bibliography{bibliography}

@article{Lorce:2013pza,
    author = "Lorc\'e, C. and Pasquini, B.",
    title = "{Structure analysis of the generalized correlator of quark and gluon for a spin-1/2 target}",
    eprint = "1307.4497",
    archivePrefix = "arXiv",
    primaryClass = "hep-ph",
    doi = "10.1007/JHEP09(2013)138",
    journal = "JHEP",
    volume = "09",
    pages = "138",
    year = "2013"
}

@article{Bertone:2025vgy,
    journal="",
    author = "Bertone, Valerio and Echevarria, Miguel G. and del Rio, \'Oscar and Rodini, Simone",
    title = "{One-loop matching for leading-twist generalised transverse-momentum-dependent distributions}",
    eprint = "2502.07576",
    archivePrefix = "arXiv",
    primaryClass = "hep-ph",
    reportNumber = "IPARCOS-UCM-25-004, DESY-25-024",
    month = "2",
    year = "2025"
}

@article{Bhattacharya:2022vvo,
    author = "Bhattacharya, Shohini and Boussarie, Renaud and Hatta, Yoshitaka",
    title = "{Signature of the Gluon Orbital Angular Momentum}",
    eprint = "2201.08709",
    archivePrefix = "arXiv",
    primaryClass = "hep-ph",
    doi = "10.1103/PhysRevLett.128.182002",
    journal = "Phys. Rev. Lett.",
    volume = "128",
    number = "18",
    pages = "182002",
    year = "2022"
}

@article{Bhattacharya:2024sck,
    author = "Bhattacharya, Shohini and Boussarie, Renaud and Hatta, Yoshitaka",
    title = "{Exploring orbital angular momentum and spin-orbit correlations for gluons at the Electron-Ion Collider}",
    eprint = "2404.04209",
    archivePrefix = "arXiv",
    primaryClass = "hep-ph",
    doi = "10.1103/PhysRevD.111.034019",
    journal = "Phys. Rev. D",
    volume = "111",
    number = "3",
    pages = "034019",
    year = "2025"
}

@article{Bhattacharya:2017bvs,
    author = "Bhattacharya, Shohini and Metz, Andreas and Zhou, Jian",
    title = "{Generalized TMDs and the exclusive double Drell{\textendash}Yan process}",
    eprint = "1702.04387",
    archivePrefix = "arXiv",
    primaryClass = "hep-ph",
    doi = "10.1016/j.physletb.2017.05.081",
    journal = "Phys. Lett. B",
    volume = "771",
    pages = "396--400",
    year = "2017",
    note = "[Erratum: Phys.Lett.B 810, 135866 (2020)]"
}

@article{Bhattacharya:2018lgm,
    author = "Bhattacharya, Shohini and Metz, Andreas and Ojha, Vikash Kumar and Tsai, Jeng-Yuan and Zhou, Jian",
    title = "{Exclusive double quarkonium production and generalized TMDs of gluons}",
    eprint = "1802.10550",
    archivePrefix = "arXiv",
    primaryClass = "hep-ph",
    doi = "10.1016/j.physletb.2022.137383",
    journal = "Phys. Lett. B",
    volume = "833",
    pages = "137383",
    year = "2022"
}

@article{Meissner:2007rx,
    author = "Meissner, S. and Metz, A. and Goeke, K.",
    title = "{Relations between generalized and transverse momentum dependent parton distributions}",
    eprint = "hep-ph/0703176",
    archivePrefix = "arXiv",
    doi = "10.1103/PhysRevD.76.034002",
    journal = "Phys. Rev. D",
    volume = "76",
    pages = "034002",
    year = "2007"
}

@article{Meissner:2009ww,
    author = "Meissner, Stephan and Metz, Andreas and Schlegel, Marc",
    title = "{Generalized parton correlation functions for a spin-1/2 hadron}",
    eprint = "0906.5323",
    archivePrefix = "arXiv",
    primaryClass = "hep-ph",
    reportNumber = "JLAB-THY-09-1018",
    doi = "10.1088/1126-6708/2009/08/056",
    journal = "JHEP",
    volume = "08",
    pages = "056",
    year = "2009"
}

@article{Boer:2021upt,
    author = {Boer, Dani{\"e}l and Setyadi, Chalis},
    title = "{GTMD model predictions for diffractive dijet production at EIC}",
    eprint = "2106.15148",
    archivePrefix = "arXiv",
    primaryClass = "hep-ph",
    doi = "10.1103/PhysRevD.104.074006",
    journal = "Phys. Rev. D",
    volume = "104",
    number = "7",
    pages = "074006",
    year = "2021"
}

@article{Boer:2023mip,
    author = {Boer, Dani{\"e}l and Setyadi, Chalis},
    title = "{Probing gluon GTMDs through exclusive coherent diffractive processes}",
    eprint = "2301.07980",
    archivePrefix = "arXiv",
    primaryClass = "hep-ph",
    doi = "10.1140/epjc/s10052-023-12040-6",
    journal = "Eur. Phys. J. C",
    volume = "83",
    number = "10",
    pages = "890",
    year = "2023"
}

@article{Benic:2026idy,
    author = "Beni{\'c}, Sanjin and Hagiwara, Yoshikazu and {\v{S}}ari{\'c}, Boris and Vivoda, Eric Andreas",
    title = "{Generalized transverse momentum distributions at small-x}",
    eprint = "2603.06092",
    archivePrefix = "arXiv",
    primaryClass = "hep-ph",
    reportNumber = "ZTF-EP-26-04",
    doi = "10.1007/JHEP09(2026)142",
    journal = "JHEP",
    volume = "09",
    pages = "142",
    year = "2026"
}

@article{Linek:2023kga,
    author = {Linek, Barbara and {\L}uszczak, Agnieszka and {\L}uszczak, Marta and Pasechnik, Roman and Sch{\"a}fer, Wolfgang and Szczurek, Antoni},
    title = "{Probing proton structure with $ \textrm{c}\overline{\textrm{c}} $ correlations in ultraperipheral pA collisions}",
    eprint = "2308.00457",
    archivePrefix = "arXiv",
    primaryClass = "hep-ph",
    doi = "10.1007/JHEP10(2023)179",
    journal = "JHEP",
    volume = "10",
    pages = "179",
    year = "2023"
}

@article{ReinkePelicer:2018gyh,
    author = {Reinke Pelicer, Mateus and Gr{\"a}ve De Oliveira, Emmanuel and Pasechnik, Roman},
    title = "{Exclusive heavy quark-pair production in ultraperipheral collisions}",
    eprint = "1811.12888",
    archivePrefix = "arXiv",
    primaryClass = "hep-ph",
    doi = "10.1103/PhysRevD.99.034016",
    journal = "Phys. Rev. D",
    volume = "99",
    number = "3",
    pages = "034016",
    year = "2019"
}

@article{Braun:2005rg,
    author = "Braun, V. M. and Ivanov, D. Yu.",
    title = "{Exclusive diffractive electroproduction of dijets in collinear factorization}",
    eprint = "hep-ph/0505263",
    archivePrefix = "arXiv",
    doi = "10.1103/PhysRevD.72.034016",
    journal = "Phys. Rev. D",
    volume = "72",
    pages = "034016",
    year = "2005"
}

@article{Hatta:2016dxp,
    author = "Hatta, Yoshitaka and Xiao, Bo-Wen and Yuan, Feng",
    title = "{Probing the Small- x Gluon Tomography in Correlated Hard Diffractive Dijet Production in Deep Inelastic Scattering}",
    eprint = "1601.01585",
    archivePrefix = "arXiv",
    primaryClass = "hep-ph",
    reportNumber = "YITP-16-1",
    doi = "10.1103/PhysRevLett.116.202301",
    journal = "Phys. Rev. Lett.",
    volume = "116",
    number = "20",
    pages = "202301",
    year = "2016"
}

@article{Altinoluk:2015dpi,
    author = "Altinoluk, Tolga and Armesto, N{\'e}stor and Beuf, Guillaume and Rezaeian, Amir H.",
    title = "{Diffractive Dijet Production in Deep Inelastic Scattering and Photon-Hadron Collisions in the Color Glass Condensate}",
    eprint = "1511.07452",
    archivePrefix = "arXiv",
    primaryClass = "hep-ph",
    doi = "10.1016/j.physletb.2016.05.032",
    journal = "Phys. Lett. B",
    volume = "758",
    pages = "373--383",
    year = "2016"
}

@article{Kowalski:2006hc,
    author = "Kowalski, H. and Motyka, L. and Watt, G.",
    title = "{Exclusive diffractive processes at HERA within the dipole picture}",
    eprint = "hep-ph/0606272",
    archivePrefix = "arXiv",
    reportNumber = "DESY-06-095",
    doi = "10.1103/PhysRevD.74.074016",
    journal = "Phys. Rev. D",
    volume = "74",
    pages = "074016",
    year = "2006"
}

@article{Boussarie:2019vmk,
    author = "Boussarie, Renaud and Hatta, Yoshitaka and Szymanowski, Lech and Wallon, Samuel",
    title = "{Probing the Gluon Sivers Function with an Unpolarized Target: GTMD Distributions and the Odderons}",
    eprint = "1912.08182",
    archivePrefix = "arXiv",
    primaryClass = "hep-ph",
    doi = "10.1103/PhysRevLett.124.172501",
    journal = "Phys. Rev. Lett.",
    volume = "124",
    number = "17",
    pages = "172501",
    year = "2020"
}

@article{Dominguez:2010xd,
    author = "Dominguez, Fabio and Xiao, Bo-Wen and Yuan, Feng",
    title = "{$k_t$-factorization for Hard Processes in Nuclei}",
    eprint = "1009.2141",
    archivePrefix = "arXiv",
    primaryClass = "hep-ph",
    doi = "10.1103/PhysRevLett.106.022301",
    journal = "Phys. Rev. Lett.",
    volume = "106",
    pages = "022301",
    year = "2011"
}

@article{Dominguez:2011wm,
    author = "Dominguez, Fabio and Marquet, Cyrille and Xiao, Bo-Wen and Yuan, Feng",
    title = "{Universality of Unintegrated Gluon Distributions at small x}",
    eprint = "1101.0715",
    archivePrefix = "arXiv",
    primaryClass = "hep-ph",
    doi = "10.1103/PhysRevD.83.105005",
    journal = "Phys. Rev. D",
    volume = "83",
    pages = "105005",
    year = "2011"
}

@article{Boer:2015pni,
    author = {Boer, Dani{\"e}l and Echevarria, Miguel G. and Mulders, Piet and Zhou, Jian},
    title = "{Single spin asymmetries from a single Wilson loop}",
    eprint = "1511.03485",
    archivePrefix = "arXiv",
    primaryClass = "hep-ph",
    doi = "10.1103/PhysRevLett.116.122001",
    journal = "Phys. Rev. Lett.",
    volume = "116",
    number = "12",
    pages = "122001",
    year = "2016"
}

@article{Boer:2016xqr,
    author = {Boer, Dani{\"e}l and Cotogno, Sabrina and van Daal, Tom and Mulders, Piet J. and Signori, Andrea and Zhou, Ya-Jin},
    title = "{Gluon and Wilson loop TMDs for hadrons of spin $\leq$ 1}",
    eprint = "1607.01654",
    archivePrefix = "arXiv",
    primaryClass = "hep-ph",
    reportNumber = "NIKHEF-2016-030",
    doi = "10.1007/JHEP10(2016)013",
    journal = "JHEP",
    volume = "10",
    pages = "013",
    year = "2016"
}

@article{Boer:2018vdi,
    author = {Boer, Dani{\"e}l and Van Daal, Tom and Mulders, Piet J. and Petreska, Elena},
    title = "{Directed flow from C-odd gluon correlations at small $x$}",
    eprint = "1805.05219",
    archivePrefix = "arXiv",
    primaryClass = "hep-ph",
    doi = "10.1007/JHEP07(2018)140",
    journal = "JHEP",
    volume = "07",
    pages = "140",
    year = "2018"
}

@article{Bertone:2022awq,
    author = "Bertone, Valerio",
    title = "{Matching generalised transverse-momentum-dependent distributions onto generalised parton distributions at one loop}",
    eprint = "2207.09526",
    archivePrefix = "arXiv",
    primaryClass = "hep-ph",
    doi = "10.1140/epjc/s10052-022-10863-3",
    journal = "Eur. Phys. J. C",
    volume = "82",
    number = "10",
    pages = "941",
    year = "2022"
}

@article{Brodsky:1997de,
    author = "Brodsky, Stanley J. and Pauli, Hans-Christian and Pinsky, Stephen S.",
    title = "{Quantum chromodynamics and other field theories on the light cone}",
    eprint = "hep-ph/9705477",
    archivePrefix = "arXiv",
    reportNumber = "SLAC-PUB-7484, MPIH-V1-1997",
    doi = "10.1016/S0370-1573(97)00089-6",
    journal = "Phys. Rept.",
    volume = "301",
    pages = "299--486",
    year = "1998"
}

@article{Pang:2026lsr,
    journal="",
    author = "Pang, Zhuoyi and Sznajder, Pawe{\l} and Szymanowski, Lech and Wagner, Jakub",
    title = "{Exclusive Quark and Gluon Dijet Production as Probes of GPDs at Collider Energies}",
    eprint = "2607.04482",
    archivePrefix = "arXiv",
    primaryClass = "hep-ph",
    month = "7",
    year = "2026"
}

@article{Chall:2026oes,
    author = {Chall, Trambak Jyoti and {\L}uszczak, Marta and Sch{\"a}fer, Wolfgang and Szczurek, Antoni},
    title = "{Probing GPDs in exclusive electroproduction of dijets}",
    eprint = "2603.09686",
    archivePrefix = "arXiv",
    primaryClass = "hep-ph",
    doi = "10.1103/p7m5-qqyx",
    journal = "Phys. Rev. D",
    volume = "113",
    number = "11",
    pages = "114012",
    year = "2026"
}

@article{Boer:2024cnw,
    author = {Boer, Dani{\"e}l and Maxia, Luca and Pisano, Cristian},
    title = "{Azimuthal asymmetries in lepton and heavy-quark pair production in UPCs}",
    eprint = "2410.23924",
    archivePrefix = "arXiv",
    primaryClass = "hep-ph",
    doi = "10.1007/JHEP01(2025)076",
    journal = "JHEP",
    volume = "01",
    pages = "076",
    year = "2025"
}

@article{Zhou:2016rnt,
    author = "Zhou, Jian",
    title = "{Elliptic gluon generalized transverse-momentum-dependent distribution inside a large nucleus}",
    eprint = "1611.02397",
    archivePrefix = "arXiv",
    primaryClass = "hep-ph",
    doi = "10.1103/PhysRevD.94.114017",
    journal = "Phys. Rev. D",
    volume = "94",
    number = "11",
    pages = "114017",
    year = "2016"
}

@article{Hagiwara:2017fye,
    author = "Hagiwara, Yoshikazu and Hatta, Yoshitaka and Pasechnik, Roman and Tasevsky, Marek and Teryaev, Oleg",
    title = "{Accessing the gluon Wigner distribution in ultraperipheral $pA$ collisions}",
    eprint = "1706.01765",
    archivePrefix = "arXiv",
    primaryClass = "hep-ph",
    reportNumber = "YITP-17-56, LU-TP-17-25",
    doi = "10.1103/PhysRevD.96.034009",
    journal = "Phys. Rev. D",
    volume = "96",
    number = "3",
    pages = "034009",
    year = "2017"
}

@article{Hagiwara:2020mqb,
    author = "Hagiwara, Yoshikazu and Hatta, Yoshitaka and Pasechnik, Roman and Zhou, Jian",
    title = "{Spin-dependent Pomeron and Odderon in elastic proton-proton scattering}",
    eprint = "2003.03680",
    archivePrefix = "arXiv",
    primaryClass = "hep-ph",
    doi = "10.1140/epjc/s10052-020-8007-6",
    journal = "Eur. Phys. J. C",
    volume = "80",
    number = "5",
    pages = "427",
    year = "2020"
}

@article{Mantysaari:2019csc,
    author = {M{\"a}ntysaari, Heikki and Mueller, Niklas and Schenke, Bj{\"o}rn},
    title = "{Diffractive Dijet Production and Wigner Distributions from the Color Glass Condensate}",
    eprint = "1902.05087",
    archivePrefix = "arXiv",
    primaryClass = "hep-ph",
    doi = "10.1103/PhysRevD.99.074004",
    journal = "Phys. Rev. D",
    volume = "99",
    number = "7",
    pages = "074004",
    year = "2019"
}

@article{Collins:2003fm,
    author = "Collins, John C.",
    title = "{What exactly is a parton density?}",
    eprint = "hep-ph/0304122",
    archivePrefix = "arXiv",
    journal = "Acta Phys. Polon. B",
    volume = "34",
    pages = "3103",
    year = "2003"
}

@book{Collins:2011zzd,
    author = "Collins, John",
    title = "{Foundations of Perturbative QCD}",
    doi = "10.1017/9781009401845",
    isbn = "978-1-009-40184-5, 978-1-009-40183-8, 978-1-009-40182-1",
    publisher = "Cambridge University Press",
    volume = "32",
    year = "2011"
}

@article{Radyushkin:2011dh,
    author = "Radyushkin, A. V.",
    title = "{Generalized Parton Distributions and Their Singularities}",
    eprint = "1101.2165",
    archivePrefix = "arXiv",
    primaryClass = "hep-ph",
    reportNumber = "JLAB-THY-11-06",
    doi = "10.1103/PhysRevD.83.076006",
    journal = "Phys. Rev. D",
    volume = "83",
    pages = "076006",
    year = "2011"
}

@article{Buffing:2013kca,
    author = "Buffing, M. G. A. and Mukherjee, A. and Mulders, P. J.",
    title = "{Generalized Universality of Definite Rank Gluon Transverse Momentum Dependent Correlators}",
    eprint = "1306.5897",
    archivePrefix = "arXiv",
    primaryClass = "hep-ph",
    reportNumber = "NIKHEF-2013-019",
    doi = "10.1103/PhysRevD.88.054027",
    journal = "Phys. Rev. D",
    volume = "88",
    pages = "054027",
    year = "2013"
}

@article{Hatta:2017cte,
    author = "Hatta, Yoshitaka and Xiao, Bo-Wen and Yuan, Feng",
    title = "{Gluon Tomography from Deeply Virtual Compton Scattering at Small-x}",
    eprint = "1703.02085",
    archivePrefix = "arXiv",
    primaryClass = "hep-ph",
    reportNumber = "YITP-17-22",
    doi = "10.1103/PhysRevD.95.114026",
    journal = "Phys. Rev. D",
    volume = "95",
    number = "11",
    pages = "114026",
    year = "2017"
}

@article{Metz:2011wb,
    author = "Metz, Andreas and Zhou, Jian",
    title = "{Distribution of linearly polarized gluons inside a large nucleus}",
    eprint = "1105.1991",
    archivePrefix = "arXiv",
    primaryClass = "hep-ph",
    doi = "10.1103/PhysRevD.84.051503",
    journal = "Phys. Rev. D",
    volume = "84",
    pages = "051503",
    year = "2011"
}

@article{Boer:2022njw,
    author = {Boer, Dani{\"e}l and Hagiwara, Yoshikazu and Zhou, Jian and Zhou, Ya-jin},
    title = "{Scale evolution of T-odd gluon TMDs at small x}",
    eprint = "2203.00267",
    archivePrefix = "arXiv",
    primaryClass = "hep-ph",
    doi = "10.1103/PhysRevD.105.096017",
    journal = "Phys. Rev. D",
    volume = "105",
    number = "9",
    pages = "096017",
    year = "2022"
}

@article{Ji:2016jgn,
    author = "Ji, Xiangdong and Yuan, Feng and Zhao, Yong",
    title = "{Hunting the Gluon Orbital Angular Momentum at the Electron-Ion Collider}",
    eprint = "1612.02438",
    archivePrefix = "arXiv",
    primaryClass = "hep-ph",
    doi = "10.1103/PhysRevLett.118.192004",
    journal = "Phys. Rev. Lett.",
    volume = "118",
    number = "19",
    pages = "192004",
    year = "2017"
}

\end{document}